\documentclass[11pt,tightenlines,superscriptaddress,notitlepage,aps,pra,longbibliography,nofootinbib]{revtex4-1}

\usepackage{color}
\usepackage[usenames,dvipsnames]{xcolor}
\usepackage{algorithm}
\usepackage{algpseudocode}

\usepackage{amsmath,amsfonts,amssymb,amsthm}
\usepackage{graphicx}
\usepackage{fullpage}
\PassOptionsToPackage{caption=false}{subfig}
\usepackage{subfig}
\usepackage{enumerate}
\usepackage{bbm}

\usepackage{tikz}
\usetikzlibrary{arrows,decorations.pathmorphing,backgrounds,positioning,fit}

\usepackage{epstopdf} 
\usepackage{bm}  

\usepackage[pdfpagelabels,pdftex,bookmarks,breaklinks]{hyperref}
\definecolor{darkblue}{RGB}{0,0,127} 
\definecolor{darkgreen}{RGB}{0,150,0}
\hypersetup{colorlinks, linkcolor=darkblue, citecolor=darkgreen, filecolor=red, urlcolor=blue}
\hypersetup{pdftitle={Thermalization, Error-Correction, and Memory Lifetime for Ising Anyon Systems}, pdfauthor={Courtney G. Brell, Simon Burton, Guillaume Dauphinais, Steven T. Flammia, David Poulin}}

\def\C{\mathbb{C}}

\def\e{\mathrm{e}}

\newcommand{\Eref}[1]{Eq.~(\ref{#1})}
\newcommand{\Sref}[1]{Sec.~\ref{#1}}
\newcommand{\Fref}[1]{Fig.~\ref{#1}}
\newcommand{\Aref}[1]{Appendix~\ref{#1}}

\newcommand{\ket}[1]{|{#1}\rangle}

\newcommand{\ketbra}[2]{|{#1}\rangle\!\langle{#2}|}
\newcommand{\braket}[2]{\langle{#1}|{#2}\rangle}
\newcommand{\proj}[1]{\ketbra{#1}{#1}}

\newcommand*{\one}{\mathbbm{1}} 

\newcommand*{\vac}{\mathbb{I}}

\newcommand*{\lbr}{\mathcal{L}}

\begin{document}

\title{Thermalization, Error-Correction, and Memory Lifetime\\for Ising Anyon Systems}

\author{Courtney G.\ Brell}
\author{Simon Burton}
\affiliation{Centre for Engineered Quantum Systems, School of Physics, The University of Sydney, Sydney, Australia}

\author{Guillaume~Dauphinais}
\affiliation{D\'{e}partement de Physique, Universit\'{e} de Sherbrooke, Sherbrooke, Qu\'{e}bec, Canada}

\author{Steven T.\ Flammia}
\affiliation{Centre for Engineered Quantum Systems, School of Physics, The University of Sydney, Sydney, Australia}

\author{David Poulin}
\affiliation{D\'{e}partement de Physique, Universit\'{e} de Sherbrooke, Sherbrooke, Qu\'{e}bec, Canada}

\begin{abstract}
We consider two-dimensional lattice models that support Ising anyonic excitations and are coupled to a thermal bath. We propose a phenomenological model for the resulting short-time dynamics that includes pair-creation, hopping, braiding, and fusion of anyons. By explicitly constructing topological quantum error-correcting codes for this class of system, we use our thermalization model to estimate the lifetime of the quantum information stored in the encoded spaces. To decode and correct errors in these codes, we adapt several existing topological decoders to the non-Abelian setting. We perform large-scale numerical simulations of these two-dimensional Ising anyon systems and  find that the thresholds of these models range between 13\% to 25\%. 
To our knowledge, these are the first numerical threshold estimates for quantum codes without explicit additive structure.
\end{abstract}

\maketitle

\section{Introduction}\label{S:intro}

One of the most interesting features of two-dimensional quantum systems with anyonic excitations~\cite{Wilczek1990} is their application to quantum information science, where the topological nature of the anyons offers some intrinsic protection of quantum coherence~\cite{Kitaev2003}. These topologically ordered systems are insensitive to local perturbations~\cite{Bravyi2010, Bravyi2011a, Michalakis2013}, have a ground-space degeneracy that is a function of the topology of the system~\cite{Wen1990, Einarsson1995} and can realize quantum gates by braiding anyons, which are naturally robust to error due to their topological nature~\cite{Kitaev2003, Nayak2008}. Many common anyon models such as the toric code~\cite{Kitaev2003} and the color codes~\cite{Bombin2006} have an Abelian structure that simplifies their analysis greatly. However, Abelian anyon models are very restricted when compared to the general class of anyonic systems. Notably, Abelian anyons lack the capacity to perform quantum computation by braiding of anyons alone. By contrast, non-Abelian anyon systems have excitations that can be used to implement topological quantum computation~\cite{Freedman2002, Freedman2002b}, and so they are of interest for quantum information processing schemes as well as quantum error correction. It is also known that the dynamics of Abelian anyons can significantly differ from those of non-Abelian anyons in some circumstances~\cite{Brennen2010, Lehman2011, Lehman2012, Zatloukal2012}.

A particular class of non-Abelian anyonic excitations called \emph{Ising anyons} are thought to be among the most promising candidates for experimental realization of topological quantum computing. Ising anyons are experimentally motivated~\cite{Willett1987} as the expected excitations of the $\nu=\frac{5}{2}$ fractional quantum Hall states~\cite{Moore1991, Nayak1996}, and also appear as the excitations of several lattice models~\cite{Kitaev2006, Levin2005a, Kapit2013, Palumbo2014}. In particular, Ising anyons can robustly perform Clifford gates via braiding and have a high noise threshold for universal quantum computation when coupled with magic state distillation~\cite{Bravyi2006b, Freedman2006}. 

These general features of anyon models, and of Ising anyons in particular, have attracted a lot of attention during the past decade. This picture of an intrinsically robust topological quantum computer is spoiled at finite temperature, however. Indeed, if left unattended, thermal excitations diffusing in the system can corrupt the quantum information stored in the ground space. While these excitations are suppressed by a mass gap (ideally) much larger than the temperature, they nonetheless appear in a finite density at any non-zero temperature, and for scalability issues it is essential to devise a scheme that can cope with their existence. One approach is to engineer a mass gap that grows with the system size, leading to a self-correcting behavior of the memory. However, there is accumulating evidence that this cannot be realized in two spatial dimensions~\cite{Bravyi2009, Bravyi2010a, Haah2012a, Landon-Cardinal2012a}. Additionally, even a self-correcting system in contact with a thermal bath will accumulate a finite density of excitations over time, and it is not clear that the quantum information stored in the ground state can be manipulated or read out without first cooling back to the correct ground state~\cite{Pastawski2010}.

Here we pursue a different approach that can be seen as a viable alternative to self-correction as well as a reliable method for readout of noisy stored quantum information. It consists of actively monitoring the presence of thermal anyons and correcting the unwanted operations resulting from their presence in the system. The Ising anyon model is an obvious candidate to benchmark such a proposal because it is numerically tractable via a mapping to free Majorana fermions~\cite{Bravyi2006b}.

The motivation behind our work stems from two distinct potential applications. On the one hand, as outlined above, anyons can arise as localized gapped excitations in two-dimensional systems, e.g.,~\cite{Moore1991, Nayak1996}. In this setting, our method is needed to eliminate thermal defects in the systems without spoiling the topologically encoded information. On the other hand, our results can also be understood from a purely coding-theoretic perspective. Some topologically ordered systems can arise in lattice models with local commuting Hamiltonians~\cite{Kitaev2003, Levin2005a, Koenig2010b}. The local terms of the Hamiltonian can be thought of as check operators defining a \emph{local commuting projector code}, a class of codes going beyond the usual stabilizer formalism that has been the object of recent study~\cite{Bravyi2009, Bravyi2010a, Haah2012a, Landon-Cardinal2012a}. Such codes allow for the implementation of quantum error-correction schemes in a geometrically local setting~\cite{Kitaev2003, Dennis2002}. The local check operators can be measured using standard circuitry~\cite{Bonesteel2012} on an ordinary, circuit-model quantum computer. Thus, this second scenario we consider is not contingent on the existence of physical systems that naturally present topological quantum order; it could in principle be used in any quantum computing architecture with nearest-neighbor interactions on a two-dimensional lattice. In this setting, our results provide an efficient decoding algorithm for the corresponding quantum error-correcting code. To our knowledge, this provides the first efficient decoding schemes for a family of non-additive quantum codes, i.e.\ codes without explicit Pauli-matrix tensor product structure.

\subsection{Summary of main results}

We begin by proposing a phenomenological model of dynamics in an Ising anyon system and several noise models. Importantly, these models encompass all of the major physical error processes associated to non-Abelian anyons, including pair-creation, hopping, braiding, and fusion of anyons. We define two distinct families of noise models, following the two physical applications of our method described in the previous paragraph. To simulate the noise affecting a topologically-ordered system at finite temperature, we use a Metropolis procedure to choose among the possible physical error processes. Because the Gibbs thermal equilibrium state is the fixed point of this local noise process, we believe that it captures the essential properties of a real thermalization process. To simulate the noise affecting a quantum computer that uses an Ising error-correcting code, we use a more generic ``white noise" model where each type of physical error process occurs at a predetermined fixed rate. This model is in essence an infinite temperature limit of the previous noise model, and so it probably also accurately captures the short-time dynamics of a topologically-ordered system at finite temperature. Note that we do not necessarily expect the physical noise occurring in a circuit-based quantum computer to be described by these physical processes. Nonetheless, in absence of detailed microscopic descriptions of the device, they provide a good starting point to benchmark the code (in the same sense that Pauli errors are used to benchmark stabilizer codes). 

We explore the effect of these noise channels on two different encoding schemes (degenerate ground spaces). The first encoding we examine stores the information in the degenerate ground space of a system with Ising anyonic excitations embedded in a torus while the second stores information in a degenerate space associated with the fusion space of several well-separated Ising anyons. Both of these encodings lend themselves equally well to the two physical scenarios described above (although embedding an actual topologically ordered system on a torus probably poses an additional challenge). These two encodings are motivated by topological quantum memories \cite{Dennis2002} and topological quantum computing \cite{Bravyi2006b,BR07a} schemes respectively. 

Following this, we adapt existing techniques from topological quantum error correction to the non-Abelian setting. The non-Abelian nature of the model significantly changes the decoding problem and presents a main challenge of our study. In an Abelian model, decoding is realized by grouping nearby excitations that fuse to the vacuum. In contrast, for a non-Abelian model, it is not possible to predict with certainty the outcome of a fusion process. Thus, decoding needs to be an \emph{iterative} process: in a first round particles are grouped in a certain way; if all groups do not fuse to the vacuum, then a second round groups the remaining particles, and so on. In particular, we adapt the clustering renormalization group decoder of Bravyi and Haah~\cite{Bravyi2011} and the perfect matching algorithm decoders~\cite{Dennis2002, Wang2010} to our non-Abelian topological codes to determine their error-correction thresholds and memory lifetimes under various noise channels. Our decoders provably run in polynomial time in the size of the lattice. 

Using these decoders, we explicitly calculate numerical error-correction threshold estimates for these codes and for various noise regimes, including those where hopping and braiding of anyons dominates pair-creation. Our results seem to indicate that the perfect matching decoder outperforms the clustering decoder, and that the thresholds are broadly comparable for both types of codes, and for various types of noise. We find numerical threshold estimates that vary between 13\% and 25\%, where the percentages are in terms of a total noise rate density which we define and discuss in detail in Sections~\ref{S:phenom} and~\ref{S:numerics}. 

These thresholds are obtained by large-scale two-dimensional simulations of the dynamics of a dense ``gas'' of non-abelian anyons on lattice sizes of up to $48 \times 48$ nodes. All previous results have focused on quasi-one-dimensional models or single-particle quantum walks~\cite{Brennen2010, Lehman2011, Lehman2012, Zatloukal2012}.

The paper is organized as follows. In \Sref{S:phenom} we introduce the Ising anyons, as well as our phenomenological models for their dynamics. Next, \Sref{S:sim} goes into some detail about the efficient simulation of these dynamics on a classical computer. Following this in \Sref{S:code} we give explicit constructions for the topological quantum codes that we study. We then detail our decoding algorithms in \Sref{S:algo} and present numerical threshold estimates in \Sref{S:numerics}. We discuss the significance of these numerical results in \Sref{S:conclusion}, and offer two appendices, the first of which summarizes the definition of the Ising anyons, and the second discusses the regime of validity of one of the noise models we analyse.

\section{Phenomenological Model for Ising Anyon Dynamics}\label{S:phenom}

An anyon model is specified by its (gapped) excitation types as well as several different ways these excitations can interact, specifically by braiding, fusion, and splitting. These three processes in fact completely characterize the universal topological properties of a system with anyonic excitations. Other processes will depend on microscopic details of the system and will be largely irrelevant for the behavior of the global topological degrees of freedom of the system that we will be interested in. For this reason they can safely be ignored in this work.

We will now describe a phenomenological model of these topological dynamical processes that allows us to simulate the dynamics of Ising anyons. We will use this to discuss several important classes of noise models and quantum codes and study the effects of these noise models on the codes we define.

The Ising anyon model consists of two distinct non-trivial excitation types (or charges) conventionally labelled $\psi$ and $\sigma$, and for convenience we label the vacuum (no particle) by $\vac$.  Since this is a non-Abelian anyon model, there are multiple ways in which two anyonic charges can combine by fusion. These are specified by the fusion rules~\cite{Rowell2009}
\begin{align}
	\psi \times \psi &= \vac\quad\quad
	&\psi \times \sigma &= \sigma\quad\quad
	&\sigma \times \sigma &= \vac+\psi \,.
\end{align}
The full data specifying Ising anyon dynamics are given in \Aref{A:isingdata} and describe the braiding, fusion, and splitting of Ising anyons. 

We consider dynamics on a discretized space defined by a graph $G = G(V,E)$. Associated with each site $i \in V$ of the graph are a set of occupation variables $n_q^i$ (taking nonnegative integer values) which specify the number of particles with charge $q\in\{\psi,\sigma\}$ located at $i$. The collection of all occupation variables specifies the charge configuration on the lattice. Although we define charge configurations with multiple particles at each site of the graph, we will mostly regard all particles at a given site as physically being fused together, though we denote them separately for convenience. 

In order to complete the specification of the total state of the system, we also require a Hilbert space, which we call the fusion space, associated with fusion outcomes of any $\sigma$ particles on the lattice ($\psi$ particles have unique fusion outcomes, and so will not contribute to this space). For a charge configuration with $2m$ $\sigma$ particles (recall that $\sigma$ particles can only be created or destroyed in pairs), the fusion space will be $2^{m}$ dimensional.

In order to model the most general allowed processes on a topological system, processes of braiding, fusion, and splitting are most easily recast as the following elementary operations on the graph $G$.
\begin{enumerate}
	\item Creation of particle-antiparticle pairs of the form $q\times \bar{q}$ for some $q\in{\psi,\sigma}$ on sites incident to edge $(i,j)\in E$;
	\item Hopping of all charges from one site $i$ to a neighboring site $j$;
	\item Exchange of all the charges from one site $i$ with those on a neighboring site $j$ either clockwise or anticlockwise;
	\item Decoherence of the total charge of a site $i$.
\end{enumerate}
Here we use the notation $\bar{q}$ for the antiparticle of $q$, i.e., the particle type such that $q\times\bar{q}=\vac+\ldots$. For Ising anyons, each particle type is its own antiparticle. The decoherence process is the only one of these four which may require some explanation: this process corresponds to projecting into a particular fusion outcome of the set of charges at a site. In general, our model allows superpositions of different fusion outcomes at a site, and depending on the physical system we are attempting to model these may tend to decohere slowly or rapidly.

Our aim is to simulate the dynamics of a system with Ising anyon excitations in contact with a thermal bath for a short period of time. In order to do this, we have two complimentary methods for sampling from the above processes.

\subsection{Fixed Rate Sampling}\label{S:fixedrate}

The first sampling mechanism is the most naive and the most general. It proceeds by taking a set of rates for each fundamental anyonic noise process: $\gamma_c^q$ for pair creation of charge $q$, $\gamma_h$ for hopping and $\gamma_e$ for exchange. Since decoherence is a slightly different type of noise process, we will treat it separately for convenience, and associate a decoherence probability  $p_d$ to decoherence events. Since we will be interested in calculating memory lifetimes for a fixed strength noise channel, any rescaling of the sum of the $\gamma$ noise rates will simply correspond to a rescaling of the memory lifetime. For this reason, the rates $\gamma$ should simply be regarded as \emph{relative} rates of the different noise processes, and we will always normalize the total rate to 1.

We simulate the dynamics of the system for $T$ time steps with $T$ drawn from a Poisson distribution of mean $T_0$. At each step a single operation from the set $\{$pair creation, hopping, exchange$\}$ will be performed. Specifically, this proceeds by randomly choosing a directed edge $e=(i,j)$ uniformly from the graph, and then selecting from the nontrivial processes allowed on that edge by their relative rates. We call a process trivial if it leaves the state invariant. The states on which each process may act nontrivially are summarized in the Table~\ref{T:sim}, where we use $q$, $q'$ to denote any non-vacuum anyonic charge.

\begin{table}[t]
\centering
\begin{tabular}{cc|c|c|c|}
\cline{2-5}
&\multicolumn{4}{|c|}{Charges at $i$, $j$}\\
\hline
\multicolumn{1}{|c|}{Process}&$\vac,\vac$&$\vac,q$&$q,\vac$&$q,q'$\\
\hline
\multicolumn{1}{|c|}{Pair Creation} &\checkmark &\checkmark &\checkmark &\checkmark\\
\multicolumn{1}{|c|}{Hopping} & & &\checkmark &\checkmark\\
\multicolumn{1}{|c|}{Exchange} & &\checkmark &\checkmark &\checkmark\\
\hline
\end{tabular}\caption{\label{T:sim}Allowed elementary noise processes for our phenomenological noise model of anyon dynamics. The check marks denote allowed processes on a directed edge $i \to j$ of the underlying graph $G$, and $q$,$q'$ denote non-vacuum anyon charges.}
\end{table}

Given the set of nontrivial processes that are allowed on the selected edge, an operation to perform is chosen according to the relative rates $\gamma$ of these nontrivial processes. Following each such operation, the decoherence process is applied to every site with probability $p_d$. After $T$ time steps, this will approximately correspond to a simulation of the system running for time $t_{\mathrm{sim}}\propto\frac{T_0}{|E|}$ for a graph with $|E|$ edges.

This fixed rate sampling mechanism is a natural noise model in a quantum computing architecture which directly implements our error correcting codes, and may also be a good approximation to short-timescale thermalization processes in Ising-anyon quantum systems for appropriately chosen rates. However, the timescale on which it is a good approximation in this latter case may be very small. For this reason, we also consider an alternative sampling mechanism.

\subsection{Metropolis Sampling}\label{S:metropolis}

In order to more faithfully capture the thermalization process in our model, we consider a second sampling mechanism based on the Metropolis sampling method. In order to specify a Metropolis procedure we give a mass to each particle type $m_{\psi}$ and $m_{\sigma}$, and then define the Hamiltonian
\begin{align}
H=\sum_{i\in V} \left(m_{\psi}P^{\psi}_i+m_{\sigma}P^{\sigma}_i\right)
\end{align}
for $P_i^q$ the projector to total charge $q$ on site $i$.

The Metropolis method proceeds by proposing an elementary local noise operation following a fixed distribution $P(A\rightarrow B)$, and then calculating the energy difference between the state before and after this operation $\Delta_E$. The operation is then accepted with probability $\mathrm{min}\{1,\e^{-\beta\Delta_E}\}$ for a system at inverse temperature $\beta = 1/k_BT$. Provided that the distribution $P(A\rightarrow B)$ is reversible and ergodic, this procedure guarantees convergence to a thermal state obeying detailed balance.

Although the process we defined is not technically ergodic over the entire Hilbert space of our system as defined earlier, it is ergodic over a restricted space accessible to topological operations that respect superselection rules, etc. This restricted space is similar in spirit to the gauge-invariant subspace of a lattice gauge theory, and is evidenced by the block-diagonal structure of the transition matrix at Table~\ref{transition}. While the system may be defined most naturally on a larger Hilbert space, the structure of the model is such that only a smaller ``physical'' Hilbert space is of interest. When we talk about ergodicity or the thermal state, we consider it to be defined only on the appropriate smaller physical space.

Unfortunately, the hopping and pair-creation processes we wish to consider will not always take energy eigenstates to energy eigenstates, and so it will be convenient to consider a restricted set of noise operations that will ensure that our system remains in an energy eigenstate at each time step of its evolution. To achieve this, after each operation is applied we decohere every site of the lattice, projecting into an energy eigenstate of the system. This will allow us to consistently define Metropolis acceptance probabilities at the cost of taking the extreme decoherence limit of our noise model. Because it is a local noise process that converges to the thermal state, satisfying the detailed balanced condition, we believe that this sampling method captures the essence of real thermalization processes.

The transition rules $P(A\rightarrow B)$ we have chosen are as follows. We pick a directed edge $E = (i, j)$ at random. We then create a pair of particles $(q,q)$ from the vacuum with some probability $P(q)$, bring them to sites $i$ and $j$, and finally measure the total charge at each site. The probability $P(q)$ is actually conditional on the charges $q_i, q_j$, so we write it $P(q|q_i,q_j)$, and it is given by
\begin{align}
P(\vac|q_i,q_j) &= \frac 12\label{eq:transition1}\\
P(\psi|q_i,q_j) &= \left\{
\begin{array}{ll}
 0 & {\rm \ if \ }q_i=q_j=\sigma \\
\frac 14 & {\rm \ else} 
\end{array}
\right.\\
\label{eq:transition}P(\sigma|q_i,q_j) &= \left\{
\begin{array}{ll}
 \frac 12 & {\rm \ if \ }q_i=q_j=\sigma \\
\frac 14 & {\rm \ else} 
\end{array}
\right.
\end{align}
These rates imply the transition matrix shown at Table~\ref{transition}. We see by inspection that $P(A\rightarrow B) = P(B\rightarrow A)$, so this choice is reversible and hence a detailed balance condition follows. We note that these are not the only choices of transition rates that lead to detailed balance; we have chosen these conventions for simplicity.

\begin{table}[t]
\centering
\begin{tabular}{c||ccc|ccc|cccc}
 & $(\vac,\vac)$ & $(\psi,\psi)$ & $(\sigma,\sigma;\vac)$ & $(\vac,\psi)$ & $(\psi,\vac)$ &  $(\sigma,\sigma;\psi)$ &  $(\vac,\sigma)$ & $(\sigma,\vac)$ & $(\psi,\sigma)$ & $(\sigma,\psi)$ \\
 \hline
 \hline
$(\vac,\vac)$ & $\frac 12$  & $\frac 14$ & $\frac 14$ & 0 & 0 & 0 &0 & 0 & 0 & 0\\ 
$(\psi,\psi)$  & $\frac 14$ & $\frac 12$ &$\frac 14$  & 0 & 0&0&0 & 0 & 0 & 0\\
$(\sigma,\sigma;\vac)$  & $\frac 14$ & $\frac 14$  & $\frac 12$  & 0 & 0 & 0 & 0 & 0 & 0 & 0 \\
\hline
$(\psi,\vac)$  & 0 &  0&  0& $\frac 12$ & $\frac 14$ &$\frac 14$ &0 & 0 &0& 0\\ 
$(\vac,\psi)$  & 0 & 0 & 0& $\frac 14$ & $\frac 12$ &$\frac 14$ &0 & 0 &0&0\\ 
$(\sigma,\sigma;\psi)$  & 0 & 0 & 0 & $\frac 14$  & $\frac 14$  & $\frac 12$  & 0 & 0 & 0 & 0 \\
\hline
$(\vac,\sigma)$  & 0 & 0 & 0 & 0& 0 & 0 & $\frac 12$ & $\frac 18$ & $\frac 14$ & $\frac 18$  \\
$(\sigma,\vac)$  & 0 & 0 & 0 & 0& 0 & 0 & $\frac 18$ & $\frac 12$ & $\frac 18$ & $\frac 14$  \\
$(\psi,\sigma)$  & 0 & 0 & 0 & 0& 0 & 0 & $\frac 14$ & $\frac 18$ & $\frac 12$ & $\frac 18$  \\
$(\sigma,\psi)$  & 0 & 0 & 0 & 0& 0 & 0 & $\frac 18$ & $\frac 14$ & $\frac 18$ & $\frac 12$  \\
\end{tabular} 
\caption{Transition matrix $P(A\rightarrow B)$ following from Eqs.~(\ref{eq:transition1}-\ref{eq:transition}).The three blocks have topological charge $\vac$, $\psi$, and $\sigma$ respectively.}
\label{transition}
\end{table}

We can consider this a \emph{semiclassical} noise model, since we effectively disallow superpositions of different total charges at each site. We discuss the limitations of this noise model in \Aref{A:semiclassical}.

\section{Simulating Ising anyon phenomenology}\label{S:sim}

It is known that the dynamics of Ising anyons can be efficiently simulated~\cite{Bravyi2006b}. By this we mean that the operations induced on the fusion space and charge configuration space by pair-creation, braiding, and fusion of $n$ Ising anyons can be simulated, modulo global phases, by a classical computer in time polynomial in $n$. We will make use of this result to simulate the dynamics of our phenomenological model under various noise channels. This will allow us to determine an error-correction threshold for our system by observing when it is able to preserve quantum information.

Recall that in the Ising anyon model, the $\psi$ particles do not contribute to the fusion space as they have unique fusion outcomes. Thus we can associate the fusion space with the set of $\sigma$ particles. In order to explicitly quantify the effects of braiding operations on this fusion space, it is convenient to assign a Majorana fermion mode to each $\sigma$ particle present in our system. This allows us to define a set of Majorana operators $\hat{c}_{\alpha}$ for each $\sigma$ particle ${\alpha}$ with
\begin{align}
	\hat{c}_{\alpha}\hat{c}_{\beta}&=2\delta_{\alpha\beta}-\hat{c}_{\beta}\hat{c}_{\alpha}\\
	\hat{c}_{\alpha}^\dagger &= \hat{c}_{\alpha}
\end{align}
For $2m$ $\sigma$ particles, this defines a $2^m$ dimensional space which will serve as the fusion space for these particles. A simulation of Ising anyon dynamics requires the simulation of both the charge configurations and their associated fusion spaces.

\subsection{Planar graph simulations}
For simplicity, let us first describe a simulation where $G$ is a planar graph. With this constraint, all physical operations will preserve global charge. That is, any physical observable $O$ satisfies
\begin{align}
	[O,Q]&=0 \,
\end{align}
where $Q$ is the total charge operator
\begin{align}
	Q &= (-i)^m\hat{c}_1\hat{c}_2\cdots\hat{c}_{2m} \,.
\end{align}
The eigenvalues of $Q$ correspond to the possible total charges of all $\sigma$ particles, and after fixing this global charge to be vacuum, the remaining fusion space is $2^{m-1}$ dimensional.

Braiding processes between anyons act as unitary transformations on the fusion space. In order to specify the transformation corresponding to a braid operation between anyons on adjacent sites of $G$, we must first specify a basis for this space by defining an ordering of the $\sigma$ particles in the system. It is convenient to choose the ordering of the $\sigma$ particles relative to a natural ordering of the sites in the graph $G$. This is given by a bijective function
	\[\#_G: V\to \{1,\ldots,|V|\}\]
over sites of $G$ (i.e.\ simply a fixed labeling of sites). We can then choose an ordering function $\#_\sigma$ for the $\sigma$ particles in our system in a way that is consistent with the $\#_G$ order. By consistent, we mean that for $\sigma$ particles $\alpha$ and $\beta$ at sites $i$ and $j$ respectively with $\#_G(i)<\#_G(j)$, we require that $\#_\sigma(\alpha)<\#_\sigma(\beta)$. For clarity, we will conventionally use Roman letters to refer to sites of $G$, while using Greek letters to denote $\sigma$ particles in the $\#_{\sigma}$ ordering.

Braiding operations involving $\psi$ particles act only on the fusion space by accumulating global phases (see \Aref{A:isingdata} for details). For this reason, we only need explicitly consider the action on the fusion space of braiding operations between $\sigma$ particles. The most convenient generating set of operators for braiding and fusion of $\sigma$ particles is that given by operations between $\#_{\sigma}$-neighboring particles. Note that we will conventionally take the subscript of the Majorana operators $\hat{c}_{\alpha}$ to refer to this ordering of the corresponding $\sigma$ particle. When performing a clockwise braid between two $\#_{\sigma}$-adjacent particles $\sigma_{\alpha}$ and $\sigma_{{\alpha}+1}$, we enact the following operation on the fusion space (up to global phases)~\cite{Bravyi2006b}
\begin{align}
	B_{\alpha} &=\frac{1}{\sqrt{2}}\left(\one-\hat{c}_{\alpha}\hat{c}_{{\alpha}+1}\right)\,.
\end{align}		
The anticlockwise exchange of these particles corresponds to the operation
\begin{align}
	B_{\alpha}^{\dagger}&=\frac{1}{\sqrt{2}}\left(\one+\hat{c}_{\alpha}\hat{c}_{{\alpha}+1}\right)\,.
\end{align}
These operators can be seen to map the Majorana operators as follows:
\begin{align}
	B_\alpha\hat{c}_\alpha B_\alpha^{\dagger}&=\hat{c}_{\alpha+1}\label{e:braid1a}\\
	B_\alpha\hat{c}_{\alpha+1}B_\alpha^{\dagger}&=-\hat{c}_{\alpha}\\
	B_\alpha^{\dagger}\hat{c}_\alpha B_\alpha&=-\hat{c}_{\alpha+1}\\
	B_\alpha^{\dagger}\hat{c}_{\alpha+1}B_\alpha&=\hat{c}_{\alpha}\label{e:braid1d} \, .
\end{align}

We track the state of our system in the fusion space in the Heisenberg representation; that is, we will represent the state by a group of (commuting) stabilizer operators whose common $+1$ eigenspace defines the desired state. For each pair of $\sigma$ particles in our system, we will include an additional generator for the stabilizer group $S$. The stabilizer group will always contain the total charge operator $Q$.

As well as the operation on the fusion space corresponding to braiding particles, we must also specify an analogous operator corresponding to fusion of neighboring particles.
The fusion product (corresponding to the combined charge) of a pair of $\#_{\sigma}$-neighboring particles is represented by the operator
\begin{align}\label{e:measure2}
	M_{\alpha}^{(2)} = -i\hat{c}_\alpha\hat{c}_{\alpha+1} \,.
\end{align}
The $+1$ eigenvalue corresponds to combined vacuum charge, while the $-1$ eigenvalue corresponds to combined $\psi$ charge. When two $\sigma$ particles are created from vacuum, two new modes are created and initialized in the $+1$ eigenstate of a corresponding $M^{(2)}$ operator.

Similarly, the fusion product of $2l$ neighboring $\sigma$ anyons is given by
\begin{align}\label{e:measurel}
	M_{\alpha}^{(2l)} = (-i)^l \hat{c}_\alpha\hat{c}_{\alpha+1}\hat{c}_{\alpha+2}\cdots\hat{c}_{\alpha+2l-1} \,.
\end{align}		
Decoherence or measurement of the total charge of a set of $\sigma$ particles corresponds to a measurement of the relevant $M$ operator.

We are able to neglect global phases acquired during braiding operations as long as we guarantee that our system remains in a common charge eigenstate at each stage of the simulation. That is, as long as the occupation variables $n_{\psi}$ and $n_{\sigma}$ can be treated classically. Our phenomenological model involves only processes which take charge eigenstates to charge eigenstates, and so the (unitary) braiding and (projective) measurement operations above suffice to generate all required dynamics of our system.

Although the braid and measurement operations introduced above are enough in principle to specify all the dynamics of our system, there is no sense in which they correspond to the elementary processes of our phenomenological model of dynamics. The phenomenological processes are operations on the graph $G$, while the braid moves described by (\ref{e:braid1a})-(\ref{e:braid1d}) corresponds to operations on the line specified by the $\#_{\sigma}$ ordering. In order to translate between these two pictures, we introduce the notion of linearized braid sequences.

\subsection{Linearized braid sequences}

In order to explicitly compute the effect on the fusion space of an operation involving $G$-neighboring sites (e.g.~exchange), we must map it to a sequence of $\#_\sigma$-neigboring operations. We will call this sequence the ``linearized'' braid sequence corresponding to a graph edge $e=(i,j)$. This will give us a prescription to take some subset of the charges at $i$ to site $j$, where another operation may be performed such as exchange or hopping.

\begin{figure}
	\centering
	\includegraphics{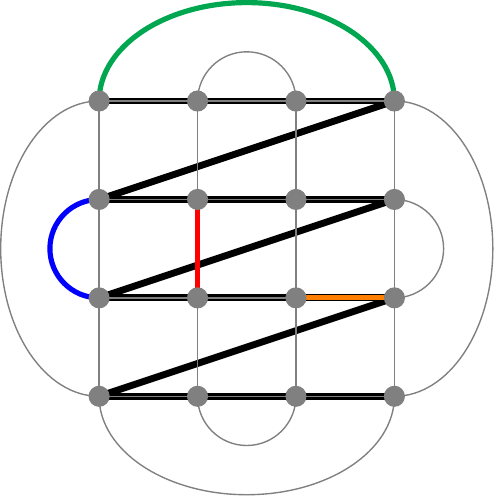}
	
	\vspace{0.75cm}
	
	\includegraphics{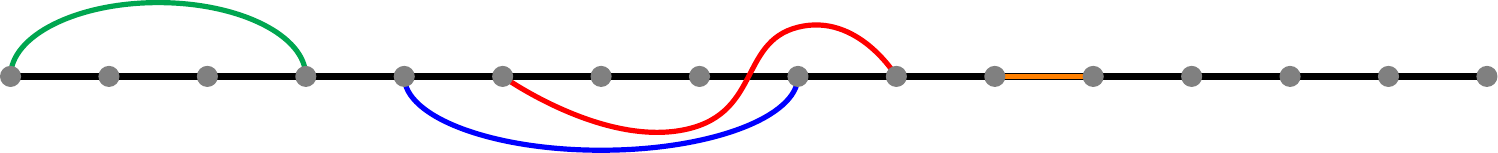}

	\caption{An example choice of $G$ (shown with gray edges) and $\#_G$ (shown with solid black edges). We also highlight several edges of $G$ to demonstrate the map between $G$-neighbors and paths over $\#_G$. After embedding $G$ in the plane, we overlay a linear graph corresponding to the $\#_G$ ordering. Following this, we can map any desired lattice edge to a braid sequence on the line smoothly.}\label{f:braidmap-sphere}
\end{figure}

The linearized braid sequence is most easily understood graphically, and is shown in \Fref{f:braidmap-sphere}. To define it, we embed $G$ in the plane, and overlay a planar path connecting each node of $G$ in the order $\#_G$. We can then smoothly deform this graph until the $\#_G$ path is a straight line, and the image of an edge $e=(i,j)\in E$ under this deformation provides us with the prescription to translate a $G$-neighboring operation on $e$ into a sequence of $\#_G$- and $\#_\sigma$-neighboring operations.
	
To move the charges at site $i$ to site $j$ along edge $e$, we follow the deformed edge as it moves along the $\#_G$ path from $i$. If the edge moves over the top of a node $k$ in $\#_G$, then moving right along that edge corresponds to a sequence of clockwise exchanges between the $\sigma$ particles from $i$ with those at $k$. Moving left corresponds to a sequence of anticlockwise exchanges. If the edge moves under a node in $\#_G$, the braid sense is reversed. The fact that the $\#_G$ and $\#_\sigma$ orderings are consistent guarantees that these operations are sequences of $\#_\sigma$-neighboring exchanges. These exchanges are implemented mathematically by conjugating the stabilizer group by the relevant braid operator $B_{\alpha}$ or $B_{\alpha}^{\dagger}$. We successively exchange the charges from $i$ along the path until they arrive at $j$, where the desired physical process can be performed.
	
Using these linearized braid sequences, we can simulate the phenomenological processes defined in~\Sref{S:phenom} as follows:
\begin{enumerate}
	\item	To pair-create $q$-type charges along edge $e=(i,j)$, we create two charges of type $q$ at site $i$, initialized in the vacuum fusion channel. Following this, we perform a linearized braid sequence to take one of these charges from $i$ to $j$, where it is added to the total charge at $j$;
	\item	To hop charges along edge $e$ from $i$ to $j$, we perform a linearized braid sequence taking charges from $i$ to $j$, where they are then simply added to the existing charges at $j$;
	\item	To exchange charges along edge $e=(i,j)$ either clockwise or anticlockwise, we perform the linearized braid sequence taking charges at site $i$ to $j$, where they can be exchanged with the $j$ charges in the relevant sense, and then return charges from site $j$ to site $i$ by the reverse linearized braid sequence;
	\item	To decohere the total charge at site $i$, we have no need to linearize the process as it acts only on a single site. We simply perform a projective measurement of the total charge at $i$. If $n_{\sigma}^i$ is odd, the total charge will necessarily be $\sigma$, otherwise the total charge can be calculated using \Eref{e:measurel} (also taking into account any existing $\psi$ particles at $i$).
\end{enumerate}	
In this way we can simulate the phenomenology of an Ising anyon system in contact with a thermal bath.

In order to perform an error-correction routine on this system, we will require one further algorithm which simulates fusion of the total charge within a region of the graph. We achieve this by using linearized braid sequences to hop all charges in the region to a common site, after which we successively projectively fuse all pairs of $\sigma$ particles and calculate the total charge on this site.

\subsection{Non-planar graph simulations}\label{S:phenomtorus}

As well as planar graphs, we will also consider anyon dynamics on non-planar graphs. Specifically, we will treat a square lattice graph with periodic boundary conditions. We will restrict discussion to this case, though discretizations of general 2-manifolds follow in a similar manner.

The behavior of anyons braiding on a torus will differ from braiding on a plane insofar as the torus admits braiding processes around non-contractible loops on the manifold. Mathematically, the braid group on a nontrivial manifold requires additional generators to describe these processes compared to the braid group on the plane, as is discussed in Ref. \cite{Birman1969}.

In contrast to a sphere, anyon models on a torus define a degenerate Hilbert space even in the absence of anyonic charges. A basis for this space can be defined by fluxes corresponding to each anyonic charge running through a nontrivial loop of the torus\footnote{Strictly speaking, this discussion assumes a \textit{modular} anyon model. Most anyon models of interest including the Ising anyons satisfy this property and so we will disregard subtleties arising for non-modular models.}. An anyon braiding around this nontrivial loop will behave as if it had performed a monodromy of the corresponding enclosed charge.

Although there are two nontrivial loops on a torus, they do not each carry an independent associated flux. The fluxes through these two loops are related by topological $S$ transformations, given for the Ising anyon model in \Aref{A:isingdata}. The behavior of anyons braiding around the nontrivial loops of the torus can also be analyzed by the extended diagrammatic calculus of Pfeifer et~al.~\cite{Pfeifer2012}.

\begin{figure}
      \centering
	\includegraphics{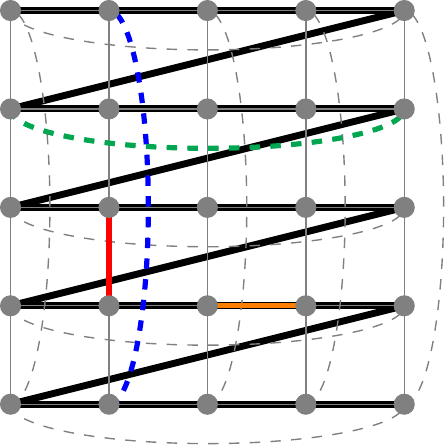}

	\vspace{0.75cm}
	
	\includegraphics{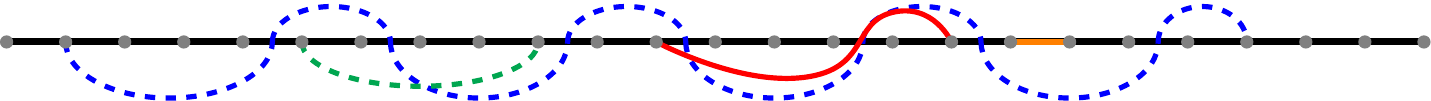}

	\caption{A non-planar graph $G$ (shown with gray edges) and $\#_G$ (shown with solid black edges). The edges along the ``seams'' of the torus are shown as dashed lines. As in the planar case of \Fref{f:braidmap-sphere}, we also show the deformation of several edges into linearized braid sequences.}\label{f:braidmap-torus}
\end{figure}

We will now describe the simulation of Ising anyon phenomenology on a torus, concentrating on the ways in which it differs from the planar case. Consider the graph in \Fref{f:braidmap-torus}. Although not planar, this graph has a natural embedding in the torus. As in the planar graph case, it is convenient to choose a linear ordering of sites of $G$ as shown. Again, this allows us to find linearized braid sequences for a given graph edge from its image under a smooth deformation of the graph to the line.

\begin{figure}
      \centering
		\includegraphics{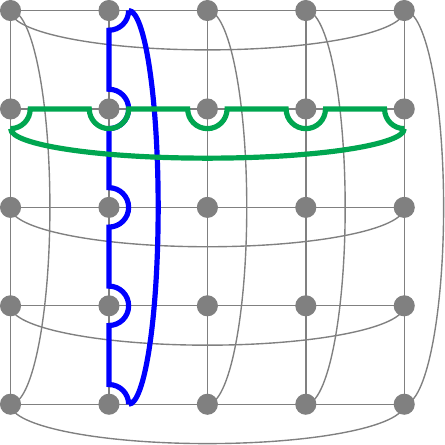}
%

	\caption{Representatives of the two classes of nontrivial loops on the non-planar graph are shown. The $h$ loop is shown in green, while the $v$ loop is shown in blue.}\label{f:nontrivial-torus}
\end{figure}

An Ising anyon system on a torus has an associated 3-dimensional ``topological'' Hilbert space in addition to the fusion space and the charge configuration space. This represents the flux running through a nontrivial loop of the manifold. There are two natural orthonormal bases for this space corresponding to the two nontrivial loops. Calling these loops $h$ and $v$ as shown in \Fref{f:nontrivial-torus}, we denote these bases as $\{\ket{\vac}_h,\ket{\psi}_h,\ket{\sigma}_h\}$ and $\{\ket{\vac}_v,\ket{\psi}_v,\ket{\sigma}_v\}$. Given the charge configuration corresponding to vacuum at all sites of $G$ except charge $q$ at a single site (all charge configurations can be transformed into a state of this form by braiding and fusion operations that do not cross the seams of the torus; see Fig.~\ref{f:braidmap-torus}), we can transform from the $h$ basis to the $v$ basis by topological $S$ transformation $S_q$ (see \Aref{A:isingdata} for details). For Ising anyons, these have the form
\begin{align}
S_{\vac}&=\frac{1}{2}\begin{pmatrix}1&1&\sqrt{2}\\1&1&-\sqrt{2}\\\sqrt{2}&-\sqrt{2}&0\end{pmatrix} \quad\quad \mbox{and} \quad\quad
S_{\psi}=\begin{pmatrix}0&0&0\\0&0&0\\0&0&e^{-i\frac{\pi}{4}}\end{pmatrix} \,.
\end{align}
Note that on a nontrivial manifold, the total charge of the system is no longer a conserved quantity, though for the Ising anyons we can guarantee that the total charge will never be $\sigma$ and so we need not consider $S_{\sigma}$.

Each time an anyon moves around a nontrivial loop of the torus, it effectively braids around the flux through that loop~\cite{Pfeifer2012}. For example, if an anyon braids along the green path in \Fref{f:nontrivial-torus} while the topological state is $\ket{q}_h$, this anyon acts precisely as if it had performed a double exchange with charge $q$.
In contrast, the blue path corresponds to performing a monodromy of the flux in the $v$ basis. In order to determine the effect of the relevant operations, we must use an $S$ transformation to change between the $h$ and $v$ bases in between these processes. The $S$-matrix required will depend on the total charge on the lattice at any given point in time.

We can derive the rules for anyons braiding around the nontrivial loops of the torus by using the Pfiefer et~al.~diagrammatic calculus~\cite{Pfeifer2012} or by simply considering the $S$ transformations and the braiding rules of the Ising anyons. Consider a process labeled $\lbr^{h}_q$, by which we create a pair of $q\times \bar{q}$ particles from vacuum, braid one around an $h$ loop, then fuse them together again. We find that the action of this operator on the various basis states is as follows:
\begin{align}
	\lbr^{h}_\psi\ket{\vac}_h\otimes\ket{\vac}_{\mathrm{bulk}}&=\ket{\vac}_h\otimes\ket{\vac}_{\mathrm{bulk}}&
	\lbr^{h}_\sigma\ket{\vac}_h\otimes\ket{\vac}_{\mathrm{bulk}}&=\ket{\vac}_h\otimes\ket{\vac}_{\mathrm{bulk}}\label{e:topocorrectbegin}\\
	\lbr^{h}_\psi\ket{\psi}_h\otimes\ket{\vac}_{\mathrm{bulk}}&=\ket{\psi}_h\otimes\ket{\vac}_{\mathrm{bulk}}&
	\lbr^{h}_\sigma\ket{\psi}_h\otimes\ket{\vac}_{\mathrm{bulk}}&=-\ket{\psi}_h\otimes\ket{\vac}_{\mathrm{bulk}}\\
	\lbr^{h}_\psi\ket{\sigma}_h\otimes\ket{\vac}_{\mathrm{bulk}}&=-\ket{\sigma}_h\otimes\ket{\vac}_{\mathrm{bulk}}&
	\lbr^{h}_\sigma\ket{\sigma}_h\otimes\ket{\vac}_{\mathrm{bulk}}&=\e^{-i\frac{\pi}{4}}\ket{\sigma}_h\otimes\ket{\psi}_{\mathrm{bulk}}\label{e:lbrsigh1}\\	
	\lbr^{h}_\psi\ket{\sigma}_h\otimes\ket{\psi}_{\mathrm{bulk}}&=-\ket{\sigma}_h\otimes\ket{\psi}_{\mathrm{bulk}}&
	\lbr^{h}_\sigma\ket{\sigma}_h\otimes\ket{\psi}_{\mathrm{bulk}}&=\e^{-i\frac{\pi}{4}}\ket{\sigma}_h\otimes\ket{\vac}_{\mathrm{bulk}}\label{e:lbrsigh2} \,,
\end{align}
where $\ket{\vac}_{\mathrm{bulk}}$ has $\vac$ charge at every site of $G$, and $\ket{\psi}_{\mathrm{bulk}}$ denotes a state with a single $\psi$ charge at one site, with all other sites taking charge $\vac$ (note that all of these states have trivial fusion spaces).

We also consider the analogous operators corresponding to braiding around the $v$ loop, $\lbr_q^v$:
\begin{align}
	\lbr^{v}_\psi\ket{\vac}_h\otimes\ket{\vac}_{\mathrm{bulk}}&=\ket{\psi}_h\otimes\ket{\vac}_{\mathrm{bulk}}&
	\lbr^{v}_\sigma\ket{\vac}_h\otimes\ket{\vac}_{\mathrm{bulk}}&=\frac{1}{\sqrt{2}}\ket{\sigma}_h\otimes\bigl[\ket{\vac}_{\mathrm{bulk}}+\ket{\psi}_{\mathrm{bulk}}\bigr]\\
	\lbr^{v}_\psi\ket{\psi}_h\otimes\ket{\vac}_{\mathrm{bulk}}&=\ket{\vac}_h\otimes\ket{\vac}_{\mathrm{bulk}}&
	\lbr^{v}_\sigma\ket{\psi}_h\otimes\ket{\vac}_{\mathrm{bulk}}&=\frac{1}{\sqrt{2}}\ket{\sigma}_h\otimes\bigl[\ket{\vac}_{\mathrm{bulk}}-\ket{\psi}_{\mathrm{bulk}}\bigr]\\
	\lbr^{v}_\psi\ket{\sigma}_h\otimes\ket{\vac}_{\mathrm{bulk}}&=\ket{\sigma}_h\otimes\ket{\vac}_{\mathrm{bulk}}&
		\lbr^{v}_\sigma\ket{\sigma}_h\otimes\ket{\vac}_{\mathrm{bulk}}&=\frac{1}{\sqrt{2}}\bigl[\ket{\vac}_h+\ket{\psi}_h\bigr]\otimes\ket{\vac}_{\mathrm{bulk}}\\
	\lbr^{v}_\psi\ket{\sigma}_h\otimes\ket{\psi}_{\mathrm{bulk}}&=-\ket{\sigma}_h\otimes\ket{\psi}_{\mathrm{bulk}}&
	\lbr^{v}_\sigma\ket{\sigma}_h\otimes\ket{\psi}_{\mathrm{bulk}}&=\frac{i}{\sqrt{2}}\bigl[\ket{\vac}_h-\ket{\psi}_h\bigr]\otimes\ket{\vac}_{\mathrm{bulk}}\label{e:topocorrectend}\,.
\end{align}

Note that these relations make explicit the connection between the $h$ and $v$ loops of the torus, insofar as the identity $(\lbr_q^v)^{\dagger}=S\lbr_q^hS^{\dagger}$ holds, where $S$ acts as $S_{\vac}$ or $S_{\psi}$ depending on the bulk charge. We have made a number of convention choices here, most significantly choosing a preferred braiding sense around each of the nontrivial loops. When an anyon braids in the preferred sense, it corresponds to performing the relevant $\lbr$ operation, while braiding in the reverse direction corresponds to the adjoint operation $\lbr^{\dagger}$.

The definitions of $\lbr^v$ and $\lbr^h$ operators allow us to not only study the specific processes by which they were defined, but also general braiding processes on the torus. A general braiding operation on the graph in \Fref{f:braidmap-torus} will perform some action on the fusion space by braiding with other anyons, as well as applying a ``topological correction'' operation on the ground space according to the relevant $\lbr$ operator. This topological correction must be performed each time an anyon moves along a ``seam'' edge in \Fref{f:braidmap-torus} (as well as the standard linearized braid operations that are required). This prescription will ensure that operations on the torus satisfy appropriate Fox braid group relations for this nontrivial manifold~\cite{Birman1969, Hatsugai1992}. 

Although we could directly simulate the processes corresponding to (\ref{e:topocorrectbegin})-(\ref{e:topocorrectend}), in practice there is a more straightforward method to simulate these processes up to global phases that takes advantage of the structure of the Ising anyons~\cite{Oshikawa2007,Nayak2008}. This corresponds to tracking the evolution of the (commuting) operators $\lbr_\psi^h$ and $\lbr_\psi^v$. Since they are self-inverse, these operators each have eigenvalues $\pm 1$. The corresponding four eigenstates span the space $\ket{\cdot}_h\otimes\ket{\cdot}_{\mathrm{bulk}}$. Noting that $\left[\lbr_{\sigma}^h,\lbr_{\psi}^h\right]=0$ and $\lbr_{\sigma}^v\lbr_{\psi}^h\left(\lbr_{\sigma}^v\right)^{\dagger}=-\lbr_{\psi}^h$ (and similarly under exchange of $h$ and $v$), this is sufficient to track the ground state of the system under topological correction operations. Finally, when a $\sigma$ particle crosses a seam for which the corresponding $\lbr_\psi$ operator has eigenvalue $-1$, the change in bulk charge that occurs causes the $\sigma$ particle to acquire an additional $\psi$ charge (mathematically, its Majorana operator's eigenvalue changes sign).

\section{Quantum error correcting codes in Ising anyon systems}\label{S:code}

In order to discuss error correction in an Ising anyon system, we must define an encoding of a logical Hilbert space in our model. In anyonic models, the construction of this encoding generally follows one of two paths. The first is motivated by topological quantum computation schemes, where quantum information is encoded in the fusion space associated with a fixed number of non-Abelian anyons. This kind of scheme is well developed for Ising anyons~\cite{Bravyi2006b}. The second is motivated by topological codes such as the toric code, where the degenerate ground space on a nontrivial manifold is used as a codespace~\cite{Dennis2002}. We present constructions for both types of code and compare their properties.

During this discussion, we will abstract away much of the explicit formalism introduced earlier for simulating Ising anyon dynamics. As before, we regard our system as being defined over a graph $G$ with charges located at each site of $G$. Here we represent the charge configuration space by $3$-dimensional Hilbert spaces for each site $s\in V$ with basis $\{\ket{\vac}_s,\ket{\psi}_s,\ket{\sigma}_s\}$ representing the charge present at $s$. We will represent the state within the fusion space (if one exists) by kets of the form $\ket{q}^{f}_{\alpha\beta}$ for the state where the $\sigma$ particles labelled $\alpha$ and $\beta$ fuse to outcome $q\in\{\vac,\psi\}$.

\begin{figure}
  	\centering
	\includegraphics{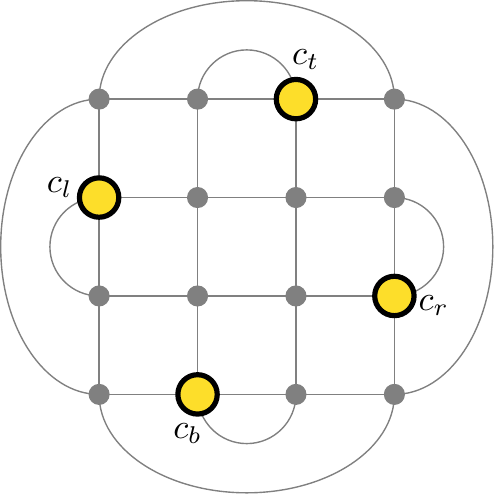}

	\caption{The Ising fusion code uses a square graph with spherical boundary conditions. Four preferred code sites are highlighted. These sites were chosen to maximize the distance between them (and hence to maximize the code distance). We label them $c_t$, $c_r$, $c_b$, and $c_l$ as shown.}\label{f:code-sphere}
\end{figure}

\subsection{Ising Fusion Code}

The first code we consider is constructed in the fusion space of several anyons placed at preferred sites of a graph $G$. We call the resulting code the Ising fusion code (IFC).
	
Consider a system of $L^2$ sites arranged in a square lattice as shown in \Fref{f:code-sphere}. The boundaries are chosen to give $G$ a spherical topology. There are four preferred sites on the lattice chosen to be as far apart as possible in graph distance (which will maximize the code distance), and we call these preferred sites ``code'' sites. We refer to all non-code sites as ``bulk'' sites. 

Since $G$ has spherical boundary conditions, it follows that all physical actions on our system will conserve global charge. We choose to work within the total vacuum charge sector of our model. The IFC is defined by vacuum charge at each bulk site, and a $\sigma$ charge at each of the code sites. Although the code space corresponds to a fixed charge configuration, a 2-dimensional fusion space is also associated with the four $\sigma$ particles in the system. We can define a basis for this space by considering the fusion product of a preferred pair of $\sigma$ particles, which we choose to be those at code sites $c_t$ and $c_r$. These basis states have the form:
\begin{align}
\ket{\vac}_C&\equiv  \ket{\sigma}_{c_t}\otimes\ket{\sigma}_{c_r}\otimes\ket{\sigma}_{c_l}\otimes\ket{\sigma}_{c_b}\otimes\ket{\vac}_{\mathrm{bulk}}\otimes\ket{
\vac}_{rt}^f\\
\ket{\psi}_C&\equiv  \ket{\sigma}_{c_t}\otimes\ket{\sigma}_{c_r}\otimes\ket{\sigma}_{c_l}\otimes\ket{\sigma}_{c_b}\otimes\ket{\vac}_{\mathrm{bulk}}\otimes\ket{
\psi}_{rt}^f
\end{align}
where we denote the $\sigma$ particles at each code site by $t$, $r$, $l$, and $b$.

Logical operators in this code can be taken to correspond physically to braiding the code site $\sigma$ anyons around one another. Equivalent operations can also be achieved e.g.~by tunneling processes between code sites. These operations have been well studied, and all Clifford group operations on one qubit can be achieved in this way~\cite{Fan2010, Ahlbrecht2009}. As an example, the logical Pauli $X$ operator (note that the codespace is a qubit) can be implemented by performing a monodromy between the $r$ and $b$ $\sigma$ particles, while a logical Pauli $Z$ operator corresponds to a monodromy between the $t$ and $r$ anyons.

We remark that we could consider a version of the code we have described here to be implemented as the ground space of a noninteracting Hamiltonian with the form
\begin{align}
	H=-\sum_{c\in\mathrm{code}\;\mathrm{sites}}\proj{\sigma}_c-\sum_{s\in\mathrm{bulk}}\proj{\vac}_{s} \,.
\end{align}
Such a Hamiltonian could emerge as a low-energy effective Hamiltonian in a many-body lattice model, for example~\cite{Kitaev2006, Levin2005a,Kapit2013, Palumbo2014}.

\subsection{Ising Topological Code}

We also consider a code defined as the ground space of an Ising anyon system embedded in a nontrivial manifold. Specifically, we choose $G$ to be an $L\times L$ square lattice with periodic boundary conditions as in~\Fref{f:braidmap-torus},  and imagine a local gapped Hamiltonian whose ground space corresponds to vacuum charge at every site of $G$. An arbitrary anyon system of this form embedded in a torus has a ground space dimension equal to the number of distinct anyon species (including vacuum), and so the ground space of our Ising anyon system will be three dimensional. We call the resulting code the Ising topological code (ITC). 

Basis states for this ground space can be labelled by fluxes taking values $\{\vac,\psi,\sigma\}$ running through one of the nontrivial loops (labelled $h$ and $v$) of the torus. We denote these basis states by
\[\ket{\vac}_\mu\otimes\ket{\vac}_{\mathrm{bulk}},\qquad\ket{\psi}_\mu\otimes\ket{\vac}_{\mathrm{bulk}},\qquad\ket{\sigma}_\mu\otimes\ket{\vac}_{\mathrm{bulk}}\]
for $\mu\in\{h,v\}$. As noted in \Sref{S:phenom}, the basis transformation corresponding to exchanging the two nontrivial loops on the torus (exchanging the $h$ and $v$ bases) is given by a topological $S$ transformation.

The logical operators of this code correspond to braiding of anyons around these nontrivial loops of the torus. Explicitly, they correspond to the operators $\lbr_q$ defined in~\Sref{S:phenomtorus}. Although $\lbr_\sigma^h$ and $\lbr_\sigma^v$ are not strictly logical operators (as they do not preserve the codespace), they correspond to an irrecoverable decoding failure for this code, as with any of the rest of these operations. The only distinction is that errors corresponding to $\lbr_\sigma^h$ and $\lbr_\sigma^v$ may be detectable. One might think that if this error is detected by a measurement of the total bulk charge we could correct it by performing the inverse braid, but by detecting it we have performed a partial measurement on the state of the system and so corrupted the encoded information. Note however that $(\lbr^h_{\sigma})^2$ and $(\lbr^v_{\sigma})^2$ are valid logical operators.

We could consider a version of the code we have described here to be implemented as the ground space of the Hamiltonian
\begin{align}
	H=-\sum_{s\in V}\proj{\vac}_s
\end{align}
or any alternative local Hamiltonian that energetically penalizes any non-vacuum charge in the bulk.
%
\subsection{Codes with boundaries}
%
When discussing topological codes, often planar codes are considered as they may offer a more feasible implementation than a code which requires a nontrivial manifold (such as the ITC). The most well known of these is the surface code, the analogue of Kitaev's toric code with boundary~\cite{Dennis2002,Bravyi1998}. However, these codes require the ability to construct a local gapped boundary at which some subset of the charges in the anyon model may condense. The construction of these boundaries is not always possible for an arbitrary anyon model, and the exact boundaries that can be constructed may even depend on whether the microscopic Hamiltonian is based on bosonic or fermionic degrees of freedom.

Abelian anyon boundaries are discussed in~\cite{Levin2013, Barkeshli2013}, and although the non-Abelian case (including Ising anyons) is not treated it seems likely that an analysis of this type could be extended to show that no gapped boundaries are possible for an Ising anyon system based on an underlying bosonic Hamiltonian. However, it may be possible to construct a gapped boundary which absorbs $\psi$ charges in an Ising anyon system based on a fermionic Hamiltonian. This would allow for the implementation of an Ising anyon code defined on an annulus (topologically equivalent to the cylinder) with a 2-dimensional codespace, or similarly a family of codes with many internal boundaries. We do not treat this case explicitly here, though the construction and simulation of these codes would follow straightforwardly from standard topological code techniques and other techniques developed here.

\subsection{Determining code thresholds}\label{S:threshold}

A threshold for a code is a critical value of some parameter of the noise channel such that  in the thermodynamic limit (i.e.~as $L\to\infty$), we can guarantee that the recovery operation calculated by the decoder will exactly reverse the effects of the noise channel (that is it will restore the encoded quantum state with unit probability) if the parameter is below the critical value. Of course, this threshold will depend on the decoder and noise channel in general.
	
To the best of our knowledge, all previous estimates of thresholds (numerical or otherwise) for quantum codes have only been made for additive codes (i.e.\ codes described by the Pauli stabilizer formalism~\cite{Gottesman1998} or the higher dimensional generalizations thereof~\cite{Gottesman1999a, Van-den-Nest2012, Bermejo-Vega2012}). In these codes, it is straightforward to calculate the precise effect of the noise and recovery operations on the codespace in generality. In our codes, this is not so trivial, and so we introduce a new tool to the study of quantum codes to calculate thresholds for the IFC and the ITC.

In order to estimate the threshold for the codes and decoders we have defined, we first benchmark our error-correction protocol by calculating quasi-thresholds\footnote{By quasi-threshold, we simply mean a noise strength below which a \emph{particular} state will be perfectly preserved under the composition of noise and recovery maps as $L\to\infty$, as opposed to a true threshold which indicates preservation of the entire logical space.}
 below which our recovery procedure will perfectly preserve a set of particular code states (as opposed to preservation of the whole codespace) in the thermodynamic limit. This can easily be done by simulating the initialization of the codespace in this state, and performing a measurement in an appropriate basis after the recovery operation has been performed.

Next we will see how we can  guarantee preservation of the entire code space by simply looking at a few quasi-thresholds for specific states, thus avoiding the problem of tracking the full logical subspace. The structure of the space that is perfectly preserved by the noise and recovery operations in the infinite limit must form a closed matrix algebra~\cite{Blume-Kohout2008,Blume-Kohout2010b}. For a $d$-dimensional codespace, if we choose $d$ non-orthogonal pure states that span $\C^d$ and guarantee that they are each preserved, it follows that the entire Hilbert space must be preserved. This is a consequence of the fact that the density matrices corresponding to these $d$ states generate all density matrices over this space as a matrix algebra. Thus, the threshold for the code will be the minimum of the $d$ quasi-thresholds arising from these states (below which, we can guarantee that each of these states are preserved in the thermodynamic limit). This cannot be achieved with fewer than $d$ pure states.

To prove these claims, define $d$ non-orthogonal states $\rho_i=\proj{\phi_i}$, and note that $\rho_i\rho_j\propto\ketbra{\phi_i}{\phi_j}$ for any $\braket{\phi_i}{\phi_j}\neq 0$. Since the $\{\ket{\phi_i}\}$ span the space, we can construct any density matrix as a linear combination of $\ketbra{\phi_i}{\phi_j}$ (since we can take linear combinations of $\ket{\phi_i}$ to form any pure state, we can take linear combinations of $\ketbra{\phi_i}{\phi_j}$ to form any density matrix). The fact that this cannot be achieved with $k<d$ pure states can be seen by noting that the set $\{\rho_i\rho_j\}$ is closed under multiplication (up to proportionality) and contains $k^2$ unique elements for $1\leq i,j\leq k$.  The Hermitian part of the span of these matrices is also spanned by $\{(\rho_i\rho_j+\rho_j\rho_i)\}$ (by linear independence of the $\rho_i\rho_j$), of which there are $\frac{k(k+1)}{2}$ unique elements. Since an arbitrary density matrix is specified by $\frac{d(d+1)}{2}-1$ parameters, the smallest $k$ to generate all density matrices as linear combinations of $\{\rho_i\rho_j\}$ is $d$.

Note that mixed states can in general reduce this upper bound below $d$ to as little as 3, but they introduce other challenges in verifying that they are preserved by the joint noise-recovery operation, so we eschew them here.

In terms of our simulations this means that since the IFC codespace is $2$-dimensional, if we calculate a quasi-thresholds for a computational basis state $\ket{0}$ and a conjugate basis state $\ket{+}$ then we can guarantee that the threshold for the code will be the minimum of these two quasi-thresholds. Similarly for the ITC, the codespace is $3$-dimensional, and so we can obtain the code threshold by taking the minimum of three quasi-thresholds corresponding to independent non-orthogonal states.

\section{Decoding Algorithms}\label{S:algo}

In order to calculate a memory lifetime for the codes
described in \Sref{S:code} under a noise channel described by the
phenomenology of \Sref{S:phenom}, we initialize in a preferred codestate $\ket{\phi_0}_C$, before sampling noise processes by either a fixed rate or Metropolis mechanism for $T$ time steps. For a lattice of linear size $L$ (where $L=\sqrt{\frac{|E|}{2}}$ for our models), we choose $T$ according to a Poisson distribution with mean $T_0=t_{\mathrm{sim}}|E|$ for some $t_{\mathrm{sim}}>0$. This $t_{\mathrm{sim}}$ represents the physical time over which we
have coupled the system to a thermal bath. If, after applying a decoding algorithm, we arrive back in the $\ket{\phi_0}_C$
state, then the decoding succeeded; otherwise, we say there was
a logical error. Loosely speaking, the threshold is the critical time
$t^*$ such that for all $t_{\mathrm{sim}} < t^*$ the logical
error rate decreases asymptotically to zero as a function of
the lattice size $L$ for all $\ket{\phi_0}_C$ in the code
space. As noted in \Sref{S:threshold}, for a $d$-dimensional codespace, the
preservation of $d$ non-orthogonal linearly independent states is sufficient to
guarantee the preservation of the entire space.

The critical lifetime $t^*$ of the code will depend not
only on the rates $\bm{\gamma}$, but also on the decoder
used to readout the quantum information.

We adapt two algorithms used in decoding Abelian anyon systems,
a clustering renormalization group (RG) decoder due to Bravyi and Haah~\cite{Bravyi2011}, and the perfect matching decoder~\cite{Dennis2002}.
These decoders rely on the metric structure of the surface
in order to make decisions about which (nearby) anyons to fuse together as part of the recovery process. Compared to the Abelian decoders, several additional complications arise from the presence of non-Abelian anyons in our system. The most obvious is the fact that in the Abelian setting, it is always possible to predict fusion outcomes of any pair of anyons with certainty. In the non-Abelian setting, this is no longer possible in general, and so decoding must proceed iteratively as the decoder proposes fusion operations and queries the system to learn the outcome of these fusions. Another complication involves the fact that the precise path taken by two anyons being fused (not just the homology class of the path) plays a role in the outcomes of the process. For Abelian anyons, braiding between two anyons simply results in a global phase that we can disregard. When dealing with non-Abelian anyons, two paths that differ by a braid around another anyon may give rise to differing fusion outcomes, and so we must be more precise when the decoder specifies the recovery operation to be performed.

\subsection{The Simple Clustering Decoder}

The first decoder we consider uses an agglomerative
clustering algorithm inspired by the clustering RG decoder 
of Bravyi and Haah~\cite{Bravyi2011}. Related decoders have also recently been proposed~\cite{Wootton2013b, AnwarInPrep}.
The Bravyi-Haah RG decoder proceeds by placing every charge in its own cluster and then iteratively: growing each cluster, merging the
overlapping clusters, and then fusing all charges within
each cluster. 

When applied to Abelian anyon systems, the cluster decoder only
needs to measure the system once to learn the initial
charge configuration, as this implies the charge configurations at all
points during the decoding and recovery routine. For the same
reason, the fusion of charges within each cluster need not
be applied in sequence, as they can be simulated on
a classical computer and applied to the system altogether in one
final step.

In contrast, when applied to a system of
non-Abelian Ising anyons, the clustering decoder must measure the charge
configuration and physically perform the fusions before each round of
clustering. In our simulation this is done by the decoherence/measurement
and fusion protocols defined in \Sref{S:phenom}. Thus, our simple adaptation
of the clustering decoder can be summarized as follows:

\begin{algorithmic}
	\Function{Cluster\_Simple}{}
	\State Measure charges on each site 
	\State Initialize a cluster for each site with non-trivial charge
	\While{more than one cluster remains}
		\State Fuse all charges within each cluster
        \State Eliminate any cluster with vacuum charge
        \State Grow each cluster by a constant length
        \State Merge overlapping clusters
	\EndWhile
	\EndFunction
\end{algorithmic}

Clusters are grown according to the graph distance metric. In order to merge overlapping clusters, we construct a spanning tree for each cluster and then fuse along the branches of the tree. The charge resulting from the fusion of all anyons is placed at the root of the tree.

When applied to the IFC we must also slightly modify
this algorithm to ensure that $\sigma$ particles are returned to
the code sites if they have been moved away by
thermal processes. This can be achieved simply by initializing clusters on the code sites and returning any fusion results from clusters containing the code sites to these sites (also in this case the termination condition for the algorithm must be modified).

The original clustering RG decoder~\cite{Bravyi2011} has
a provable threshold and we expect that their analysis can be extended to
the non-Abelian case, as it relies primarily only on the separation of defect clusters 
(though we note that the proof of threshold for the clustering RG decoder in 
Ref.~\cite{Bravyi2011} crucially depends on an exponential schedule of cluster size 
increases, as compared to the linear schedule of our algorithm). We also note that our 
adapted clustering decoder inherits its polynomial runtime guarantee from the arguments in 
Ref.~\cite{Bravyi2011}. 

\subsection{The Fusion-Aware Clustering Decoder}

The simple clustering algorithm defined above does not take
advantage of all the structure of the anyonic excitations present
in our system. In particular, we can attempt to optimize
this decoder by making use of details of the anyon
fusion algebra. Instead of clustering together all nearby charges on
the lattice, we can choose to cluster together only
particular types of charge at each iteration. Although we will
only describe this process explicitly for our Ising anyon system,
the ideas extend naturally to arbitrary anyon models.

The algorithm is based on the following subroutine.
\begin{algorithmic}
	\Function{Cluster\_Typed}{set of charge types $Q_i$}
	\State Measure charges on each site
	\State Initialize cluster for each site with non-trivial charge in $Q_i$
	\While{more than one cluster remains}
		\State Fuse all charges in $Q_i$ within each cluster
                \State Eliminate any cluster without charges in $Q_i$
                \State Grow each cluster by a constant length
                \State Merge overlapping clusters
	\EndWhile
	\EndFunction
\end{algorithmic}

The decoder then proceeds by performing \textsc{Cluster\_Typed} on several sets of charges $Q_i$ for $i=1$ to $\chi$. Explicitly, we define the following function.
\begin{algorithmic}
	\Function{Cluster\_Aware}{}
		\For{$i=1$ to $\chi$}
			\State \Call{Cluster\_Typed}{$Q_i$}
		\EndFor
	\EndFunction
\end{algorithmic}

We can see that the \textsc{Cluster\_Simple} decoding routine simply corresponds
to choosing $\chi=1$ and $Q_1=\{\vac,\psi,\sigma\}$. We customize this process for
the Ising anyons by choosing $\chi=2$ and $Q_1=\{\vac,\sigma\}$, $Q_2=\{\vac,\psi\}$. This
amounts to fusing all pairs of $\sigma$ particles into either
vacuum or $\psi$, and then fusing all remaining $\psi$ particles
separately. It generally produces slightly higher thresholds than the simple clustering
decoder in the regimes of interest.

\subsection{The Perfect Matching Decoder}

The second kind of decoder we use is based on the perfect matching algorithm (PMA) of Edmonds~\cite{Edmonds1965}
and its descendants~\cite{Micali1980}. The PMA decoder is typically used for
Abelian anyon systems whose charges are self-inverse (i.e.\ $q=\bar{q}$ for
all charges $q$) such as the toric code~\cite{Kitaev2003}. It proceeds
by calculating the minimum-weight matching between anyons that can fuse
to vacuum.

The naive generalization of the PMA decoder to
non-self-inverse anyon models would calculate 
minimum weight matchings between \emph{hyperedges}
of three or more particles; deciding the existence of such
hypergraph perfect matchings is NP-complete in general~\cite{Karp1972} as is the
minimization problem. It is therefore very unlikely to be efficient
(though heuristic methods have been used to approximate the solution
to this problem~\cite{Wang2010a}).

Fortunately, the Ising anyon model has only
self-inverse charges, and so we need not consider the hypergraph
matching problem here, though similar methods could be applied to
arbitrary anyon models if a heuristic hypergraph matching algorithm were
used in place of the PMA.

Our adaptation is algorithmically similar to the standard implementation of PMA-based decoders for Abelian
anyon systems, but there are qualitative distinctions in its implementation
due to the indeterminacy of the fusion outcomes of non-Abelian
anyons. The PMA decoder is based on the following subroutine:
\begin{algorithmic}
	\Function{PMA\_Single}{charge $q$, configuration of $q$-type anyons $n_q^s$}
	\State Construct a complete graph $K_{q}$ on the set of $q$-type anyons, with edge weights given by the minimum length path between anyons
	\State Use the PMA to compute a minimum-weight perfect matching of this graph
	\State Fuse pairs of anyons that are matched in $K_{q}$
	\EndFunction
\end{algorithmic}

In our implementation of this algorithm, the path along which pairs are fused is taken to be
any shortest length path between the two anyons.

In the case of the IFC, this algorithm must again
be modified slightly to take into account the fact that
the codespace is not the vacuum state.   This can be
achieved by considering the code sites as boundaries and calculating
a matching over a modified graph as in Ref.~\cite{Wang2010}. Note that
when applied to the ITC, the fact that the global
charge is not conserved may lead to a graph without
a perfect matching. This corresponds to decoding failure.

When applied to Ising anyons, we make use of the
structure of the fusion algebra as in the fusion-aware cluster
decoding algorithm \textsc{Cluster\_Aware}. A more general algorithm could be devised
for arbitrary self-inverse anyon models (or arbitrary anyon models if
a heuristic hypergraph matching algorithm replaces PMA), but we restrict discussion
to the specific Ising anyon decoder. This has the form
\begin{algorithmic}
	\Function{PMA\_Ising}{}
		\State Measure charges on each site $s$ 
		\State \Call{PMA\_Single}{$\sigma$, $n_{\sigma}^s$}
		\State Measure charges on each site $s$
		\State \Call{PMA\_Single}{$\psi$, $n_{\psi}^s$}
	\EndFunction
\end{algorithmic}
The PMA decoder has complexity $O(N^3)$ for a $N$ site
system~\cite{Duclos-Cianci2010}
(again assuming constant-time fusion and measurement protocols), though related
decoders~\cite{Fowler2012} achieve average complexity $O(N)$ and are parallelizable to average
time $O(1)$. The thresholds given by the PMA decoder are
typically higher than those obtained from the clustering decoders.

\section{Numerical Results}\label{S:numerics}

To estimate error threshold values for our codes, we use Monte Carlo sampling with various noise models and decoders. The four main questions we pose are:
\begin{enumerate}
	\item What are the thresholds/memory lifetimes of Ising anyon codes?
	\item How do these values vary between different Ising anyon codes?
	\item How do different decoding schemes perform on these codes?
	\item How do different noise models affect the thresholds/lifetimes of these codes?
\end{enumerate}

While the first question is the most natural, it is the least clearly posed and so we will postpone discussion of it until~\Sref{s:threshanal}. We will first discuss questions $2$-$4$ in turn.

In each of our threshold plots, the $x$-axis is given by $t_{\mathrm{sim}}$, or equivalently the average number of errors per edge $\frac{T_0}{|E|}$. This is the natural measure of noise strength in our system and is equivalent to the time spent in contact with the thermal bath before decoding. We discuss this measure further in \Sref{s:threshanal}. The quasi-thresholds and thresholds we extract from our data will thus be given as critical values $t^*$, where of course the specific value of $t^*$ depends on the choice of code, noise model, and decoder.

\subsection{Comparing code performance}

In order to compare the performance of the IFC and the ITC, we restrict to a particular simple noise channel. Specifically, we consider the case where $\gamma_c^{\psi}=\gamma_c^{\sigma}\neq 0$, all other $\gamma=0$, and $p_d=0$. Although we present explicit data only for this choice of parameters, further numerics suggest that the results we find are fairly insensitive to variation of the noise channel. In particular, under no noise model does the \emph{existence} of a threshold seem to change. For simplicity, we will also exclusively restrict to use of the PMA decoder in this subsection. The implementation of PMA we use is Blossom~V~\cite{Kolmogorov2009}.

\subsubsection{Ising Fusion Code}\label{s:pmadata}

As described in \Sref{S:threshold}, in order to determine a threshold for the IFC we first determine quasi-thresholds below which we can guarantee the preservation of the $\ket{0}$ and $\ket{+}$ states respectively in the thermodynamic limit. The minimum of these two quasi-thresholds is the threshold for this code, below which we can guarantee the preservation of the entire logical qubit in the thermodynamic limit.

\begin{figure}
  	\centering
	a) \includegraphics[width=0.47\textwidth]{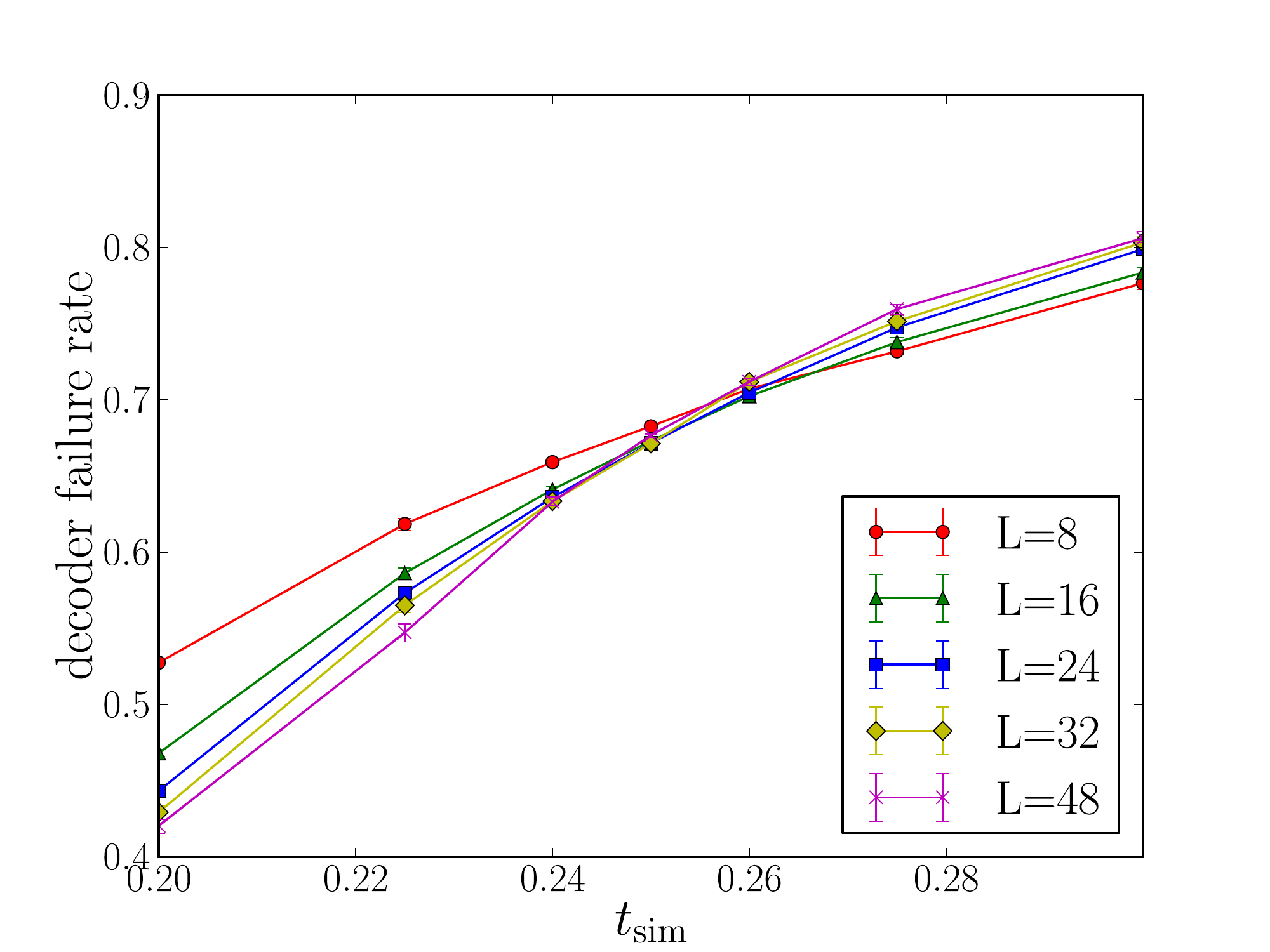}
	b) \includegraphics[width=0.47\textwidth]{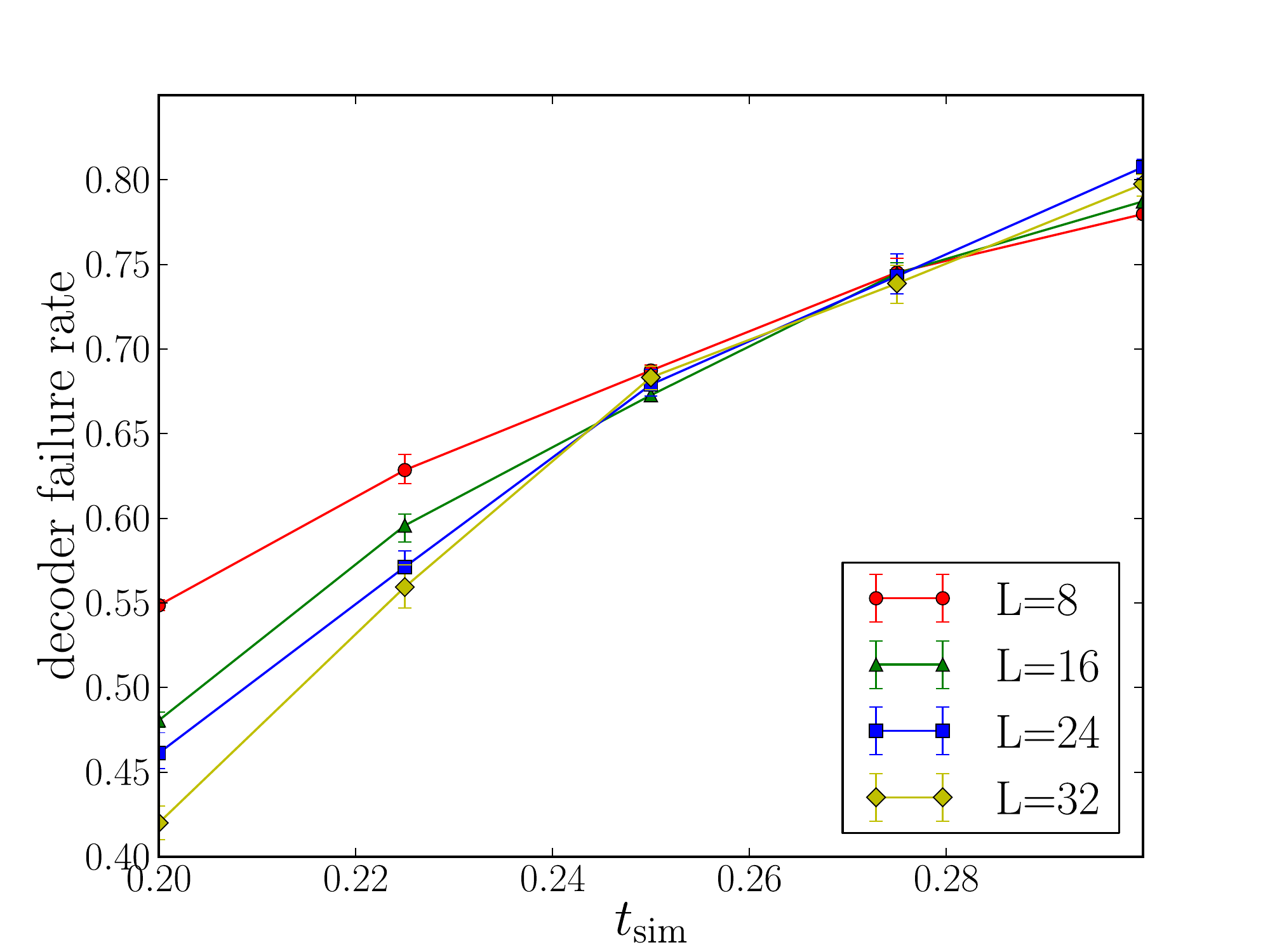}
	\caption{a) Logical $\ket{0}$ state failure probability versus simulation time (average errors per edge) for $\gamma_c^{\psi}=\gamma_c^{\sigma}$ using the PMA decoder on the Ising Fusion Code. b) Logical $\ket{+}$ state failure probability for the same decoder/code/noise combination. Both quasi-thresholds are equal to within sampling and finite-size errors.}\label{f:pma}
\end{figure}

Preservation probabilities for the $\ket{0}$ and $\ket{+}$ states are shown in \Fref{f:pma}. We find that the two quasi-thresholds for this system are indistinguishable up to numerical uncertainty, and thus find the threshold for the IFC as $t^*_{\mathrm{IFC}}\approx 0.24$.

\subsubsection{Ising Topological Code}

\begin{figure}
  	\centering
	\includegraphics[width=0.47\textwidth]{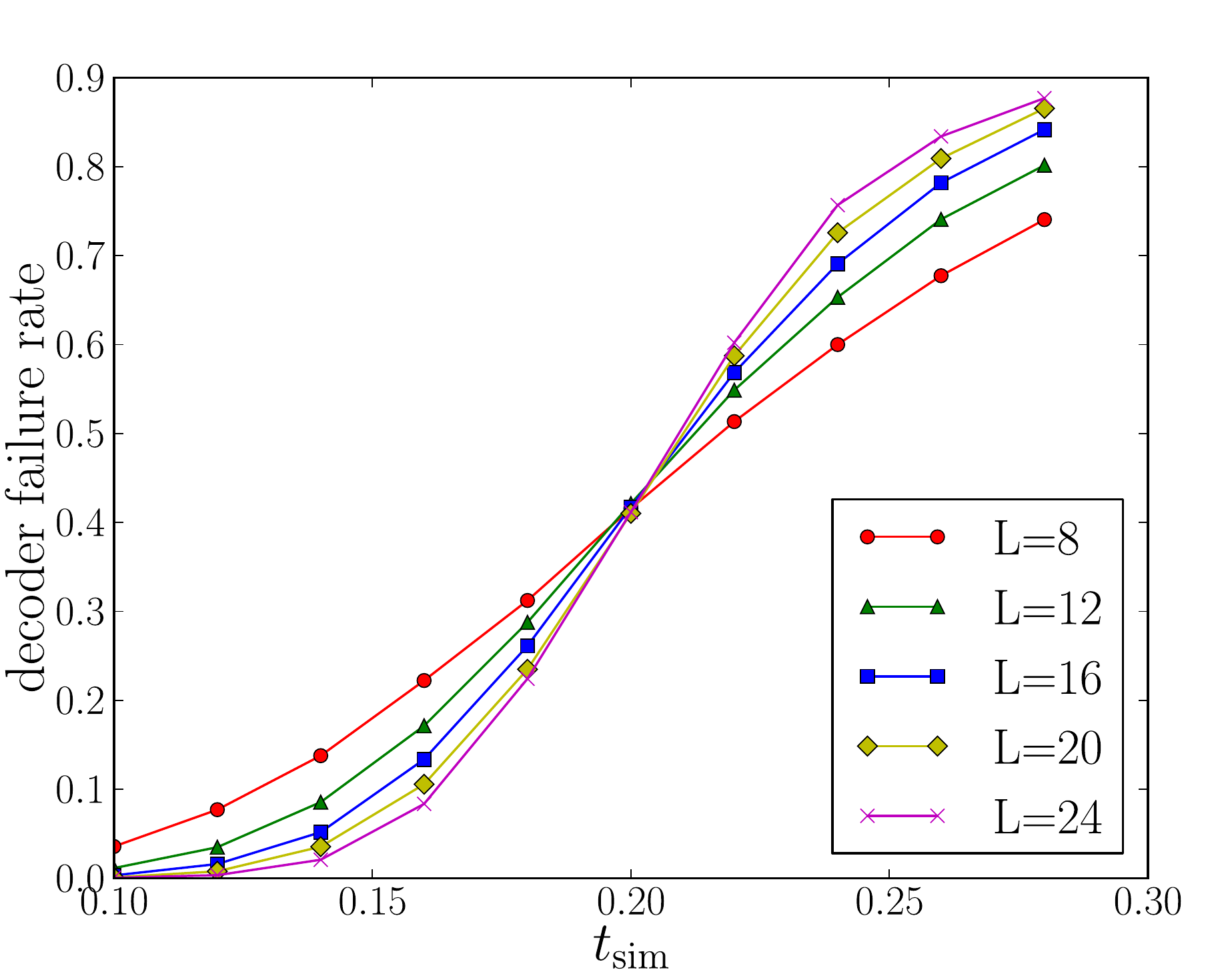}
	\caption{Failure probability versus simulation time (average errors per edge) for $\gamma_c^{\psi}=\gamma_c^{\sigma}$ using the PMA decoder on the Ising Topological Code.}\label{f:pma_torus}
\end{figure}

Our simulations of the system embedded on a torus \Fref{f:pma_torus} yield a threshold of $t^*_{\rm ITC} \approx 0.20$, lower than the threshold obtained on the sphere with four localized anyons. We note however that the decoding failure rate drops faster on the torus for $t_{sim}$ between $0.20$ and $0.28$, a phenomenon that we attribute to the fact that the code's minimum distance is essentially doubled on the torus.

\subsection{Comparing decoder performance}

In order to compare the different decoders described in \Sref{S:algo}, we will again restrict to the simple noise model described by $\gamma_c^{\psi}=\gamma_c^{\sigma}\neq 0$, all other $\gamma=0$, and $p_d=0$. As in the previous section, the results we find seem insensitive to this choice. We will use the IFC threshold as a benchmark to compare the relative performance of our alternative decoding algorithms. We will present only plots for preservation of the $\ket{0}$ state, as plots for the $\ket{+}$ state are quite similar and give identical threshold estimates. 

\begin{figure}
  	\centering
	a) \includegraphics[width=0.47\textwidth]{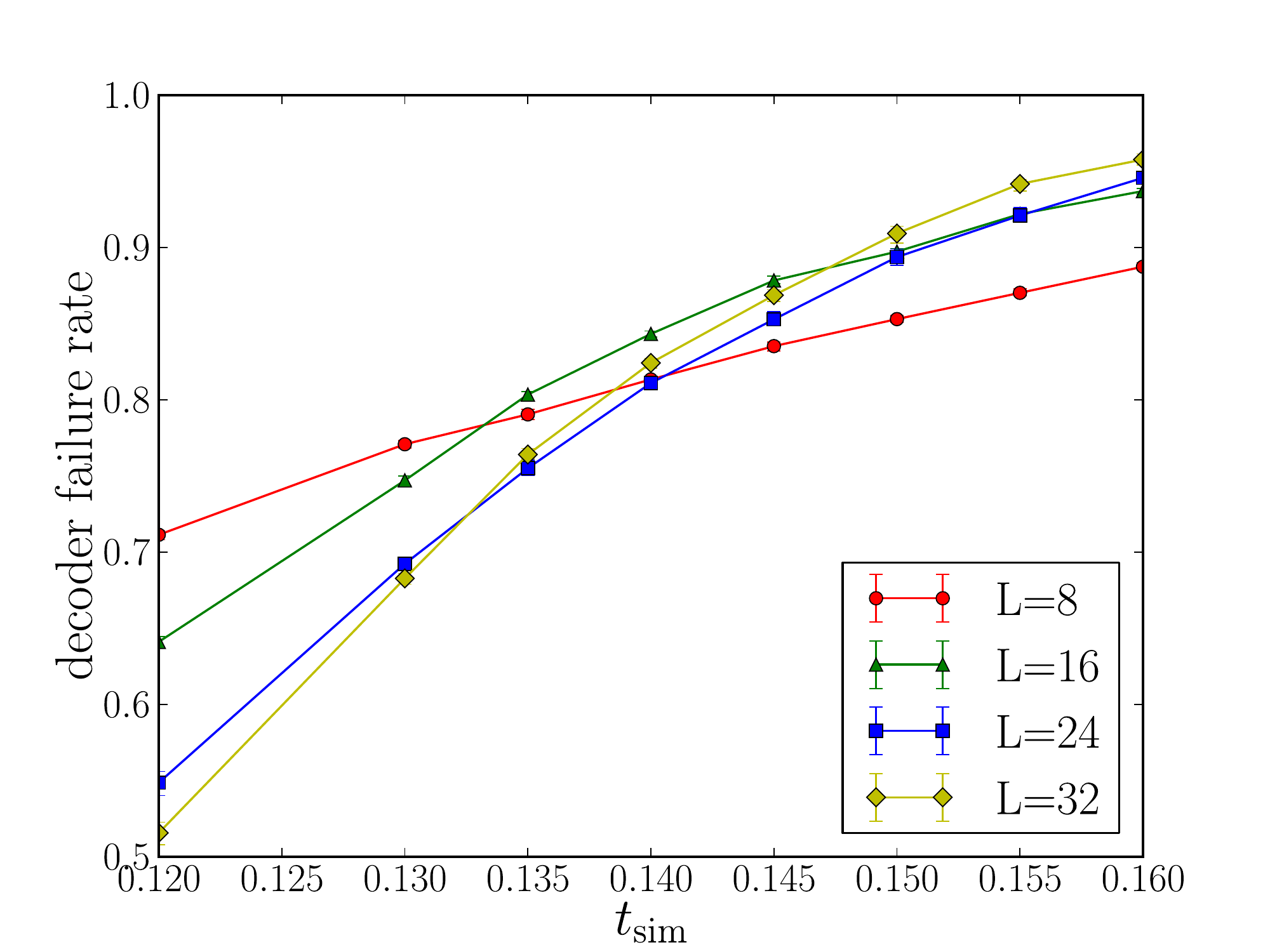}
	b) \includegraphics[width=0.47\textwidth]{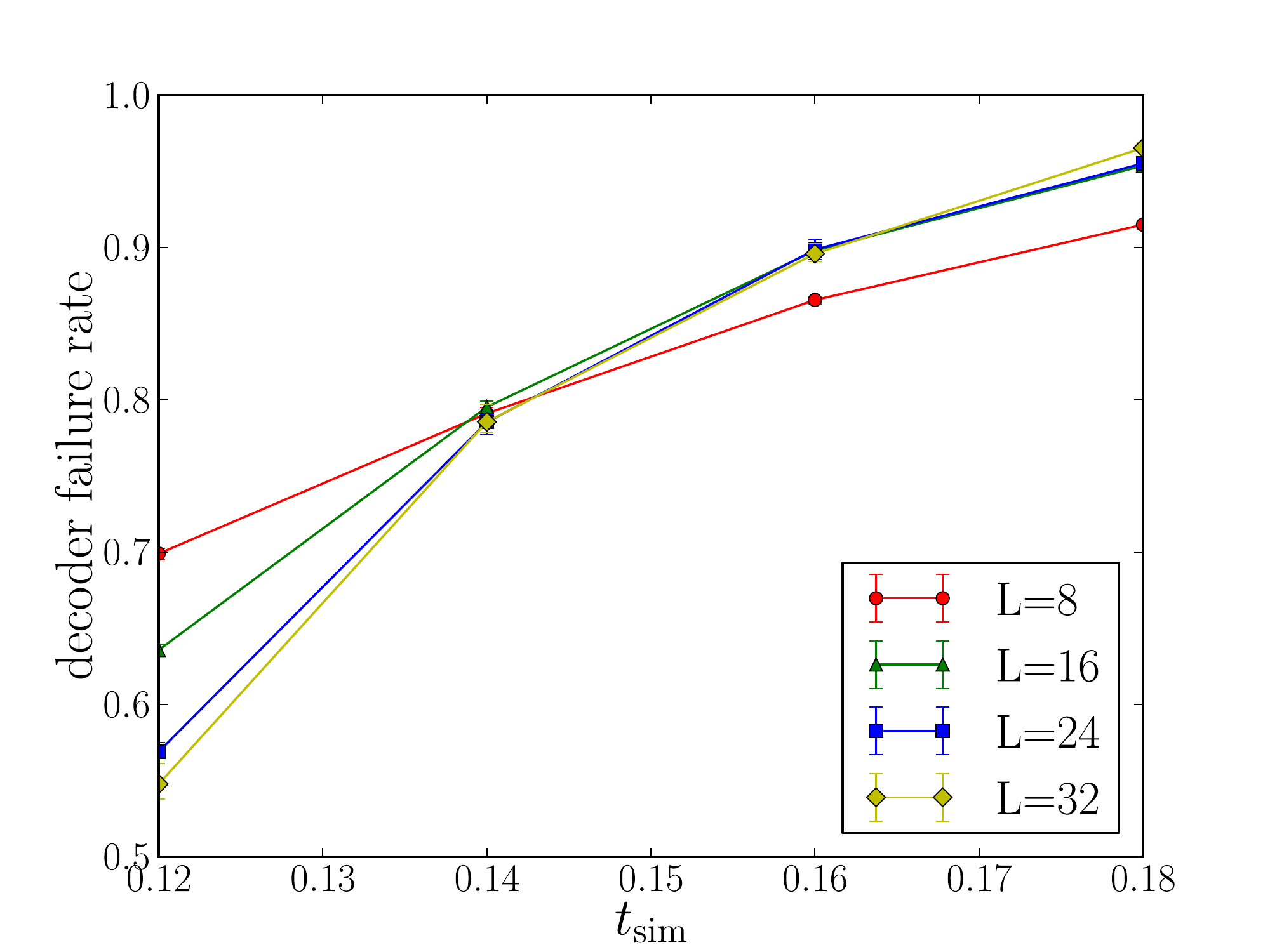}
	\caption{a) Logical $\ket{0}$ state failure probability versus simulation time (average errors per edge) for $\gamma_c^{\psi}=\gamma_c^{\sigma}$ using the simple clustering decoder. b) Logical $\ket{0}$ state failure probability versus simulation time for $\gamma_c^{\psi}=\gamma_c^{\sigma}$ using the fusion-aware clustering decoder. These plots both use the Ising Fusion Code. We see that the performance of the clustering-type decoders is poorer than the PMA-type decoder for the same noise models, though they still exhibit a threshold.}\label{f:bh}
\end{figure}

The threshold estimate obtained from the PMA decoder was determined in the previous section to be $t^*_{\mathrm{IFC}} \approx 0.24$. This compares to the numerical results for the simple clustering decoder and fusion aware clustering decoder shown in Figure~\ref{f:bh}. These can be seen to demonstrate threshold values of $t^*_{\mathrm{SC}}\approx 0.14$ for the simple clustering decoder and $t^*_{\mathrm{FAC}}\approx 0.15$ for the fusion aware clustering decoder.

Note that as well as giving a much higher threshold than the clustering decoders, the PMA decoder also produces numerical data in which the threshold is much more easily identifiable and the uncertainty in the threshold value is clearly lower. This gives a justification for our use of the PMA decoder in the comparison of the different codes and noise models above and in subsequent sections.

\subsection{Variation with relative (fixed) rates}

Here we consider the fixed rate sampling mechanism described in \Sref{S:phenom}. By varying the parameters of our noise model, we can vary the relative size of effects from the non-Abelian nature of the Ising anyons. We can interpolate between a purely Abelian anyon model ($\gamma_c^{\sigma}=0$) and one with non-Abelian charges, as well as varying the relative strengths of processes that contribute to the non-Abelian nature of the anyon dynamics. In particular, setting $\gamma_h$ very high would allow braiding processes to easily perform non-local unitary gates on the fusion space (the hallmark of a non-Abelian anyon system), while setting $p_d$ to be very high suppresses some effects of having a non-trivial fusion space (as discussed in \Aref{A:semiclassical}). Thus by studying the variation of the threshold value for our system under different noise channels, we can infer the relative effects (if any) that the non-Abelian nature of our anyons has on this quantity.

To this end, we consider the following three distinct scenarios to compare to that studied in \Sref{s:pmadata}. We use the IFC and the PMA decoder to study these noise channels.

\subsubsection{\texorpdfstring{$\psi$}{psi}-only pair creation}

We first consider the case where $\gamma_c^{\psi}\neq 0$, all other $\gamma=0$, and $p_d=0$. This situation disallows the presence of $\sigma$ anyons, and so never allows a non-trivial fusion space to arise. The resulting model is completely analogous to the toric code under the bit-flip channel, and this correspondence will be made explicit in \Sref{s:threshanal}.

\begin{figure}
  	\centering
	a) \includegraphics[height=0.3\textheight]{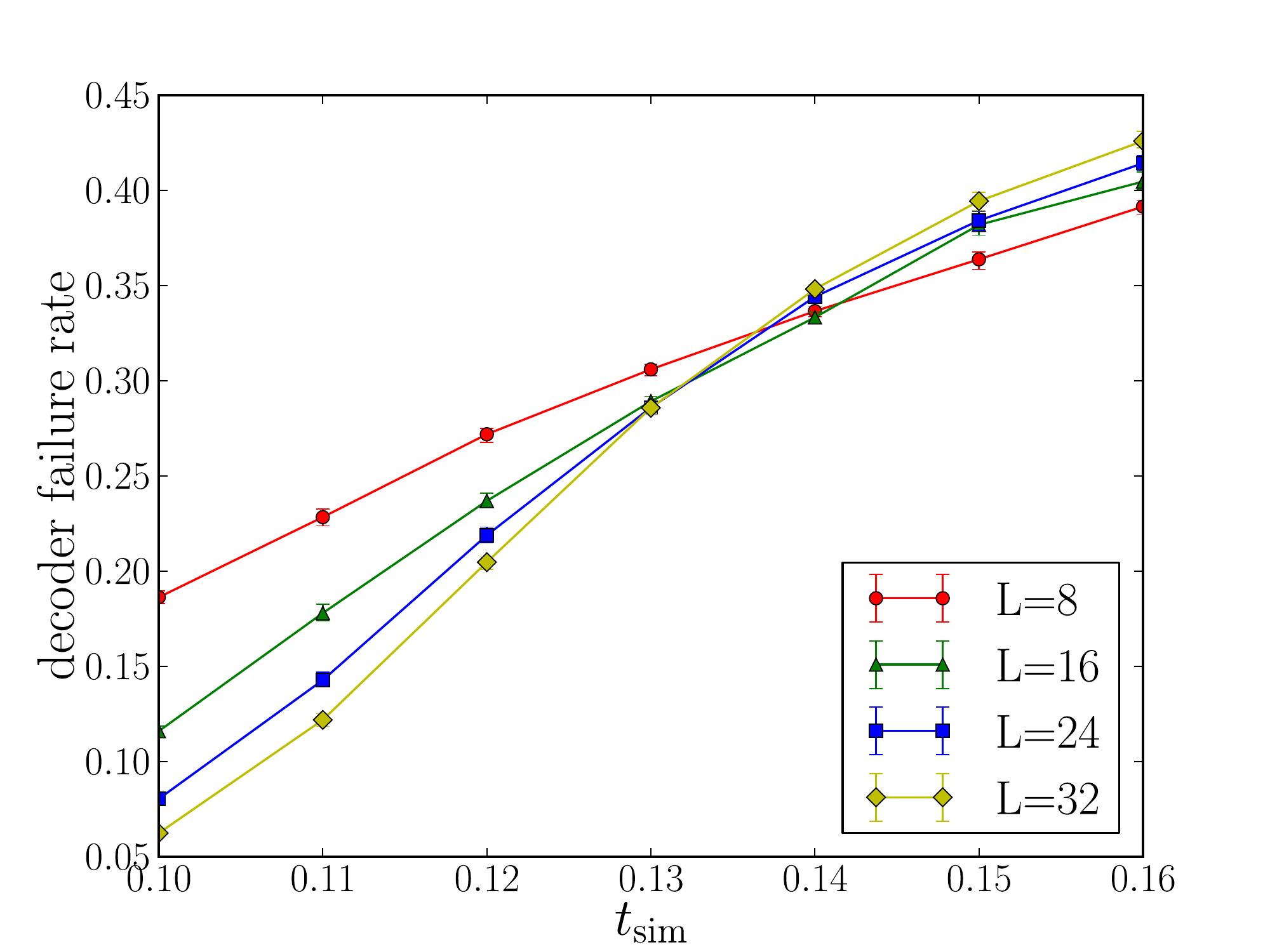}\\
	b) \includegraphics[height=0.3\textheight]{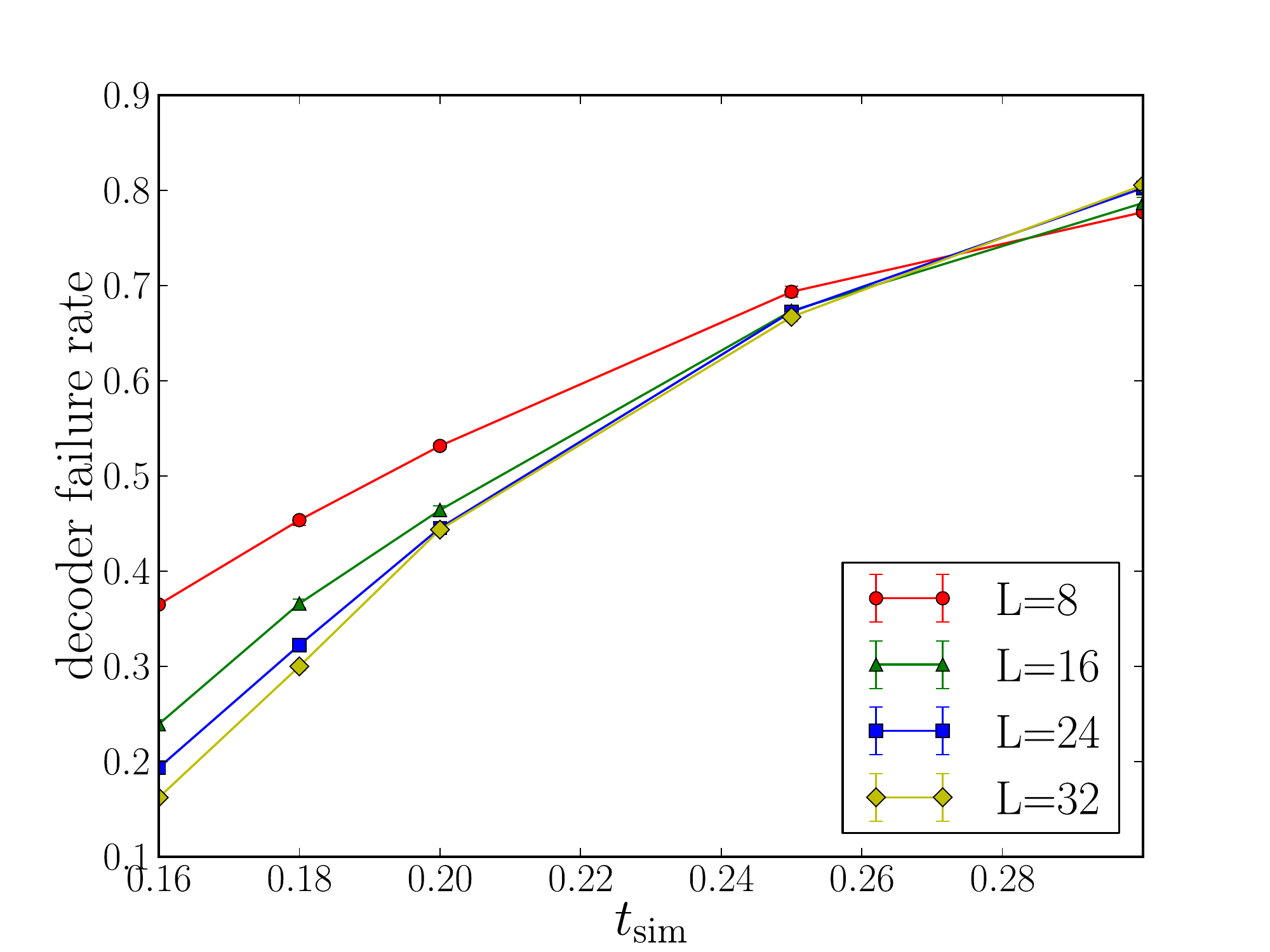}\\
	c) \includegraphics[height=0.3\textheight]{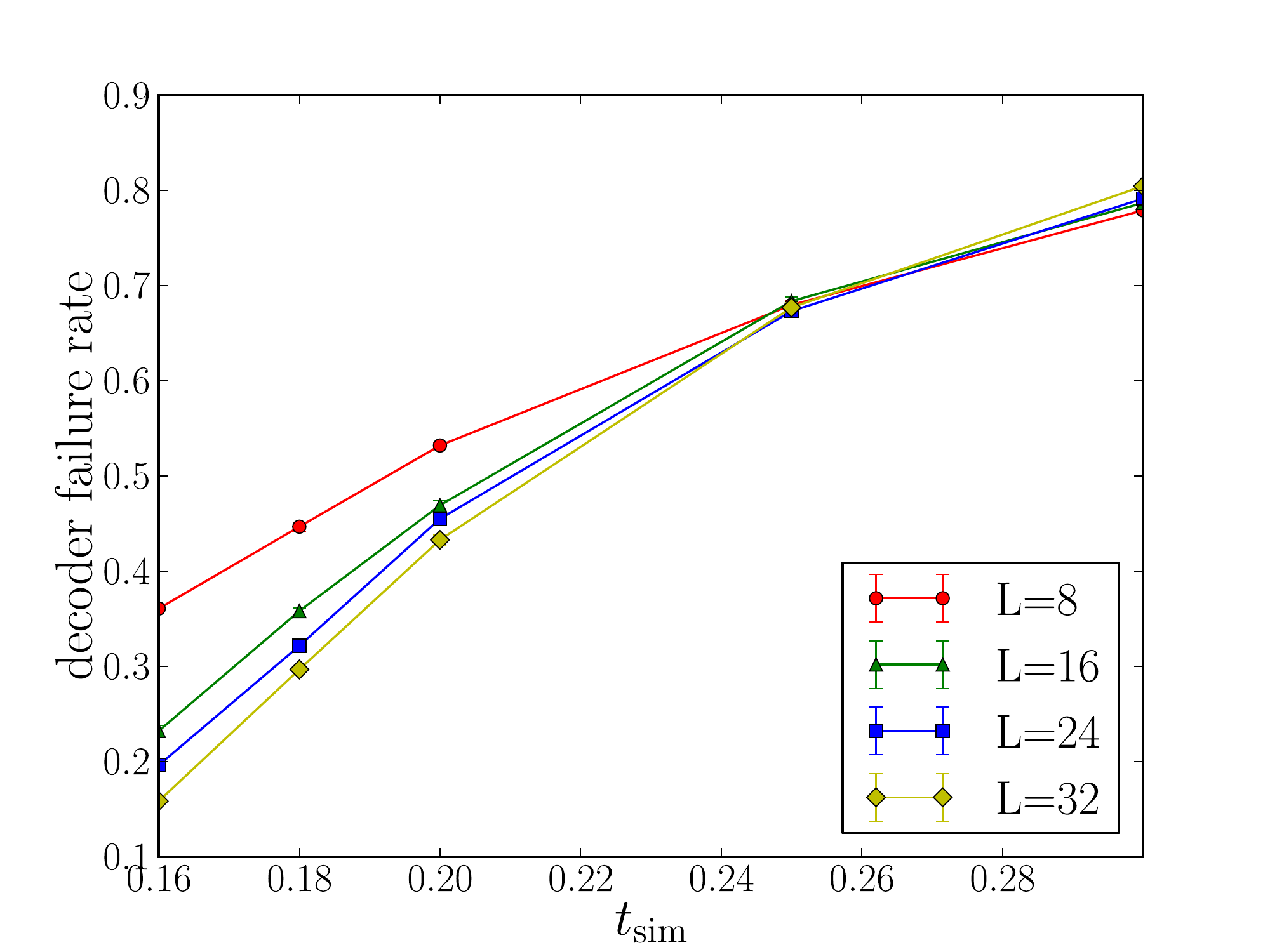}
	\caption{Logical $\ket{0}$ state failure probability against simulation time (average errors per edge) for three different noise models: a) $\psi$ particle creation only, $\gamma_c^{\psi}\neq 0,p_d=0$; b) decoherence-dominated noise, $\gamma_c^{\psi}=\gamma_c^{\sigma}\neq 0, p_d=1$; c) hopping-dominated noise $\gamma_h\gg\gamma_c^{\psi}=\gamma_c^{\sigma}\neq 0, p_d=0$. Each plot is for the Ising Fusion Code using the PMA-type decoder.}\label{f:pma-psi}
\end{figure}

The data for this noise model is shown in \Fref{f:pma-psi}a. We estimate the threshold for this system as $t^*_{\psi}\approx 0.13$. This is around half of the value obtained when \emph{both} the $\sigma$ and $\psi$ anyons are created, as might be expected from similar study of e.g.~the toric code under the bit-flip channel compared to the combined bit and phase-flip channel. The fact that the sum of the $\gamma$ rates are always normalized to one means that by introducing two almost independent types of errors, we can preserve our quantum code space for around twice as many time steps.

\subsubsection{Pair creation and decoherence}

We now study the system in a regime where decoherence is the dominant mechanism. Explicitly, we have $\gamma_c^{\psi}=\gamma_c^{\sigma}\neq 0$, all other $\gamma=0$, and $p_d=1$. The results are shown in \Fref{f:pma-psi}b. We find a threshold of $t^*_{\mathrm{dec}}\approx 0.25$, consistent with values where no decoherence occurs. The fact that local charge decoherence has little effect on information storage supports the idea that Metropolis sampling accurately models thermalization despite the fact that it imposes complete charge decoherence at every time step, which may not be true in real thermalization processes.

\subsubsection{Pair creation dominated by hopping}

Finally, we study a regime where instead of pair-creation, hopping is the dominant mechanism for transporting charge around the lattice. We take $\gamma_h\gg\gamma_c^{\psi}=\gamma_c^{\sigma}\neq 0$, all other $\gamma=0$, and $p_d=0$. We find a threshold of $t^*_{\mathrm{h}}\approx0.25$, again consistent with the situation presented in \Fref{f:pma}.

We conclude that the threshold values are largely insensitive to variation of the $\gamma$ rate parameters, except in predictable ways such as the threshold decreasing by up to a factor of $\frac{1}{2}$ as $\frac{\gamma_{c}^{\psi}}{\gamma_c^{\sigma}}$ varies away from 1. We do not present data for non-zero values of $\gamma_e$, as any braiding process can also be viewed as a higher-order process consisting of many hopping operations, and we again find the threshold values to be insensitive to variation in this parameter.

\subsection{Metropolis sampling}

\begin{figure}
\includegraphics[height=0.4\textheight]{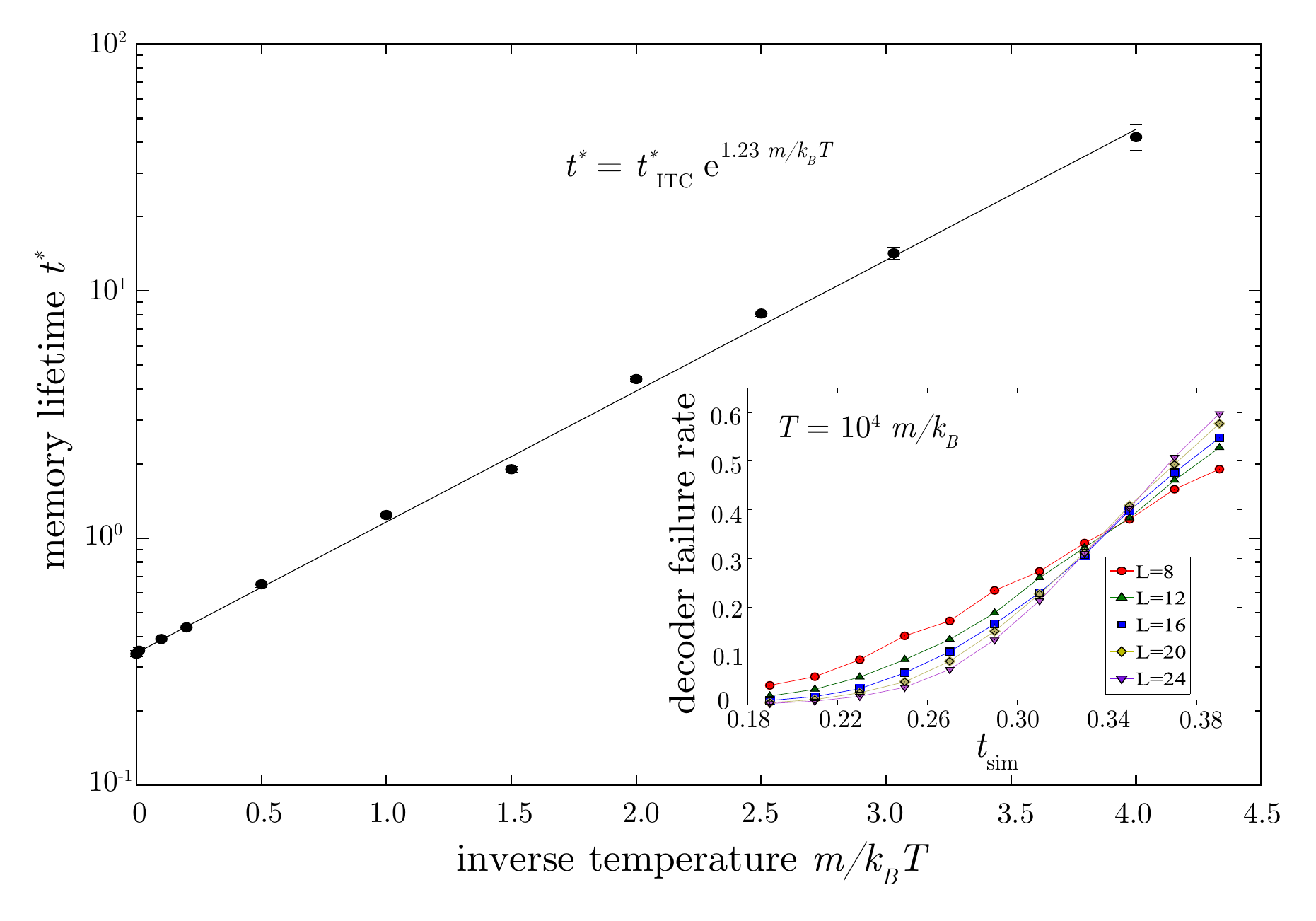}
\caption{Memory lifetime $t^*$ scales exponentially with the inverse temperature. The linear fit is $t_{\rm mem} = 0.343 e^{1.23m/k_BT}$. Inset: Thermal simulations used to determine the memory lifetime $t^*_{\rm ITC} \approx 0.343$ of the system at temperature $T = 10^{4} m/ k_B$. Similar simulations (with $L=20$, $16$, $12$ \& $8$) were performed for different values of $T=10^{4},\ 100,\ 10,\ 5,\ 2,\ 1,\ 0.75,\ 0.5,\ 0.4,\ 0.33$ and $0.25$ to determine the memory lifetime as a function of temperature. \label{F:phase}}
\end{figure}

Here we consider the Metropolis sampling mechanism described in Sec.~\ref{S:metropolis} for the Ising Topological Code. In this case, the rates of the various noise  mechanism are not fixed, but instead are given by a rate equation that depends on the state of the system and a temperature. We have set the mass of both the fermion and anyon to $m=1$, which fixes the energy units.  We estimate the threshold $t^*$ at a fixed temperature $T$ by varying the lattice size $L$ and finding the intersection point as in the previous sections. By repeating for different temperatures, we obtained the memory phase diagram of Fig.~\ref{F:phase}. 

We find that the memory  lifetime scales exponentially with the inverse temperature. We expect the density of particles to be exponentially suppressed at low temperature, and the noise to be dominated by diffusive processes (exchange, hoping and fusion). Assuming that the particle density is low, at every time step the Metropolis rule causes the creation of a pair of particles to be accepted with probability $e^{-2m/k_BT}$. The memory lifetime being given by $t^*=0.343 e^{1.23m/k_BT}$ (see Fig.~\ref{F:phase}), we conclude that the average particle density present at time $t^*$ is proportional to $e^{-0.77m/k_BT}$.


\subsection{Threshold analysis}\label{s:threshanal}

In comparing our results to existing topological code thresholds (e.g.\ with the toric code~\cite{Wang2010, Bravyi2011, Duclos-Cianci2010, Duclos-Cianci2010a, Duclos-Cianci2013}), one should note that we use a slightly different scale on the $x$-axis of our plots than is conventional. In previous analyses, the $x$-axis has typically been represented by some form of iid noise strength. Since the non-Abelian nature of our noise processes means that the noise model \emph{cannot} obviously be recast as an iid error model, we instead use the related measure of average error operations per edge. As noted previously, our fixed rate sampling (see \Sref{S:fixedrate}) simulations proceed by applying a number $T$ of error operations where $T$ is drawn from a Poisson distribution of mean $T_0$, and we define the average error operations per edge, or simulation time, as $t_{\mathrm{sim}}\equiv \frac{T_0}{|E|}=t$ as our measure of noise strength.

We can directly compare this value to the equivalent iid noise parameter in some cases where both are equally valid. Specifically, consider the toric code under the iid bitflip error channel with strength $p_{\mathrm{iid}}$. Comparing this with our noise model for rates such that $\gamma_c^{\sigma}=p_d=0$, we can directly compute a mapping between the Poisson and iid binomial noise models to obtain an equivalence of the parameters given by
\begin{align}
	p_{\mathrm{iid}}=\frac{1-\e^{-2 t_{\mathrm{sim}}}}{2} \,.
\end{align}
Thus we have for $t_{\mathrm{sim}}\ll 1$ that $p_{\mathrm{iid}}\approx t_{\mathrm{sim}}$.

This error model coincides with that of our Ising anyons in the case where $\sigma$ anyons are excluded. Since $\psi$ anyons are self-inverse Abelian anyons, they share all the features which are expected to affect the threshold with one species of toric code anyons. For this reason, we can directly compare the thresholds we find for our model in the $\psi$-only case to those for the toric code (note that without $\sigma$ anyons present, the decoherence mechanism is trivial, and the pair-creation and hopping operations are identical).

As is seen in \Fref{f:pma-psi}a, we find a threshold for the IFC of around $t^*_{\psi}\approx 0.13$. This corresponds to an iid threshold of $p^{\psi}_{\mathrm{iid}} \approx 0.11$. This may seem surprising as the theoretical maximum achievable threshold for the toric code under this error model is estimated as $0.1094$~\cite{Dennis2002}. We believe that the discrepancy in this case lies in an additional finite-size effect which is not present in the toric code, as well as statistical errors that contribute to the uncertainty in our threshold estimates. 

The lowest-order logical operators for the IFC correspond to tunneling processes between code sites. These operators have fixed end points, in comparison to the logical operators for the toric code which consist of any closed loops. It is reasonable to suspect that this fact may affect the likelihood of a logical operator occurring, particularly on finite size lattices. Indeed, we obtained initial numerical estimates of these effects though a simple percolation model, and found evidence that they indeed contribute to an overestimate of the threshold. A precise and rigorous accounting of all systematic and statistical error sources would allow for a more precise determination of the thresholds for our codes and decoders, but is beyond the scope of this paper.

Note that these extra sources of finite-size scaling effects do not apply to the Ising Topological Code, and so we expect that the threshold values quoted above for the ITC apply in the thermodynamic limit.

\section{Discussion}\label{S:conclusion}

Our results demonstrate that error correction is possible in non-Abelian anyon systems and indicate that the non-Abelian nature of the excitations do not have a significant effect on the memory lifetime of stored quantum information. In fact, the various values of the error correction thresholds were remarkably insensitive to the details of the noise processes, including the details of the noise rate equation.

Our results have application to both topological quantum memories and topological quantum computing schemes based on systems with Ising anyon excitations, but more broadly they demonstrate the feasibility in more general 2-dimensional topologically ordered systems. Any such scheme of sufficient size would require some form of error correction protocol in order to succeed. We show how an appropriate error correction scheme may be implemented and how it may behave under a range of noise models.

In so doing, we have constructed several topological codes of interest. These codes are to our knowledge the first non-additive codes to be numerically simulated, and as such we develop a number of new tools to the study of quantum error-correcting codes and in particular topological quantum codes. Most notably, the decoding algorithms we describe may be more broadly applicable to other similar decoding problems.

Based on our numerics, whether or not the system rapidly decoheres in the charge basis plays little role in the memory lifetime of these systems. This can be understood as follows: although the correspondence between the decoding problem and percolation in topological codes is not exact~\cite{Hastings2013}, there is an intuitive correlation between percolation events and decoding failures in these systems. That is, for logical errors to occur, anyonic charge must be transported along a non-local path. Decoherence processes do not affect this intuitive picture at all, and so we might naively expect that they will not significantly alter the nature of the logical noise channel.

The fact that decoherence does not significantly affect the nature of the error-correction problem in our Ising anyon system indicates that the semiclassical noise model used in our Metropolis sampling simulations is justified, and the restriction of the noise model in this case is not expected to significantly affect the results.

Our work lends itself to extension in several directions. Firstly, it would be interesting to apply a similar analysis to other numerically tractable anyon models and/or to concrete microscopic Hamiltonians with anyonic excitations. During preparation of this manuscript, some work in this direction has been done~\cite{Wootton2013}. It is also an open problem as to how our error correction scheme may be leveraged into a demonstration of fault tolerant error correction, which must be the true goal of any theoretical error correction protocol.

\acknowledgments
We thank Guillaume Duclos-Cianci, Sergey Bravyi, Stephen Bartlett, Gavin Brennen, Ben Brown, and Andrew Doherty for valuable discussions. This work was supported by the Australian Research Council Centre of Excellence for Engineered Quantum Systems CE110001013, by the IARPA MQCO program, by Qu\'{e}bec's FQRNT, and Canada's NSERC. Computational resources were provided by Calcul Qu\'{e}bec and Compute Canada.

%

\appendix

\section{Ising Anyon Data}\label{A:isingdata}


Here we summarize the basic data that define the Ising anyon model~\cite{Rowell2009}. For a general mathematical introduction to anyonic theories, see Ref.~\cite[Appendix~E]{Kitaev2006}. 

There are three particle types in the Ising anyon theory: $\vac$ (vacuum), $\psi$, and $\sigma$. The nontrivial fusion rules are
\begin{align}
	\psi \times \psi &= \vac\qquad\qquad
	&\psi \times \sigma &= \sigma\qquad\qquad
	&\sigma \times \sigma &= \vac+\psi \,.
\end{align}
The topological spin for each charge is given by
\begin{align}
	\theta_\vac &=1\qquad\qquad
	&\theta_{\psi} &=-1\qquad\qquad
	&\theta_{\sigma}&=\e^{i\frac{\pi}{8}} \,.
\end{align}
The quantum dimensions are given by
\begin{align}
	d_{\vac}&=1 \qquad\qquad
	&d_{\psi}&=1 \qquad\qquad
	&d_{\sigma}&=\sqrt{2} \,.
\end{align}
The braid matrices are given by
\begin{align}
	R_\vac^{\vac\vac}&=R_{\psi}^{\psi \vac}=R_{\psi}^{\vac\psi}=R_{\sigma}^{\sigma \vac}=R_{\sigma}^{\vac\sigma}=1\qquad
	&R_\vac^{\sigma\sigma}&=\e^{-i\frac{\pi}{8}}\qquad\qquad
	&R_{\vac}^{\psi\psi}&=-1\nonumber\\
	R_\psi^{\sigma\sigma}&=\e^{3i\frac{\pi}{8}}\qquad\qquad
	&R_{\sigma}^{\sigma\psi}&=R_{\sigma}^{\psi\sigma} =-i \,.
\end{align}
The nontrivial $F$ moves are given by the matrices
\begin{align}
    F_{\psi}^{\psi\psi\psi} &=1\qquad\qquad
	&F_{\psi}^{\sigma\psi\sigma} =F_{\sigma}^{\psi\sigma\psi}&=-1\qquad\qquad
	&F_{\sigma}^{\sigma\sigma\sigma} &=\frac{1}{\sqrt{2}}\begin{pmatrix}1&1\\1&-1\end{pmatrix} \,,
\end{align}
where in the last expression the matrix is written in the basis ordered $(\vac,\psi)$ with all other elements vanishing.

The dynamics of anyons can conveniently be expressed by fusion and braiding diagrams. These can be interpreted as (1+1)D spacetime diagrams showing the worldlines of anyons. The data above correspond to the following manipulations of these diagrams. For the topological spin,
\begin{align}
\begin{tikzpicture}[baseline=0.5cm,scale=0.33]		
	\draw plot [smooth, tension=1] coordinates { (1,1.5)(0.9,1.10) (0.5,1) (0.05,2) (0,3)};
		\draw[color=white, line width=5] plot [smooth, tension=1] coordinates { (1,1.5)(0.9,1.90) (0.5,2) (0.05,1) (0,0)};
	\draw plot [smooth, tension=1] coordinates { (1,1.5)(0.9,1.90) (0.5,2) (0.05,1) (0,0)};
	\node at (0,0) [anchor=north] {$c$};
	\node at (0,3) [anchor=south] {$c$};
\end{tikzpicture}
&\;\;=\;\;\theta_c\cdot\;
\begin{tikzpicture}[baseline=0.5cm,scale=0.33]		
	\draw(0,0)--(0,3);
	\node at (0,0) [anchor=north] {$c$};
	\node at (0,3) [anchor=south] {$c$};
\end{tikzpicture}
\end{align}
and for the quantum dimension,
\begin{align}
\begin{tikzpicture}[baseline=0.5cm,scale=0.33]		
	\draw(0,0)--(0,3);
	\draw (-2,1.5) circle (5mm);
	\node at (0,0) [anchor=north] {$a$};
	\node at (-2.5,1.5) [anchor=east] {$c$};
	\node at (0,3) [anchor=south] {$a$};
\end{tikzpicture}
&\;\;=\;\;d_c\cdot\;
\begin{tikzpicture}[baseline=0.5cm,scale=0.33]		
	\draw(0,0)--(0,3);
	\node at (0,0) [anchor=north] {$a$};
	\node at (0,3) [anchor=south] {$a$};
\end{tikzpicture}
\end{align}
The $R$ matrices correspond to braid moves, denoted in the diagrammatic calculus as
	\begin{align}
		\begin{tikzpicture}[baseline=0.5cm,scale=0.33]		
			\draw(0,0)--(0,1);
			\draw plot [smooth, tension=1] coordinates { (0,1) (0.75,1.8) (-0.6,2.45) (-1,3)};
			\draw[color=white, line width=7] plot [smooth, tension=1] coordinates { (0,1) (-0.75,1.8) (0.6,2.45) (1,3)};
			\draw plot [smooth, tension=1] coordinates { (0,1) (-0.75,1.8) (0.6,2.45) (1,3)};
			\node at (0,0) [anchor=north] {$c$};
			\node at (-1,3) [anchor=south] {$b$};
			\node at (1,3) [anchor=south] {$a$};
		\end{tikzpicture}
		&\;\;=\;\;R^{ab}_c\cdot\;
		\begin{tikzpicture}[baseline=0.5cm,scale=0.33]		
			\draw(0,0)--(0,1);
			\draw(0,1)--(-1,2)--(-1,3);
			\draw(0,1)--(1,2)--(1,3);
			\node at (0,0) [anchor=north] {$c$};
			\node at (-1,3) [anchor=south] {$b$};
			\node at (1,3) [anchor=south] {$a$};
		\end{tikzpicture}
	\end{align}
while $F$ moves correspond to the associator for fusion processes, denoted as
	\begin{align}
		\begin{tikzpicture}[baseline=0.6cm,scale=0.4]
			\draw(0,0)--(0,1)--(-2,3);
			\draw(0,3)--(-1,2);
			\draw(2,3)--(0,1);
			\node at (-0.45,1.6) [anchor=north east] {$i$};
			\node at (0,0) [anchor=north] {$d$};
			\node at (2,3) [anchor=south] {$c$};
			\node at (0,3) [anchor=south] {$b$};
			\node at (-2,3) [anchor=south] {$a$};
		\end{tikzpicture}
		&\;\;=\;\;\sum_j[F^{abc}_d]_{ij}\cdot\;
		\begin{tikzpicture}[baseline=0.6cm,scale=0.4]		
			\draw(0,0)--(0,1)--(-2,3);
			\draw(0,3)--(1,2);
			\draw(2,3)--(0,1);
			\node at (0.42,1.63) [anchor=north west] {$j$};
			\node at (0,0) [anchor=north] {$d$};
			\node at (2,3) [anchor=south] {$c$};
			\node at (0,3) [anchor=south] {$b$};
			\node at (-2,3) [anchor=south] {$a$};
		\end{tikzpicture}
	\end{align}

Finally, it will be convenient to define the topological S-matrices of this theory. If we were to embed an anyon model in the torus, then these S-matrices would allow us to relate the charges measured by the two homologically non-trivial loops over the surface. That is, it would allow us to relate the charge around a circle of latitude to the charge around a circle of longitude. There is an $S_q$ matrix defined for each charge $q$ corresponding to the total charge on the manifold, with matrix elements defined by the evaluation of the diagrams
\begin{align}
	(S_q)_{ab}&= \frac{1}{D}\cdot
		\begin{tikzpicture}[baseline=-0cm,scale=1]		
			\draw(0,0)circle(0.5cm);
			\draw[line width=6, color=white](1.2,0)arc(0:180:0.5cm);
			\draw(0.7,0)circle(0.5cm);
			\draw[line width=6, color=white](0.5,0)arc(0:-180:0.5cm);
			\draw(-0.5,0)arc(180:360:0.5cm);
			\draw(0,0.5)arc(135:45:0.5cm and 0.8cm);
			\node at (0.35,0.7) [anchor=south] {$q$};
			\node at (-0.5,0) [anchor=east] {$a$};
			\node at (1.2,0) [anchor=west] {$b$};
		\end{tikzpicture}
\end{align}
with $D=\sqrt{\sum_{q}d_q^2}$ the total quantum dimension of the model. Explicitly, this gives for the Ising anyons
\begin{align}
	S_{\vac}&=\frac{1}{2}\begin{pmatrix}1&1&\sqrt{2}\\1&1&-\sqrt{2}\\\sqrt{2}&-\sqrt{2}&0\end{pmatrix}\\
	S_{\psi}&=\begin{pmatrix}0&0&0\\0&0&0\\0&0&e^{-i\frac{\pi}{4}}\end{pmatrix}\\
	S_{\sigma}&=0
\end{align}
written in a basis ordered $(\vac,\psi,\sigma)$.


\section{Validity of semiclassical noise simulations}\label{A:semiclassical}

In \Sref{S:phenom} we introduced a semiclassical noise model for Ising anyon dynamics, i.e.~one which disallows any superpositions of total charge on a given site. Here we will discuss how realistic this noise model is and where it may lead to incorrect conclusions. Note that a similar semiclassical model is also used in Ref.~\cite{Wootton2013}.

In a semiclassical model, at each time step of the simulation the system is in an energy eigenstate. The fusion space of two separated anyons is still simulated but the fusion space of two anyons on the same site is traced over in this approximation. This might seem physically reasonable if we expect fusion outcomes of nearby anyons to decohere on a timescale that is very short compared to the timescales associated to other dynamical processes. For this reason, the semiclassical noise model is most appropriate to model systems at low temperatures over short timescales.

At higher temperatures or longer timescales, this approximation may produce deviations from the true physics. One example of a process whose outcomes will be simulated incorrectly by the semiclassical model is as follows:
\begin{align}
\centering
\begin{tikzpicture}[baseline=0.5cm,scale=0.33]		
	\draw(-2,0)--(-2,1);
	\draw(-2,1)--(-3,2)--(-3,5);
	\draw(-2,1)--(-1,2)--(-1,3);
	\node at (-2,0) [anchor=north] {$\vac$};
	\node at (-3,2) [anchor=east] {$\sigma$};
	\node at (-1.3,2.2) [anchor=north west] {$\sigma$};
	\draw(2,0)--(2,1);
	\draw(2,1)--(1,2)--(1,3);
	\draw(2,1)--(3,2)--(3,5);
	\node at (2,0) [anchor=north] {$\vac$};
	\node at (1.3,2.2) [anchor=north east] {$\sigma$};
	\node at (3,2) [anchor=west] {$\sigma$};
	\begin{scope}[yshift=7cm]
	\draw(-2,0)--(-2,1);
	\draw(-2,1)--(-3,2)--(-3,5);
	\draw(-2,1)--(-1,2)--(-1,3);
	\draw(2,0)--(2,1);
	\draw(2,1)--(1,2)--(1,3);
	\draw(2,1)--(3,2)--(3,5);
	\end{scope}
	\begin{scope}[yshift=8cm, rotate=180]
	\draw(-2,1)--(-3,2)--(-3,3);
	\draw(-2,1)--(-1,2)--(-1,3);
	\draw(2,1)--(1,2)--(1,3);
	\draw(2,1)--(3,2)--(3,3);
	\end{scope}
	\begin{scope}[yshift=15cm, rotate=180]
	\draw(-2,0)--(-2,1);
	\draw(-2,1)--(-3,2)--(-3,3);
	\draw(-2,1)--(-1,2)--(-1,3);
	\draw(2,0)--(2,1);
	\draw(2,1)--(1,2)--(1,3);
	\draw(2,1)--(3,2)--(3,3);
	\end{scope}
	\draw plot [smooth, tension=1] coordinates { (1,3) (0.75,3.8) (-0.6,4.45) (-1,5)};
	\draw[color=white, line width=7] plot [smooth, tension=1] coordinates { (-1,3) (-0.75,3.8) (0.6,4.45) (1,5)};
	\draw plot [smooth, tension=1] coordinates { (-1,3) (-0.75,3.8) (0.6,4.45) (1,5)};
	\begin{scope}[yshift=7cm]
	\draw plot [smooth, tension=1] coordinates { (1,3) (0.75,3.8) (-0.6,4.45) (-1,5)};
	\draw[color=white, line width=7] plot [smooth, tension=1] coordinates { (-1,3) (-0.75,3.8) (0.6,4.45) (1,5)};
	\draw plot [smooth, tension=1] coordinates { (-1,3) (-0.75,3.8) (0.6,4.45) (1,5)};
	\end{scope}
	\draw[dashed](-5,7.5)--(5,7.5);
	\node at (-5,7.5) [anchor=east] {$t_1$};
	\draw[dashed](-5,15)--(5,15);
	\node at (-5,15) [anchor=east] {$t_2$};
\end{tikzpicture}
\end{align}

If charge measurements were made at times $t_1$, each fusion product has equal probability of being observed as $\vac$ or $\psi$. If this measurement is made, then made again at time $t_2$, the fusion outcomes are again equally likely to be observed as $\vac$ or $\psi$.

However, if the measurement is \emph{not} made at time $t_1$, and the superposition of charges is allowed to persist, then at time $t_2$ the fusion outcomes measured will deterministically be $\psi$.

Thus we see that using a semiclassical simulation that collapses local charge superpositions at each time step may misclassify some phenomena that would otherwise occur. A similar situation can be constructed for other anyon models, such as that studied in~\cite{Wootton2013}. For this reason, it is in general necessary to perform a full simulation of the fusion space to ensure that the true behavior of braiding anyons is captured. Our simulations achieve this full simulation for the special case where the only errors are topological-type errors, which allows us to have an \emph{efficient} simulation for Ising anyons. More general error models are possible and would in general lead to universal quantum computation~\cite{Bravyi2006b}, and hence they are unlikely to be efficient. Fortunately, the results of~\Sref{S:numerics} demonstrate that the appearance and value of an error-correction threshold are relatively insensitive to these details, and so a semiclassical simulation may give results that are close to the true threshold values under reasonable circumstances, a situation which is not obvious \emph{a priori}.


\begin{thebibliography}{63}%
\makeatletter
\providecommand \@ifxundefined [1]{%
 \@ifx{#1\undefined}
}%
\providecommand \@ifnum [1]{%
 \ifnum #1\expandafter \@firstoftwo
 \else \expandafter \@secondoftwo
 \fi
}%
\providecommand \@ifx [1]{%
 \ifx #1\expandafter \@firstoftwo
 \else \expandafter \@secondoftwo
 \fi
}%
\providecommand \natexlab [1]{#1}%
\providecommand \enquote  [1]{``#1''}%
\providecommand \bibnamefont  [1]{#1}%
\providecommand \bibfnamefont [1]{#1}%
\providecommand \citenamefont [1]{#1}%
\providecommand \href@noop [0]{\@secondoftwo}%
\providecommand \href [0]{\begingroup \@sanitize@url \@href}%
\providecommand \@href[1]{\@@startlink{#1}\@@href}%
\providecommand \@@href[1]{\endgroup#1\@@endlink}%
\providecommand \@sanitize@url [0]{\catcode `\\12\catcode `\$12\catcode
  `\&12\catcode `\#12\catcode `\^12\catcode `\_12\catcode `\%12\relax}%
\providecommand \@@startlink[1]{}%
\providecommand \@@endlink[0]{}%
\providecommand \url  [0]{\begingroup\@sanitize@url \@url }%
\providecommand \@url [1]{\endgroup\@href {#1}{\urlprefix }}%
\providecommand \urlprefix  [0]{URL }%
\providecommand \Eprint [0]{\href }%
\providecommand \doibase [0]{http://dx.doi.org/}%
\providecommand \selectlanguage [0]{\@gobble}%
\providecommand \bibinfo  [0]{\@secondoftwo}%
\providecommand \bibfield  [0]{\@secondoftwo}%
\providecommand \translation [1]{[#1]}%
\providecommand \BibitemOpen [0]{}%
\providecommand \bibitemStop [0]{}%
\providecommand \bibitemNoStop [0]{.\EOS\space}%
\providecommand \EOS [0]{\spacefactor3000\relax}%
\providecommand \BibitemShut  [1]{\csname bibitem#1\endcsname}%
\let\auto@bib@innerbib\@empty
\bibitem [{\citenamefont {Wilczek}(1990)}]{Wilczek1990}%
  \BibitemOpen
  \bibfield  {author} {\bibinfo {author} {\bibfnamefont {F.}~\bibnamefont
  {Wilczek}},\ }\href@noop {} {\emph {\bibinfo {title} {Fractional Statistics
  and Anyon Superconductivity}}}\ (\bibinfo  {publisher} {World Scientific},\
  \bibinfo {address} {Singapore},\ \bibinfo {year} {1990})\BibitemShut
  {NoStop}%
\bibitem [{\citenamefont {Kitaev}(2003)}]{Kitaev2003}%
  \BibitemOpen
  \bibfield  {author} {\bibinfo {author} {\bibfnamefont {A.~Yu.}\ \bibnamefont
  {Kitaev}},\ }\bibfield  {title} {\enquote {\bibinfo {title} {Fault-tolerant
  quantum computation by anyons},}\ }\href {\doibase
  10.1016/S0003-4916(02)00018-0} {\bibfield  {journal} {\bibinfo  {journal}
  {Ann. Phys.}\ }\textbf {\bibinfo {volume} {303}},\ \bibinfo {pages} {2--30}
  (\bibinfo {year} {2003})},\ \Eprint {http://arxiv.org/abs/quant-ph/9707021}
  {arXiv:quant-ph/9707021} \BibitemShut {NoStop}%
\bibitem [{\citenamefont {Bravyi}\ \emph
  {et~al.}(2010{\natexlab{a}})\citenamefont {Bravyi}, \citenamefont
  {Hastings},\ and\ \citenamefont {Michalakis}}]{Bravyi2010}%
  \BibitemOpen
  \bibfield  {author} {\bibinfo {author} {\bibfnamefont {Sergey}\ \bibnamefont
  {Bravyi}}, \bibinfo {author} {\bibfnamefont {Matthew}\ \bibnamefont
  {Hastings}}, \ and\ \bibinfo {author} {\bibfnamefont {Spyridon}\ \bibnamefont
  {Michalakis}},\ }\bibfield  {title} {\enquote {\bibinfo {title} {Topological
  quantum order: stability under local perturbations},}\ }\href {\doibase
  10.1063/1.3490195} {\bibfield  {journal} {\bibinfo  {journal} {J. Math.
  Phys.}\ }\textbf {\bibinfo {volume} {51}},\ \bibinfo {pages} {093512}
  (\bibinfo {year} {2010}{\natexlab{a}})},\ \Eprint
  {http://arxiv.org/abs/1001.0344} {arXiv:1001.0344 [quant-ph]} \BibitemShut
  {NoStop}%
\bibitem [{\citenamefont {Bravyi}\ and\ \citenamefont
  {Hastings}(2011)}]{Bravyi2011a}%
  \BibitemOpen
  \bibfield  {author} {\bibinfo {author} {\bibfnamefont {S.}~\bibnamefont
  {Bravyi}}\ and\ \bibinfo {author} {\bibfnamefont {M.~B.}\ \bibnamefont
  {Hastings}},\ }\bibfield  {title} {\enquote {\bibinfo {title} {A short proof
  of stability of topological order under local perturbations},}\ }\href
  {\doibase 10.1007/s00220-011-1346-2} {\bibfield  {journal} {\bibinfo
  {journal} {Comm. Math. Phys.}\ }\textbf {\bibinfo {volume} {307}},\ \bibinfo
  {pages} {609--627} (\bibinfo {year} {2011})},\ \Eprint
  {http://arxiv.org/abs/arXiv:1001.4363} {arXiv:1001.4363} \BibitemShut
  {NoStop}%
\bibitem [{\citenamefont {Michalakis}\ and\ \citenamefont
  {Zwolak}(2013)}]{Michalakis2013}%
  \BibitemOpen
  \bibfield  {author} {\bibinfo {author} {\bibfnamefont {Spyridon}\
  \bibnamefont {Michalakis}}\ and\ \bibinfo {author} {\bibfnamefont
  {Justyna~P.}\ \bibnamefont {Zwolak}},\ }\bibfield  {title} {\enquote
  {\bibinfo {title} {Stability of frustration-free {H}amiltonians},}\ }\href
  {\doibase 10.1007/s00220-013-1762-6} {\bibfield  {journal} {\bibinfo
  {journal} {Comm. Math. Phys.}\ }\textbf {\bibinfo {volume} {322}},\ \bibinfo
  {pages} {277--302} (\bibinfo {year} {2013})},\ \Eprint
  {http://arxiv.org/abs/1109.1588} {arXiv:1109.1588} \BibitemShut {NoStop}%
\bibitem [{\citenamefont {Wen}\ and\ \citenamefont {Niu}(1990)}]{Wen1990}%
  \BibitemOpen
  \bibfield  {author} {\bibinfo {author} {\bibfnamefont {X.~G.}\ \bibnamefont
  {Wen}}\ and\ \bibinfo {author} {\bibfnamefont {Q.}~\bibnamefont {Niu}},\
  }\bibfield  {title} {\enquote {\bibinfo {title} {Ground-state degeneracy of
  the fractional quantum {H}all states in the presence of a random potential
  and on high-genus {R}iemann surfaces},}\ }\href {\doibase
  10.1103/PhysRevB.41.9377} {\bibfield  {journal} {\bibinfo  {journal} {Phys.
  Rev. B}\ }\textbf {\bibinfo {volume} {41}},\ \bibinfo {pages} {9377--9396}
  (\bibinfo {year} {1990})}\BibitemShut {NoStop}%
\bibitem [{\citenamefont {Einarsson}(1990)}]{Einarsson1995}%
  \BibitemOpen
  \bibfield  {author} {\bibinfo {author} {\bibfnamefont {Torbj\"{o}rn}\
  \bibnamefont {Einarsson}},\ }\bibfield  {title} {\enquote {\bibinfo {title}
  {Fractional statistics on a torus},}\ }\href {\doibase
  10.1103/PhysRevLett.64.1995} {\bibfield  {journal} {\bibinfo  {journal}
  {Phys. Rev. Lett.}\ }\textbf {\bibinfo {volume} {64}},\ \bibinfo {pages}
  {1995--1998} (\bibinfo {year} {1990})}\BibitemShut {NoStop}%
\bibitem [{\citenamefont {Nayak}\ \emph {et~al.}(2008)\citenamefont {Nayak},
  \citenamefont {Simon}, \citenamefont {Stern}, \citenamefont {Freedman},\ and\
  \citenamefont {Das~Sarma}}]{Nayak2008}%
  \BibitemOpen
  \bibfield  {author} {\bibinfo {author} {\bibfnamefont {Chetan}\ \bibnamefont
  {Nayak}}, \bibinfo {author} {\bibfnamefont {Steven~H.}\ \bibnamefont
  {Simon}}, \bibinfo {author} {\bibfnamefont {Ady}\ \bibnamefont {Stern}},
  \bibinfo {author} {\bibfnamefont {Michael}\ \bibnamefont {Freedman}}, \ and\
  \bibinfo {author} {\bibfnamefont {Sankar}\ \bibnamefont {Das~Sarma}},\
  }\bibfield  {title} {\enquote {\bibinfo {title} {Non-{A}belian anyons and
  topological quantum computation},}\ }\href {\doibase
  10.1103/RevModPhys.80.1083} {\bibfield  {journal} {\bibinfo  {journal} {Rev.
  Mod. Phys.}\ }\textbf {\bibinfo {volume} {80}},\ \bibinfo {pages} {1083}
  (\bibinfo {year} {2008})},\ \Eprint {http://arxiv.org/abs/0707.1889}
  {arXiv:0707.1889} \BibitemShut {NoStop}%
\bibitem [{\citenamefont {Bombin}\ and\ \citenamefont
  {Martin-Delgado}(2006)}]{Bombin2006}%
  \BibitemOpen
  \bibfield  {author} {\bibinfo {author} {\bibfnamefont {H.}~\bibnamefont
  {Bombin}}\ and\ \bibinfo {author} {\bibfnamefont {M.~A.}\ \bibnamefont
  {Martin-Delgado}},\ }\bibfield  {title} {\enquote {\bibinfo {title}
  {Topological quantum distillation},}\ }\href {\doibase
  10.1103/PhysRevLett.97.180501} {\bibfield  {journal} {\bibinfo  {journal}
  {Phys. Rev. Lett.}\ }\textbf {\bibinfo {volume} {97}},\ \bibinfo {pages}
  {180501} (\bibinfo {year} {2006})},\ \Eprint
  {http://arxiv.org/abs/quant-ph/0605138} {quant-ph/0605138} \BibitemShut
  {NoStop}%
\bibitem [{\citenamefont {Freedman}\ \emph
  {et~al.}(2002{\natexlab{a}})\citenamefont {Freedman}, \citenamefont
  {Larsen},\ and\ \citenamefont {Wang}}]{Freedman2002}%
  \BibitemOpen
  \bibfield  {author} {\bibinfo {author} {\bibfnamefont {Michael~H.}\
  \bibnamefont {Freedman}}, \bibinfo {author} {\bibfnamefont {Michael}\
  \bibnamefont {Larsen}}, \ and\ \bibinfo {author} {\bibfnamefont {Zhenghan}\
  \bibnamefont {Wang}},\ }\bibfield  {title} {\enquote {\bibinfo {title} {A
  modular functor which is universal for quantum computation},}\ }\href
  {\doibase 10.1007/s002200200645} {\bibfield  {journal} {\bibinfo  {journal}
  {Comm. Math. Phys.}\ }\textbf {\bibinfo {volume} {227}},\ \bibinfo {pages}
  {605--622} (\bibinfo {year} {2002}{\natexlab{a}})},\ \Eprint
  {http://arxiv.org/abs/quant-ph/0001108} {arXiv:quant-ph/0001108} \BibitemShut
  {NoStop}%
\bibitem [{\citenamefont {Freedman}\ \emph
  {et~al.}(2002{\natexlab{b}})\citenamefont {Freedman}, \citenamefont
  {Kitaev},\ and\ \citenamefont {Wang}}]{Freedman2002b}%
  \BibitemOpen
  \bibfield  {author} {\bibinfo {author} {\bibfnamefont {M.~H.}\ \bibnamefont
  {Freedman}}, \bibinfo {author} {\bibfnamefont {A.}~\bibnamefont {Kitaev}}, \
  and\ \bibinfo {author} {\bibfnamefont {Z.}~\bibnamefont {Wang}},\ }\bibfield
  {title} {\enquote {\bibinfo {title} {Simulation of topological field theories
  by quantum computers},}\ }\href {\doibase 10.1007/s002200200635} {\bibfield
  {journal} {\bibinfo  {journal} {Comm. Math. Phys.}\ }\textbf {\bibinfo
  {volume} {227}},\ \bibinfo {pages} {587--603} (\bibinfo {year}
  {2002}{\natexlab{b}})},\ \Eprint {http://arxiv.org/abs/quant-ph/0001071}
  {arXiv:quant-ph/0001071} \BibitemShut {NoStop}%
\bibitem [{\citenamefont {Brennen}\ \emph {et~al.}(2010)\citenamefont
  {Brennen}, \citenamefont {Ellinas}, \citenamefont {Kendon}, \citenamefont
  {Pachos}, \citenamefont {Tsohantjis},\ and\ \citenamefont
  {Wang}}]{Brennen2010}%
  \BibitemOpen
  \bibfield  {author} {\bibinfo {author} {\bibfnamefont {Gavin~K.}\
  \bibnamefont {Brennen}}, \bibinfo {author} {\bibfnamefont {Demosthenes}\
  \bibnamefont {Ellinas}}, \bibinfo {author} {\bibfnamefont {Viv}\ \bibnamefont
  {Kendon}}, \bibinfo {author} {\bibfnamefont {Jiannis~K.}\ \bibnamefont
  {Pachos}}, \bibinfo {author} {\bibfnamefont {Ioannis}\ \bibnamefont
  {Tsohantjis}}, \ and\ \bibinfo {author} {\bibfnamefont {Zhenghan}\
  \bibnamefont {Wang}},\ }\bibfield  {title} {\enquote {\bibinfo {title}
  {Anyonic quantum walks},}\ }\href {\doibase 10.1016/j.aop.2009.12.001}
  {\bibfield  {journal} {\bibinfo  {journal} {Ann. Phys.}\ }\textbf {\bibinfo
  {volume} {325}},\ \bibinfo {pages} {664--681} (\bibinfo {year} {2010})},\
  \Eprint {http://arxiv.org/abs/arXiv:0910.2974} {arXiv:0910.2974} \BibitemShut
  {NoStop}%
\bibitem [{\citenamefont {Lehman}\ \emph {et~al.}(2011)\citenamefont {Lehman},
  \citenamefont {Zatloukal}, \citenamefont {Brennen}, \citenamefont {Pachos},\
  and\ \citenamefont {Wang}}]{Lehman2011}%
  \BibitemOpen
  \bibfield  {author} {\bibinfo {author} {\bibfnamefont {Lauri}\ \bibnamefont
  {Lehman}}, \bibinfo {author} {\bibfnamefont {Vaclav}\ \bibnamefont
  {Zatloukal}}, \bibinfo {author} {\bibfnamefont {Gavin~K.}\ \bibnamefont
  {Brennen}}, \bibinfo {author} {\bibfnamefont {Jiannis~K.}\ \bibnamefont
  {Pachos}}, \ and\ \bibinfo {author} {\bibfnamefont {Zhenghan}\ \bibnamefont
  {Wang}},\ }\bibfield  {title} {\enquote {\bibinfo {title} {Quantum walks with
  non-{A}belian anyons},}\ }\href {\doibase 10.1103/PhysRevLett.106.230404}
  {\bibfield  {journal} {\bibinfo  {journal} {Phys. Rev. Lett.}\ }\textbf
  {\bibinfo {volume} {106}},\ \bibinfo {pages} {230404} (\bibinfo {year}
  {2011})},\ \Eprint {http://arxiv.org/abs/1009.0813} {arXiv:1009.0813}
  \BibitemShut {NoStop}%
\bibitem [{\citenamefont {Lehman}\ \emph {et~al.}(2012)\citenamefont {Lehman},
  \citenamefont {Ellinas},\ and\ \citenamefont {Brennen}}]{Lehman2012}%
  \BibitemOpen
  \bibfield  {author} {\bibinfo {author} {\bibfnamefont {L.}~\bibnamefont
  {Lehman}}, \bibinfo {author} {\bibfnamefont {D.}~\bibnamefont {Ellinas}}, \
  and\ \bibinfo {author} {\bibfnamefont {G.K.}\ \bibnamefont {Brennen}},\
  }\bibfield  {title} {\enquote {\bibinfo {title} {Quantum walks of {$SU(2)_k$}
  anyons on a ladder},}\ }\href{\doibase 10.1166/jctn.2013.3102} {\bibfield {journal}
  {\bibinfo {journal}{J. Comput. Theor. Nanosci.}\ } \textbf{\bibinfo {volume}{10}}, 
  \bibinfo{pages}{1634--1643} } \href {http://arxiv.org/abs/1203.1999} {\  (\bibinfo
  {year} {2013})},\ \Eprint {http://arxiv.org/abs/1203.1999} {arXiv:1203.1999}
  \BibitemShut {NoStop}%
\bibitem [{\citenamefont {Zatloukal}\ \emph {et~al.}(2012)\citenamefont
  {Zatloukal}, \citenamefont {Lehman}, \citenamefont {Singh}, \citenamefont
  {Pachos},\ and\ \citenamefont {Brennen}}]{Zatloukal2012}%
  \BibitemOpen
  \bibfield  {author} {\bibinfo {author} {\bibfnamefont {V.}~\bibnamefont
  {Zatloukal}}, \bibinfo {author} {\bibfnamefont {L.}~\bibnamefont {Lehman}},
  \bibinfo {author} {\bibfnamefont {S.}~\bibnamefont {Singh}}, \bibinfo
  {author} {\bibfnamefont {J.~K.}\ \bibnamefont {Pachos}}, \ and\ \bibinfo
  {author} {\bibfnamefont {G.~K.}\ \bibnamefont {Brennen}},\ }\bibfield
  {title} {\enquote {\bibinfo {title} {Transport properties of anyons in random
  topological environments},}\ }\href {http://arxiv.org/abs/1207.5000} {\
  (\bibinfo {year} {2012})},\ \Eprint {http://arxiv.org/abs/1207.5000}
  {arXiv:1207.5000} \BibitemShut {NoStop}%
\bibitem [{\citenamefont {Willett}\ \emph {et~al.}(1987)\citenamefont
  {Willett}, \citenamefont {Eisenstein}, \citenamefont {St\"ormer},
  \citenamefont {Tsui}, \citenamefont {Gossard},\ and\ \citenamefont
  {English}}]{Willett1987}%
  \BibitemOpen
  \bibfield  {author} {\bibinfo {author} {\bibfnamefont {R.}~\bibnamefont
  {Willett}}, \bibinfo {author} {\bibfnamefont {J.~P.}\ \bibnamefont
  {Eisenstein}}, \bibinfo {author} {\bibfnamefont {H.~L.}\ \bibnamefont
  {St\"ormer}}, \bibinfo {author} {\bibfnamefont {D.~C.}\ \bibnamefont {Tsui}},
  \bibinfo {author} {\bibfnamefont {A.~C.}\ \bibnamefont {Gossard}}, \ and\
  \bibinfo {author} {\bibfnamefont {J.~H.}\ \bibnamefont {English}},\
  }\bibfield  {title} {\enquote {\bibinfo {title} {Observation of an
  even-denominator quantum number in the fractional quantum {H}all effect},}\
  }\href {\doibase 10.1103/PhysRevLett.59.1776} {\bibfield  {journal} {\bibinfo
   {journal} {Phys. Rev. Lett.}\ }\textbf {\bibinfo {volume} {59}},\ \bibinfo
  {pages} {1776--1779} (\bibinfo {year} {1987})}\BibitemShut {NoStop}%
\bibitem [{\citenamefont {Moore}\ and\ \citenamefont {Read}(1991)}]{Moore1991}%
  \BibitemOpen
  \bibfield  {author} {\bibinfo {author} {\bibfnamefont {Gregory}\ \bibnamefont
  {Moore}}\ and\ \bibinfo {author} {\bibfnamefont {Nicholas}\ \bibnamefont
  {Read}},\ }\bibfield  {title} {\enquote {\bibinfo {title} {Nonabelions in the
  fractional quantum {H}all effect},}\ }\href {\doibase
  10.1016/0550-3213(91)90407-O} {\bibfield  {journal} {\bibinfo  {journal}
  {Nucl. Phys. B}\ }\textbf {\bibinfo {volume} {360}},\ \bibinfo {pages} {362
  -- 396} (\bibinfo {year} {1991})}\BibitemShut {NoStop}%
\bibitem [{\citenamefont {Nayak}\ and\ \citenamefont
  {Wilczek}(1996)}]{Nayak1996}%
  \BibitemOpen
  \bibfield  {author} {\bibinfo {author} {\bibfnamefont {Chetan}\ \bibnamefont
  {Nayak}}\ and\ \bibinfo {author} {\bibfnamefont {Frank}\ \bibnamefont
  {Wilczek}},\ }\bibfield  {title} {\enquote {\bibinfo {title}
  {{$2n$}-quasihole states realize {$2n-1$}-dimensional spinor braiding
  statistics in paired quantum {H}all states},}\ }\href {\doibase
  10.1016/0550-3213(96)00430-0} {\bibfield  {journal} {\bibinfo  {journal}
  {Nucl. Phys. B}\ }\textbf {\bibinfo {volume} {479}},\ \bibinfo {pages}
  {529--553} (\bibinfo {year} {1996})},\ \Eprint
  {http://arxiv.org/abs/cond-mat/9605145} {cond-mat/9605145} \BibitemShut
  {NoStop}%
\bibitem [{\citenamefont {Kitaev}(2006)}]{Kitaev2006}%
  \BibitemOpen
  \bibfield  {author} {\bibinfo {author} {\bibfnamefont {Alexei}\ \bibnamefont
  {Kitaev}},\ }\bibfield  {title} {\enquote {\bibinfo {title} {Anyons in an
  exactly solved model and beyond},}\ }\href {\doibase
  10.1016/j.aop.2005.10.005} {\bibfield  {journal} {\bibinfo  {journal} {Ann.
  Phys.}\ }\textbf {\bibinfo {volume} {321}},\ \bibinfo {pages} {2--111}
  (\bibinfo {year} {2006})},\ \Eprint {http://arxiv.org/abs/cond-mat/0506438}
  {arXiv:cond-mat/0506438} \BibitemShut {NoStop}%
\bibitem [{\citenamefont {Levin}\ and\ \citenamefont {Wen}(2005)}]{Levin2005a}%
  \BibitemOpen
  \bibfield  {author} {\bibinfo {author} {\bibfnamefont {Michael~A.}\
  \bibnamefont {Levin}}\ and\ \bibinfo {author} {\bibfnamefont {Xiao-Gang}\
  \bibnamefont {Wen}},\ }\bibfield  {title} {\enquote {\bibinfo {title}
  {String-net condensation: A physical mechanism for topological phases},}\
  }\href {\doibase 10.1103/PhysRevB.71.045110} {\bibfield  {journal} {\bibinfo
  {journal} {Phys. Rev. B}\ }\textbf {\bibinfo {volume} {71}},\ \bibinfo
  {pages} {045110} (\bibinfo {year} {2005})},\ \Eprint
  {http://arxiv.org/abs/cond-mat/0404617} {arXiv:cond-mat/0404617} \BibitemShut
  {NoStop}%
\bibitem [{\citenamefont {Kapit}\ and\ \citenamefont
  {Simon}(2013)}]{Kapit2013}%
  \BibitemOpen
  \bibfield  {author} {\bibinfo {author} {\bibfnamefont {Eliot}\ \bibnamefont
  {Kapit}}\ and\ \bibinfo {author} {\bibfnamefont {Steven~H}\ \bibnamefont
  {Simon}},\ }\bibfield  {title} {\enquote {\bibinfo {title} {3-and 4-body
  interactions from 2-body interactions in spin models: A route to {A}belian
  and non-{A}belian fractional {C}hern insulators},}\ }\href@noop {} {\
  (\bibinfo {year} {2013})},\ \Eprint {http://arxiv.org/abs/1307.3485}
  {arXiv:1307.3485} \BibitemShut {NoStop}%
\bibitem [{\citenamefont {Palumbo}\ and\ \citenamefont
  {Pachos}(2014)}]{Palumbo2014}%
  \BibitemOpen
  \bibfield  {author} {\bibinfo {author} {\bibfnamefont {Giandomenico}\
  \bibnamefont {Palumbo}}\ and\ \bibinfo {author} {\bibfnamefont {Jiannis~K.}\
  \bibnamefont {Pachos}},\ }\bibfield  {title} {\enquote {\bibinfo {title}
  {Non-abelian chern-simons theory from a hubbard-like model},}\ }\href
  {\doibase 10.1103/PhysRevD.90.027703} {\bibfield  {journal} {\bibinfo
  {journal} {Phys. Rev. D}\ }\textbf {\bibinfo {volume} {90}},\ \bibinfo
  {pages} {027703} (\bibinfo {year} {2014})}\BibitemShut {NoStop}%
\bibitem [{\citenamefont {Bravyi}(2006)}]{Bravyi2006b}%
  \BibitemOpen
  \bibfield  {author} {\bibinfo {author} {\bibfnamefont {Sergey}\ \bibnamefont
  {Bravyi}},\ }\bibfield  {title} {\enquote {\bibinfo {title} {Universal
  quantum computation with the {$\nu = \frac{5}{2}$} fractional quantum {H}all
  state},}\ }\href {\doibase 10.1103/PhysRevA.73.042313} {\bibfield  {journal}
  {\bibinfo  {journal} {Phys. Rev. A}\ }\textbf {\bibinfo {volume} {73}},\
  \bibinfo {pages} {042313} (\bibinfo {year} {2006})},\ \Eprint
  {http://arxiv.org/abs/quant-ph/0511178} {arXiv:quant-ph/0511178} \BibitemShut
  {NoStop}%
\bibitem [{\citenamefont {Freedman}\ \emph {et~al.}(2006)\citenamefont
  {Freedman}, \citenamefont {Nayak},\ and\ \citenamefont
  {Walker}}]{Freedman2006}%
  \BibitemOpen
  \bibfield  {author} {\bibinfo {author} {\bibfnamefont {Michael}\ \bibnamefont
  {Freedman}}, \bibinfo {author} {\bibfnamefont {Chetan}\ \bibnamefont
  {Nayak}}, \ and\ \bibinfo {author} {\bibfnamefont {Kevin}\ \bibnamefont
  {Walker}},\ }\bibfield  {title} {\enquote {\bibinfo {title} {Towards
  universal topological quantum computation in the {$\nu =\frac{5}{2}$}
  fractional quantum {H}all state},}\ }\href {\doibase
  10.1103/PhysRevB.73.245307} {\bibfield  {journal} {\bibinfo  {journal} {Phys.
  Rev. B}\ }\textbf {\bibinfo {volume} {73}},\ \bibinfo {pages} {245307}
  (\bibinfo {year} {2006})},\ \Eprint {http://arxiv.org/abs/cond-mat/0512066}
  {cond-mat/0512066} \BibitemShut {NoStop}%
\bibitem [{\citenamefont {Bravyi}\ and\ \citenamefont
  {Terhal}(2009)}]{Bravyi2009}%
  \BibitemOpen
  \bibfield  {author} {\bibinfo {author} {\bibfnamefont {Sergey}\ \bibnamefont
  {Bravyi}}\ and\ \bibinfo {author} {\bibfnamefont {Barbara}\ \bibnamefont
  {Terhal}},\ }\bibfield  {title} {\enquote {\bibinfo {title} {A no-go theorem
  for a two-dimensional self-correcting quantum memory based on stabilizer
  codes},}\ }\href {\doibase 10.1088/1367-2630/11/4/043029} {\bibfield
  {journal} {\bibinfo  {journal} {New J. Phys.}\ }\textbf {\bibinfo {volume}
  {11}},\ \bibinfo {pages} {043029} (\bibinfo {year} {2009})},\ \Eprint
  {http://arxiv.org/abs/0810.1983} {arXiv:0810.1983} \BibitemShut {NoStop}%
\bibitem [{\citenamefont {Bravyi}\ \emph
  {et~al.}(2010{\natexlab{b}})\citenamefont {Bravyi}, \citenamefont {Poulin},\
  and\ \citenamefont {Terhal}}]{Bravyi2010a}%
  \BibitemOpen
  \bibfield  {author} {\bibinfo {author} {\bibfnamefont {Sergey}\ \bibnamefont
  {Bravyi}}, \bibinfo {author} {\bibfnamefont {David}\ \bibnamefont {Poulin}},
  \ and\ \bibinfo {author} {\bibfnamefont {Barbara}\ \bibnamefont {Terhal}},\
  }\bibfield  {title} {\enquote {\bibinfo {title} {Tradeoffs for reliable
  quantum information storage in 2d systems},}\ }\href {\doibase
  10.1103/PhysRevLett.104.050503} {\bibfield  {journal} {\bibinfo  {journal}
  {Phys. Rev. Lett.}\ }\textbf {\bibinfo {volume} {104}},\ \bibinfo {pages}
  {050503} (\bibinfo {year} {2010}{\natexlab{b}})},\ \Eprint
  {http://arxiv.org/abs/0909.5200} {arXiv:0909.5200} \BibitemShut {NoStop}%
\bibitem [{\citenamefont {Haah}\ and\ \citenamefont
  {Preskill}(2012)}]{Haah2012a}%
  \BibitemOpen
  \bibfield  {author} {\bibinfo {author} {\bibfnamefont {Jeongwan}\
  \bibnamefont {Haah}}\ and\ \bibinfo {author} {\bibfnamefont {John}\
  \bibnamefont {Preskill}},\ }\bibfield  {title} {\enquote {\bibinfo {title}
  {Logical-operator tradeoff for local quantum codes},}\ }\href {\doibase
  10.1103/PhysRevA.86.032308} {\bibfield  {journal} {\bibinfo  {journal} {Phys.
  Rev. A}\ }\textbf {\bibinfo {volume} {86}},\ \bibinfo {pages} {032308}
  (\bibinfo {year} {2012})},\ \Eprint {http://arxiv.org/abs/1011.3529}
  {1011.3529} \BibitemShut {NoStop}%
\bibitem [{\citenamefont {Landon-Cardinal}\ and\ \citenamefont
  {Poulin}(2013)}]{Landon-Cardinal2012a}%
  \BibitemOpen
  \bibfield  {author} {\bibinfo {author} {\bibfnamefont {Olivier}\ \bibnamefont
  {Landon-Cardinal}}\ and\ \bibinfo {author} {\bibfnamefont {David}\
  \bibnamefont {Poulin}},\ }\bibfield  {title} {\enquote {\bibinfo {title}
  {Local topological order inhibits thermal stability in {2D}},}\ }\href
  {http://arxiv.org/abs/1209.5750} {\bibfield  {journal} {\bibinfo  {journal}
  {Phys. Rev. Lett.}\ }\textbf {\bibinfo {volume} {110}},\ \bibinfo {pages}
  {090502} (\bibinfo {year} {2013})}\BibitemShut {NoStop}%
\bibitem [{\citenamefont {Pastawski}\ \emph {et~al.}(2010)\citenamefont
  {Pastawski}, \citenamefont {Kay}, \citenamefont {Schuch},\ and\ \citenamefont
  {Cirac}}]{Pastawski2010}%
  \BibitemOpen
  \bibfield  {author} {\bibinfo {author} {\bibfnamefont {Fernando}\
  \bibnamefont {Pastawski}}, \bibinfo {author} {\bibfnamefont {Alastair}\
  \bibnamefont {Kay}}, \bibinfo {author} {\bibfnamefont {Norbert}\ \bibnamefont
  {Schuch}}, \ and\ \bibinfo {author} {\bibfnamefont {J.~Ignacio}\ \bibnamefont
  {Cirac}},\ }\bibfield  {title} {\enquote {\bibinfo {title} {{Limitations of
  Passive Protection of Quantum Informaiton}},}\ }\href@noop {} {\bibfield
  {journal} {\bibinfo  {journal} {Quantum Inf. Comput.}\ }\textbf {\bibinfo
  {volume} {10}},\ \bibinfo {pages} {580} (\bibinfo {year} {2010})},\ \Eprint
  {http://arxiv.org/abs/0911.3843} {arXiv:0911.3843} \BibitemShut {NoStop}%
\bibitem [{\citenamefont {Koenig}\ \emph {et~al.}(2010)\citenamefont {Koenig},
  \citenamefont {Kuperberg},\ and\ \citenamefont {Reichardt}}]{Koenig2010b}%
  \BibitemOpen
  \bibfield  {author} {\bibinfo {author} {\bibfnamefont {Robert}\ \bibnamefont
  {Koenig}}, \bibinfo {author} {\bibfnamefont {Greg}\ \bibnamefont
  {Kuperberg}}, \ and\ \bibinfo {author} {\bibfnamefont {Ben~W.}\ \bibnamefont
  {Reichardt}},\ }\bibfield  {title} {\enquote {\bibinfo {title} {Quantum
  computation with {T}uraev-{V}iro codes},}\ }\href {\doibase
  http://dx.doi.org/10.1016/j.aop.2010.08.001} {\bibfield  {journal} {\bibinfo
  {journal} {Ann. Phys.}\ }\textbf {\bibinfo {volume} {325}},\ \bibinfo {pages}
  {2707--2749} (\bibinfo {year} {2010})},\ \Eprint
  {http://arxiv.org/abs/1002.2816} {arXiv:1002.2816} \BibitemShut {NoStop}%
\bibitem [{\citenamefont {Dennis}\ \emph {et~al.}(2002)\citenamefont {Dennis},
  \citenamefont {Kitaev}, \citenamefont {Landahl},\ and\ \citenamefont
  {Preskill}}]{Dennis2002}%
  \BibitemOpen
  \bibfield  {author} {\bibinfo {author} {\bibfnamefont {Eric}\ \bibnamefont
  {Dennis}}, \bibinfo {author} {\bibfnamefont {Alexei}\ \bibnamefont {Kitaev}},
  \bibinfo {author} {\bibfnamefont {Andrew}\ \bibnamefont {Landahl}}, \ and\
  \bibinfo {author} {\bibfnamefont {John}\ \bibnamefont {Preskill}},\
  }\bibfield  {title} {\enquote {\bibinfo {title} {Topological quantum
  memory},}\ }\href {\doibase 10.1063/1.1499754} {\bibfield  {journal}
  {\bibinfo  {journal} {J. Math. Phys.}\ }\textbf {\bibinfo {volume} {43}},\
  \bibinfo {pages} {4452--4505} (\bibinfo {year} {2002})},\ \Eprint
  {http://arxiv.org/abs/quant-ph/0110143} {arXiv:quant-ph/0110143} \BibitemShut
  {NoStop}%
\bibitem [{\citenamefont {Bonesteel}\ and\ \citenamefont
  {DiVincenzo}(2012)}]{Bonesteel2012}%
  \BibitemOpen
  \bibfield  {author} {\bibinfo {author} {\bibfnamefont {N.~E.}\ \bibnamefont
  {Bonesteel}}\ and\ \bibinfo {author} {\bibfnamefont {D.~P.}\ \bibnamefont
  {DiVincenzo}},\ }\bibfield  {title} {\enquote {\bibinfo {title} {Quantum
  circuits for measuring {L}evin-{W}en operators},}\ }\href {\doibase
  10.1103/PhysRevB.86.165113} {\bibfield  {journal} {\bibinfo  {journal} {Phys.
  Rev. B}\ }\textbf {\bibinfo {volume} {86}},\ \bibinfo {pages} {165113}
  (\bibinfo {year} {2012})},\ \Eprint {http://arxiv.org/abs/1206.6048}
  {arXiv:1206.6048} \BibitemShut {NoStop}%
\bibitem [{\citenamefont {Bravyi}\ and\ \citenamefont
  {Raussendorf}(2007)}]{BR07a}%
  \BibitemOpen
  \bibfield  {author} {\bibinfo {author} {\bibfnamefont {Sergey}\ \bibnamefont
  {Bravyi}}\ and\ \bibinfo {author} {\bibfnamefont {Robert}\ \bibnamefont
  {Raussendorf}},\ }\bibfield  {title} {\enquote {\bibinfo {title}
  {Measurement-based quantum computation with the toric code states},}\
  }\href {\doibase 10.1103/PhysRevA.76.022304}  {\bibfield  {journal} {\bibinfo  
  {journal} {Phys. Rev. A}\
  }\textbf {\bibinfo {volume} {76}},\ \bibinfo {pages} {022304} (\bibinfo
  {year} {2007})},\ \Eprint {http://arxiv.org/abs/quant-ph/0610162}
  {arXiv:quant-ph/0610162} \BibitemShut {NoStop}%
\bibitem [{\citenamefont {Bravyi}\ and\ \citenamefont
  {Haah}(2013)}]{Bravyi2011}%
  \BibitemOpen
  \bibfield  {author} {\bibinfo {author} {\bibfnamefont {Sergey}\ \bibnamefont
  {Bravyi}}\ and\ \bibinfo {author} {\bibfnamefont {Jeongwan}\ \bibnamefont
  {Haah}},\ }\bibfield  {title} {\enquote {\bibinfo {title} {Quantum
  self-correction in the 3d cubic code model},}\ }\href {\doibase
  10.1103/PhysRevLett.111.200501} {\bibfield  {journal} {\bibinfo  {journal}
  {Phys. Rev. Lett.}\ }\textbf {\bibinfo {volume} {111}},\ \bibinfo {pages}
  {200501} (\bibinfo {year} {2013})},\ \Eprint {http://arxiv.org/abs/1112.3252} {arXiv:1112.3252}
  \BibitemShut {NoStop}%
\bibitem [{\citenamefont {Wang}\ \emph
  {et~al.}(2010{\natexlab{a}})\citenamefont {Wang}, \citenamefont {Fowler},
  \citenamefont {Stephens},\ and\ \citenamefont {Hollenberg}}]{Wang2010}%
  \BibitemOpen
  \bibfield  {author} {\bibinfo {author} {\bibfnamefont {D.~S.}\ \bibnamefont
  {Wang}}, \bibinfo {author} {\bibfnamefont {A.~G.}\ \bibnamefont {Fowler}},
  \bibinfo {author} {\bibfnamefont {A.~M.}\ \bibnamefont {Stephens}}, \ and\
  \bibinfo {author} {\bibfnamefont {L.~C.~L.}\ \bibnamefont {Hollenberg}},\
  }\bibfield  {title} {\enquote {\bibinfo {title} {Threshold error rates for
  the toric and surface codes},}\ }\href@noop {} {\bibfield  {journal}
  {\bibinfo  {journal} {Quant. Info. Comput.}\ }\textbf {\bibinfo {volume}
  {10}},\ \bibinfo {pages} {456} (\bibinfo {year} {2010}{\natexlab{a}})},\
  \Eprint {http://arxiv.org/abs/0905.0531} {arXiv:0905.0531} \BibitemShut
  {NoStop}%
\bibitem [{\citenamefont {Rowell}\ \emph {et~al.}(2009)\citenamefont {Rowell},
  \citenamefont {Stong},\ and\ \citenamefont {Wang}}]{Rowell2009}%
  \BibitemOpen
  \bibfield  {author} {\bibinfo {author} {\bibfnamefont {Eric}\ \bibnamefont
  {Rowell}}, \bibinfo {author} {\bibfnamefont {Richard}\ \bibnamefont {Stong}},
  \ and\ \bibinfo {author} {\bibfnamefont {Zhenghan}\ \bibnamefont {Wang}},\
  }\bibfield  {title} {\enquote {\bibinfo {title} {On classification of modular
  tensor categories},}\ }\href {\doibase 10.1007/s00220-009-0908-z} {\bibfield
  {journal} {\bibinfo  {journal} {Comm. Math. Phys.}\ }\textbf {\bibinfo
  {volume} {292}},\ \bibinfo {pages} {343--389} (\bibinfo {year} {2009})},\
  \Eprint {http://arxiv.org/abs/0712.1377} {arXiv:0712.1377} \BibitemShut
  {NoStop}%
\bibitem [{\citenamefont {Birman}(1969)}]{Birman1969}%
  \BibitemOpen
  \bibfield  {author} {\bibinfo {author} {\bibfnamefont {Joan~S.}\ \bibnamefont
  {Birman}},\ }\bibfield  {title} {\enquote {\bibinfo {title} {On braid
  groups},}\ }\href {\doibase 10.1002/cpa.3160220104} {\bibfield  {journal}
  {\bibinfo  {journal} {Comm. Pure App. Math.}\ }\textbf {\bibinfo {volume}
  {22}},\ \bibinfo {pages} {41--72} (\bibinfo {year} {1969})}\BibitemShut
  {NoStop}%
\bibitem [{\citenamefont {Pfeifer}\ \emph {et~al.}(2012)\citenamefont
  {Pfeifer}, \citenamefont {Buerschaper}, \citenamefont {Trebst}, \citenamefont
  {Ludwig}, \citenamefont {Troyer},\ and\ \citenamefont {Vidal}}]{Pfeifer2012}%
  \BibitemOpen
  \bibfield  {author} {\bibinfo {author} {\bibfnamefont {Robert N.~C.}\
  \bibnamefont {Pfeifer}}, \bibinfo {author} {\bibfnamefont {Oliver}\
  \bibnamefont {Buerschaper}}, \bibinfo {author} {\bibfnamefont {Simon}\
  \bibnamefont {Trebst}}, \bibinfo {author} {\bibfnamefont {Andreas W.~W.}\
  \bibnamefont {Ludwig}}, \bibinfo {author} {\bibfnamefont {Matthias}\
  \bibnamefont {Troyer}}, \ and\ \bibinfo {author} {\bibfnamefont {Guifre}\
  \bibnamefont {Vidal}},\ }\bibfield  {title} {\enquote {\bibinfo {title}
  {Translation invariance, topology, and protection of criticality in chains of
  interacting anyons},}\ }\href {\doibase 10.1103/PhysRevB.86.155111}
  {\bibfield  {journal} {\bibinfo  {journal} {Phys. Rev. B}\ }\textbf {\bibinfo
  {volume} {86}},\ \bibinfo {pages} {155111} (\bibinfo {year} {2012})},\
  \Eprint {http://arxiv.org/abs/1005.5486} {arXiv:1005.5486} \BibitemShut
  {NoStop}%
\bibitem [{\citenamefont {Hatsugai}\ \emph {et~al.}(1992)\citenamefont
  {Hatsugai}, \citenamefont {Kohmoto},\ and\ \citenamefont
  {Wu}}]{Hatsugai1992}%
  \BibitemOpen
  \bibfield  {author} {\bibinfo {author} {\bibfnamefont {Yasuhiro}\
  \bibnamefont {Hatsugai}}, \bibinfo {author} {\bibfnamefont {Mahito}\
  \bibnamefont {Kohmoto}}, \ and\ \bibinfo {author} {\bibfnamefont {Yong-Shi}\
  \bibnamefont {Wu}},\ }\bibfield  {title} {\enquote {\bibinfo {title} {Braid
  groups, anyons and gauge invariance: On topologically nontrivial surfaces},}\
  }\href {\doibase 10.1143/PTPS.107.101} {\bibfield  {journal} {\bibinfo
  {journal} {Prog. Theor. Phys. Supp.}\ }\textbf {\bibinfo {volume} {107}},\
  \bibinfo {pages} {101--119} (\bibinfo {year} {1992})}\BibitemShut {NoStop}%
\bibitem [{\citenamefont {Oshikawa}\ \emph {et~al.}(2007)\citenamefont
  {Oshikawa}, \citenamefont {Kim}, \citenamefont {Shtengel}, \citenamefont
  {Nayak},\ and\ \citenamefont {Tewari}}]{Oshikawa2007}%
  \BibitemOpen
  \bibfield  {author} {\bibinfo {author} {\bibfnamefont {Masaki}\ \bibnamefont
  {Oshikawa}}, \bibinfo {author} {\bibfnamefont {Yong~Baek}\ \bibnamefont
  {Kim}}, \bibinfo {author} {\bibfnamefont {Kirill}\ \bibnamefont {Shtengel}},
  \bibinfo {author} {\bibfnamefont {Chetan}\ \bibnamefont {Nayak}}, \ and\
  \bibinfo {author} {\bibfnamefont {Sumanta}\ \bibnamefont {Tewari}},\
  }\bibfield  {title} {\enquote {\bibinfo {title} {Topological degeneracy of
  non-{A}belian states for dummies},}\ }\href {\doibase
  10.1016/j.aop.2006.08.001} {\bibfield  {journal} {\bibinfo  {journal} {Ann.
  Phys}\ }\textbf {\bibinfo {volume} {322}},\ \bibinfo {pages} {1477--1498}
  (\bibinfo {year} {2007})},\ \Eprint {http://arxiv.org/abs/cond-mat/0607743}
  {arXiv:cond-mat/0607743} \BibitemShut {NoStop}%
\bibitem [{\citenamefont {Fan}\ and\ \citenamefont {de~Garis}(2010)}]{Fan2010}%
  \BibitemOpen
  \bibfield  {author} {\bibinfo {author} {\bibfnamefont {Z.}~\bibnamefont
  {Fan}}\ and\ \bibinfo {author} {\bibfnamefont {H.}~\bibnamefont {de~Garis}},\
  }\bibfield  {title} {\enquote {\bibinfo {title} {Braid matrices and quantum
  gates for {I}sing anyons topological quantum computation},}\ }\href {\doibase
  10.1140/epjb/e2010-00087-4} {\bibfield  {journal} {\bibinfo  {journal} {Euro.
  Phys. J. B}\ }\textbf {\bibinfo {volume} {74}},\ \bibinfo {pages} {419--427}
  (\bibinfo {year} {2010})},\ \Eprint {http://arxiv.org/abs/1003.1253}
  {arXiv:1003.1253} \BibitemShut {NoStop}%
\bibitem [{\citenamefont {Ahlbrecht}\ \emph {et~al.}(2009)\citenamefont
  {Ahlbrecht}, \citenamefont {Georgiev},\ and\ \citenamefont
  {Werner}}]{Ahlbrecht2009}%
  \BibitemOpen
  \bibfield  {author} {\bibinfo {author} {\bibfnamefont {Andre}\ \bibnamefont
  {Ahlbrecht}}, \bibinfo {author} {\bibfnamefont {Lachezar~S.}\ \bibnamefont
  {Georgiev}}, \ and\ \bibinfo {author} {\bibfnamefont {Reinhard~F.}\
  \bibnamefont {Werner}},\ }\bibfield  {title} {\enquote {\bibinfo {title}
  {{Implementation of Clifford gates in the Ising-anyon topological quantum
  computer}},}\ }\href {\doibase 10.1103/PhysRevA.79.032311} {\bibfield
  {journal} {\bibinfo  {journal} {Phys. Rev. A}\ }\textbf {\bibinfo {volume}
  {79}},\ \bibinfo {pages} {032311} (\bibinfo {year} {2009})},\ \Eprint
  {http://arxiv.org/abs/0812.2338} {arXiv:0812.2338} \BibitemShut {NoStop}%
\bibitem [{\citenamefont {Bravyi}\ and\ \citenamefont
  {Kitaev}(1998)}]{Bravyi1998}%
  \BibitemOpen
  \bibfield  {author} {\bibinfo {author} {\bibfnamefont {S.~B.}\ \bibnamefont
  {Bravyi}}\ and\ \bibinfo {author} {\bibfnamefont {A.~Yu.}\ \bibnamefont
  {Kitaev}},\ }\bibfield  {title} {\enquote {\bibinfo {title} {Quantum codes on
  a lattice with boundary},}\ }\href {http://arxiv.org/abs/quant-ph/9811052} {\
   (\bibinfo {year} {1998})},\ \Eprint {http://arxiv.org/abs/quant-ph/9811052}
  {quant-ph/9811052} \BibitemShut {NoStop}%
\bibitem [{\citenamefont {Levin}(2013)}]{Levin2013}%
  \BibitemOpen
  \bibfield  {author} {\bibinfo {author} {\bibfnamefont {Michael}\ \bibnamefont
  {Levin}},\ }\bibfield  {title} {\enquote {\bibinfo {title} {Protected edge
  modes without symmetry},}\ }\href {\doibase 10.1103/PhysRevX.3.021009}
  {\bibfield  {journal} {\bibinfo  {journal} {Phys. Rev. X}\ }\textbf {\bibinfo
  {volume} {3}},\ \bibinfo {pages} {021009} (\bibinfo {year} {2013})},\ \Eprint
  {http://arxiv.org/abs/1301.7355} {arXiv:1301.7355} \BibitemShut {NoStop}%
\bibitem [{\citenamefont {Barkeshli}\ \emph {et~al.}(2013)\citenamefont
  {Barkeshli}, \citenamefont {Jian},\ and\ \citenamefont {Qi}}]{Barkeshli2013}%
  \BibitemOpen
  \bibfield  {author} {\bibinfo {author} {\bibfnamefont {Maissam}\ \bibnamefont
  {Barkeshli}}, \bibinfo {author} {\bibfnamefont {Chao-Ming}\ \bibnamefont
  {Jian}}, \ and\ \bibinfo {author} {\bibfnamefont {Xiao-Liang}\ \bibnamefont
  {Qi}},\ }\bibfield  {title} {\enquote {\bibinfo {title} {Classification of
  topological defects in {A}belian topological states},}\ }\href 
  {\doibase 10.1103/PhysRevB.88.241103 } {\bibfield {journal} {\bibinfo{journal}
  {Phys. Rev. B}\ } \textbf {\bibinfo {volume}{88}},\ \bibinfo{pages}{241103} 
  (\bibinfo {year} {2013})},\ \Eprint {http://arxiv.org/abs/1304.7579}
  {arXiv:1304.7579} \BibitemShut {NoStop}%
\bibitem [{\citenamefont {Gottesman}(1999{\natexlab{a}})}]{Gottesman1998}%
  \BibitemOpen
  \bibfield  {author} {\bibinfo {author} {\bibfnamefont {Daniel}\ \bibnamefont
  {Gottesman}},\ }\bibfield  {title} {\enquote {\bibinfo {title} {The
  {H}eisenberg representation of quantum computers},}\ }in\ \href@noop {}
  {\emph {\bibinfo {booktitle} {{Group22: Proceedings of the XXII International
  Colloquium on Group Theoretical Methods in Physics}}}},\ \bibinfo {editor}
  {edited by\ \bibinfo {editor} {\bibfnamefont {S.~P.}\ \bibnamefont {Corney}},
  \bibinfo {editor} {\bibfnamefont {R.}~\bibnamefont {Delbourgo}}, \ and\
  \bibinfo {editor} {\bibfnamefont {P.~D.}\ \bibnamefont {Jarvis}}}\ (\bibinfo
  {publisher} {International Press},\ \bibinfo {address} {Cambridge, MA},\
  \bibinfo {year} {1999})\ pp.\ \bibinfo {pages} {32--43},\ \Eprint
  {http://arxiv.org/abs/quant-ph/9807006} {arXiv:quant-ph/9807006} \BibitemShut
  {NoStop}%
\bibitem [{\citenamefont {Gottesman}(1999{\natexlab{b}})}]{Gottesman1999a}%
  \BibitemOpen
  \bibfield  {author} {\bibinfo {author} {\bibfnamefont {Daniel}\ \bibnamefont
  {Gottesman}},\ }\bibfield  {title} {\enquote {\bibinfo {title}
  {Fault-tolerant quantum computation with higher-dimensional systems},}\
  }\href {\doibase 10.1016/S0960-0779(98)00218-5} {\bibfield  {journal}
  {\bibinfo  {journal} {Chaos, Solitons and Fractals}\ }\textbf {\bibinfo
  {volume} {10}},\ \bibinfo {pages} {1749} (\bibinfo {year}
  {1999}{\natexlab{b}})},\ \Eprint {http://arxiv.org/abs/quant-ph/9802007}
  {arXiv:quant-ph/9802007} \BibitemShut {NoStop}%
\bibitem [{\citenamefont {Van Den~Nest}(2013)}]{Van-den-Nest2012}%
  \BibitemOpen
  \bibfield  {author} {\bibinfo {author} {\bibfnamefont {Maarten}\ \bibnamefont
  {Van Den~Nest}},\ }\bibfield  {title} {\enquote {\bibinfo {title} {Efficient
  classical simulations of quantum fourier transforms and normalizer circuits
  over abelian groups},}\ }\href@noop {} {\bibfield  {journal} {\bibinfo
  {journal} {Quantum Information \& Computation}\ }\textbf {\bibinfo {volume}
  {13}},\ \bibinfo {pages} {1007--1037} (\bibinfo {year} {2013})},\
  \Eprint {http://arxiv.org/abs/1201.4867} {arXiv:1201.4867} \BibitemShut
  {NoStop}%
\bibitem [{\citenamefont {Bermejo-Vega}\ and\ \citenamefont {Van~den
  Nest}(2014)}]{Bermejo-Vega2012}%
  \BibitemOpen
  \bibfield  {author} {\bibinfo {author} {\bibfnamefont {Juan}\ \bibnamefont
  {Bermejo-Vega}}\ and\ \bibinfo {author} {\bibfnamefont {Maarten}\
  \bibnamefont {Van~den Nest}},\ }\bibfield  {title} {\enquote {\bibinfo
  {title} {Classical simulations of abelian-group normalizer circuits with
  intermediate measurements},}\ }\href@noop {} {\bibfield  {journal} {\bibinfo
  {journal} {Quantum Information \& Computation}\ }\textbf {\bibinfo {volume}
  {14}},\ \bibinfo {pages} {181--216} (\bibinfo {year} {2014})},\ \Eprint {http://arxiv.org/abs/1210.3637} {arXiv:1210.3637}
  \BibitemShut {NoStop}%
\bibitem [{\citenamefont {Blume-Kohout}\ \emph {et~al.}(2008)\citenamefont
  {Blume-Kohout}, \citenamefont {Ng}, \citenamefont {Poulin},\ and\
  \citenamefont {Viola}}]{Blume-Kohout2008}%
  \BibitemOpen
  \bibfield  {author} {\bibinfo {author} {\bibfnamefont {Robin}\ \bibnamefont
  {Blume-Kohout}}, \bibinfo {author} {\bibfnamefont {Hui~Khoon}\ \bibnamefont
  {Ng}}, \bibinfo {author} {\bibfnamefont {David}\ \bibnamefont {Poulin}}, \
  and\ \bibinfo {author} {\bibfnamefont {Lorenza}\ \bibnamefont {Viola}},\
  }\bibfield  {title} {\enquote {\bibinfo {title} {Characterizing the structure
  of preserved information in quantum processes},}\ }\href {\doibase
  10.1103/PhysRevLett.100.030501} {\bibfield  {journal} {\bibinfo  {journal}
  {Phys. Rev. Lett.}\ }\textbf {\bibinfo {volume} {100}},\ \bibinfo {pages}
  {030501} (\bibinfo {year} {2008})},\ \Eprint {http://arxiv.org/abs/0705.4282}
  {arXiv:0705.4282} \BibitemShut {NoStop}%
\bibitem [{\citenamefont {Blume-Kohout}\ \emph {et~al.}(2010)\citenamefont
  {Blume-Kohout}, \citenamefont {Ng}, \citenamefont {Poulin},\ and\
  \citenamefont {Viola}}]{Blume-Kohout2010b}%
  \BibitemOpen
  \bibfield  {author} {\bibinfo {author} {\bibfnamefont {Robin}\ \bibnamefont
  {Blume-Kohout}}, \bibinfo {author} {\bibfnamefont {Hui~Khoon}\ \bibnamefont
  {Ng}}, \bibinfo {author} {\bibfnamefont {David}\ \bibnamefont {Poulin}}, \
  and\ \bibinfo {author} {\bibfnamefont {Lorenza}\ \bibnamefont {Viola}},\
  }\bibfield  {title} {\enquote {\bibinfo {title} {Information-preserving
  structures: A general framework for quantum zero-error information},}\ }\href
  {\doibase 10.1103/PhysRevA.82.062306} {\bibfield  {journal} {\bibinfo
  {journal} {Phys. Rev. A}\ }\textbf {\bibinfo {volume} {82}},\ \bibinfo
  {pages} {062306} (\bibinfo {year} {2010})},\ \Eprint
  {http://arxiv.org/abs/1006.1358} {arXiv:1006.1358} \BibitemShut {NoStop}%
\bibitem [{\citenamefont {Wootton}(2013)}]{Wootton2013b}%
  \BibitemOpen
  \bibfield  {author} {\bibinfo {author} {\bibfnamefont {James~R}\ \bibnamefont
  {Wootton}},\ }\bibfield  {title} {\enquote {\bibinfo {title} {A simple
  decoder for topological codes},}\ }\href@noop {} {\  (\bibinfo {year}
  {2013})},\ \Eprint {http://arxiv.org/abs/1310.2393} {arXiv:1310.2393}
  \BibitemShut {NoStop}%
\bibitem [{\citenamefont {H.~Anwar}\ and\ \citenamefont
  {Browne}()}]{AnwarInPrep}%
  \BibitemOpen
  \bibfield  {author} {\bibinfo {author} {
  \bibnamefont {H.~Anwar}, \bibfnamefont {B.~Brown} \bibfnamefont 
  {E.~Campbell}}\ and\ \bibinfo {author}
  {\bibfnamefont {D.}~\bibnamefont {Browne}},\ } \bibfield {title}{\enquote {\bibinfo {title}
  {Fast decoders for qudit topological codes},}\ }\href 
  {\doibase 10.1088/1367-2630/16/6/063038} {\bibfield {journal} {\bibinfo
  {journal} {New J. Phys.}\ } \textbf {\bibinfo {volume}{16}},\ \bibinfo{pages}{063038} 
  (\bibinfo{year}{2014})} \BibitemShut {NoStop}%
\bibitem [{\citenamefont {Edmonds}(1965)}]{Edmonds1965}%
  \BibitemOpen
\bibfield  {journal} {  }\bibfield  {author} {\bibinfo {author} {\bibfnamefont
  {Jack}\ \bibnamefont {Edmonds}},\ }\bibfield  {title} {\enquote {\bibinfo
  {title} {Paths, trees, and flowers},}\ }\href {\doibase
  10.4153/CJM-1965-045-4} {\bibfield  {journal} {\bibinfo  {journal} {Canad. J.
  Math.}\ }\textbf {\bibinfo {volume} {17}},\ \bibinfo {pages} {449--467}
  (\bibinfo {year} {1965})}\BibitemShut {NoStop}%
\bibitem [{\citenamefont {Micali}\ and\ \citenamefont
  {Vazirani}(1980)}]{Micali1980}%
  \BibitemOpen
  \bibfield  {author} {\bibinfo {author} {\bibfnamefont {S.}~\bibnamefont
  {Micali}}\ and\ \bibinfo {author} {\bibfnamefont {V.}~\bibnamefont
  {Vazirani}},\ }\bibfield  {title} {\enquote {\bibinfo {title} {An
  {$O(\sqrt{|V|} |E|)$} algorithm for finding maximum matching in general
  graphs},}\ }in\ \href {\doibase 10.1109/SFCS.1980.12} {\emph {\bibinfo
  {booktitle} {Proc. 21st Ann. Symp. Found. Comp. Sci.}}}\ (\bibinfo
  {publisher} {IEEE},\ \bibinfo {year} {1980})\ pp.\ \bibinfo {pages}
  {17--27}\BibitemShut {NoStop}%
\bibitem [{\citenamefont {Karp}(1972)}]{Karp1972}%
  \BibitemOpen
  \bibfield  {author} {\bibinfo {author} {\bibfnamefont {Richard~M.}\
  \bibnamefont {Karp}},\ }\bibfield  {title} {\enquote {\bibinfo {title}
  {Reducibility among combinatorial problems},}\ }in\ \href {\doibase
  10.1007/978-1-4684-2001-2_9} {\emph {\bibinfo {booktitle} {Complexity of
  Computer Computations, Proc. Sympos. IBM Thomas J. Watson Res. Center}}}\
  (\bibinfo  {publisher} {Plenum},\ \bibinfo {address} {New York},\ \bibinfo
  {year} {1972})\ pp.\ \bibinfo {pages} {85--103}\BibitemShut {NoStop}%
\bibitem [{\citenamefont {Wang}\ \emph
  {et~al.}(2010{\natexlab{b}})\citenamefont {Wang}, \citenamefont {Fowler},
  \citenamefont {Hill},\ and\ \citenamefont {Hollenberg}}]{Wang2010a}%
  \BibitemOpen
  \bibfield  {author} {\bibinfo {author} {\bibfnamefont {David~S}\ \bibnamefont
  {Wang}}, \bibinfo {author} {\bibfnamefont {Austin~G}\ \bibnamefont {Fowler}},
  \bibinfo {author} {\bibfnamefont {Charles~D}\ \bibnamefont {Hill}}, \ and\
  \bibinfo {author} {\bibfnamefont {Lloyd~CL}\ \bibnamefont {Hollenberg}},\
  }\bibfield  {title} {\enquote {\bibinfo {title} {Graphical algorithms and
  threshold error rates for the 2d color code},}\ }\href@noop {} {\bibfield
  {journal} {\bibinfo  {journal} {Quantum Information \& Computation}\ }\textbf
  {\bibinfo {volume} {10}},\ \bibinfo {pages} {780--802} (\bibinfo {year}
  {2010}{\natexlab{b}})},\ \Eprint {http://arxiv.org/abs/0907.1708}
  {arXiv:0907.1708} \BibitemShut {NoStop}%
\bibitem [{\citenamefont {Duclos-Cianci}\ and\ \citenamefont
  {Poulin}(2010{\natexlab{a}})}]{Duclos-Cianci2010}%
  \BibitemOpen
  \bibfield  {author} {\bibinfo {author} {\bibfnamefont {Guillaume}\
  \bibnamefont {Duclos-Cianci}}\ and\ \bibinfo {author} {\bibfnamefont {David}\
  \bibnamefont {Poulin}},\ }\bibfield  {title} {\enquote {\bibinfo {title}
  {Fast decoders for topological quantum codes},}\ }\href {\doibase
  10.1103/PhysRevLett.104.050504} {\bibfield  {journal} {\bibinfo  {journal}
  {Phys. Rev. Lett.}\ }\textbf {\bibinfo {volume} {104}},\ \bibinfo {pages}
  {050504} (\bibinfo {year} {2010}{\natexlab{a}})},\ \Eprint
  {http://arxiv.org/abs/0911.0581} {arXiv:0911.0581} \BibitemShut {NoStop}%
\bibitem [{\citenamefont {Fowler}\ \emph {et~al.}(2012)\citenamefont {Fowler},
  \citenamefont {Whiteside},\ and\ \citenamefont {Hollenberg}}]{Fowler2012}%
  \BibitemOpen
  \bibfield  {author} {\bibinfo {author} {\bibfnamefont {Austin~G.}\
  \bibnamefont {Fowler}}, \bibinfo {author} {\bibfnamefont {Adam~C.}\
  \bibnamefont {Whiteside}}, \ and\ \bibinfo {author} {\bibfnamefont {Lloyd
  C.~L.}\ \bibnamefont {Hollenberg}},\ }\bibfield  {title} {\enquote {\bibinfo
  {title} {Towards practical classical processing for the surface code},}\
  }\href {\doibase 10.1103/PhysRevLett.108.180501} {\bibfield  {journal}
  {\bibinfo  {journal} {Phys. Rev. Lett.}\ }\textbf {\bibinfo {volume} {108}},\
  \bibinfo {pages} {180501} (\bibinfo {year} {2012})},\ \Eprint
  {http://arxiv.org/abs/1110.5133} {arXiv:1110.5133} \BibitemShut {NoStop}%
\bibitem [{\citenamefont {Kolmogorov}(2009)}]{Kolmogorov2009}%
  \BibitemOpen
  \bibfield  {author} {\bibinfo {author} {\bibfnamefont {Vladimir}\
  \bibnamefont {Kolmogorov}},\ }\bibfield  {title} {\enquote {\bibinfo {title}
  {Blossom {V}: a new implementation of a minimum cost perfect matching
  algorithm},}\ }\href {\doibase 10.1007/s12532-009-0002-8} {\bibfield
  {journal} {\bibinfo  {journal} {Mathematical Programming Computation}\
  }\textbf {\bibinfo {volume} {1}},\ \bibinfo {pages} {43--67} (\bibinfo {year}
  {2009})}\BibitemShut {NoStop}%
\bibitem [{\citenamefont {Duclos-Cianci}\ and\ \citenamefont
  {Poulin}(2010{\natexlab{b}})}]{Duclos-Cianci2010a}%
  \BibitemOpen
  \bibfield  {author} {\bibinfo {author} {\bibfnamefont {G.}~\bibnamefont
  {Duclos-Cianci}}\ and\ \bibinfo {author} {\bibfnamefont {D.}~\bibnamefont
  {Poulin}},\ }\bibfield  {title} {\enquote {\bibinfo {title} {A
  renormalization group decoding algorithm for topological quantum codes},}\
  }in\ \href {\doibase 10.1109/CIG.2010.5592866} {\emph {\bibinfo {booktitle}
  {Information Theory Workshop (ITW), 2010 IEEE}}}\ (\bibinfo {year} {2010})\
  pp.\ \bibinfo {pages} {1--5},\ \Eprint {http://arxiv.org/abs/1006.1362}
  {arXiv:1006.1362} \BibitemShut {NoStop}%
\bibitem [{\citenamefont {Duclos-Cianci}\ and\ \citenamefont
  {Poulin}(2013)}]{Duclos-Cianci2013}%
  \BibitemOpen
  \bibfield  {author} {\bibinfo {author} {\bibfnamefont {Guillaume}\
  \bibnamefont {Duclos-Cianci}}\ and\ \bibinfo {author} {\bibfnamefont {David}\
  \bibnamefont {Poulin}},\ }\bibfield  {title} {\enquote {\bibinfo {title}
  {Fault-tolerant renormalization group decoder for {A}belian topological
  codes},}\ } {\bibfield {journal} {\bibinfo {journal}{Quantum Inf. Comput.} \textbf
  {\bibinfo {volume}{14}},\ \bibinfo{pages}{0721--0740} (\bibinfo  {year}{2014})}},
  \href {http://arxiv.org/abs/1304.6100}\ \Eprint {http://arxiv.org/abs/1304.6100} {arXiv:1304.6100}
  \BibitemShut {NoStop}%
\bibitem [{\citenamefont {Hastings}\ \emph {et~al.}(2013)\citenamefont
  {Hastings}, \citenamefont {Watson},\ and\ \citenamefont
  {Melko}}]{Hastings2013}%
  \BibitemOpen
  \bibfield  {author} {\bibinfo {author} {\bibfnamefont {Matthew~B}\
  \bibnamefont {Hastings}}, \bibinfo {author} {\bibfnamefont {Grant~H}\
  \bibnamefont {Watson}}, \ and\ \bibinfo {author} {\bibfnamefont {Roger~G}\
  \bibnamefont {Melko}},\ }\bibfield  {title} {\enquote {\bibinfo {title}
  {Self-correcting quantum memories beyond the percolation threshold},}\
  }\href{ \doibase 10.1103/PhysRevLett.112.070501} {\bibfield {journal} {\bibinfo
  {journal}{Phys. Rev. Lett.}\ } \textbf {\bibinfo {volume}{112}},\ \bibinfo {pages}
  {070501}  (\bibinfo {year}{2014})}, \Eprint
  {http://arxiv.org/abs/1309.2680} {arXiv:1309.2680} \BibitemShut {NoStop}%
\bibitem [{\citenamefont {Wootton}\ \emph {et~al.}(2013)\citenamefont
  {Wootton}, \citenamefont {Burri}, \citenamefont {Iblisdir},\ and\
  \citenamefont {Loss}}]{Wootton2013}%
  \BibitemOpen
  \bibfield  {author} {\bibinfo {author} {\bibfnamefont {James~R}\ \bibnamefont
  {Wootton}}, \bibinfo {author} {\bibfnamefont {Jan}\ \bibnamefont {Burri}},
  \bibinfo {author} {\bibfnamefont {Sofyan}\ \bibnamefont {Iblisdir}}, \ and\
  \bibinfo {author} {\bibfnamefont {Daniel}\ \bibnamefont {Loss}},\ }\bibfield
  {title} {\enquote {\bibinfo {title} {Decoding non-{A}belian topological
  quantum memories},}\ }\href {\doibase 10.1103/PhysRevX.4.011051} 
  {\bibfield {journal}{\bibinfo {journal}{Phys. Rev. X}\ } \textbf {\bibinfo {volume}{4}}, 
  \ \bibinfo {pages}{011051} (\bibinfo {year}{2014})}, \Eprint
  {http://arxiv.org/abs/1310.3846} {arXiv:1310.3846} \BibitemShut {NoStop}%
\end{thebibliography}
\end{document}